\documentclass[twocolumn]{aastex62}
\usepackage {graphicx}
\usepackage[caption=false]{subfig}
\usepackage{amsmath,amsfonts,amsthm,bm}
\usepackage[thinc]{esdiff}
\usepackage{hyperref}
\usepackage{relsize}
\usepackage{mathrsfs}
\usepackage{xcolor}
\usepackage{natbib}
\usepackage{needspace}
\usepackage{tabularx}
\usepackage{float}
\usepackage{booktabs}
\usepackage{verbatim}
\usepackage{tabularx}
\definecolor{medium-blue}{rgb}{0,0,1}
\hypersetup{colorlinks, urlcolor={medium-blue}}

\graphicspath{{pics/}}

\newcommand{\Kepler}{\textit{Kepler}}
\newcommand{\TESS}{\textit{TESS}}
\newcommand{\Gaia}{\textit{Gaia}}

\shorttitle{Alignment of Wide Binaries With Planets}
\shortauthors{Christian et al.}

\begin{document}

%

\title{A Possible Alignment Between the Orbits of Planetary Systems and their Visual Binary Companions}

\correspondingauthor{Sam Christian}
\email{samchristian@mit.edu}


\author[0000-0003-0046-2494]{Sam Christian}
\affiliation{Department of Physics and Kavli Institute for Astrophysics and Space Research, Massachusetts Institute of Technology, Cambridge, MA 02139, USA}
\affiliation{Liberal Arts and Science Academy, Austin, Texas 78724, USA}
\author[0000-0001-7246-5438]{Andrew Vanderburg}
\affiliation{Department of Physics and Kavli Institute for Astrophysics and Space Research, Massachusetts Institute of Technology, Cambridge, MA 02139, USA}
\affiliation{Department of Astronomy, University of Wisconsin-Madison, Madison, WI 53706, USA}
\author[0000-0002-7733-4522]{Juliette Becker}
\affiliation{Division of Geological and Planetary Sciences, California Institute of Technology, Pasadena, CA 91125, USA}
\affiliation{51 Pegasi b Fellow}
\author[0000-0003-4755-584X]{Daniel A. Yahalomi}
\affiliation{Department of Astronomy, Columbia University, 550 W 120th Street, New York, NY 10027, USA}
\author[0000-0003-3904-7378]{Logan Pearce}
\affiliation{Steward Observatory, University of Arizona, Tucson, AZ 85721, USA}
\affiliation{NSF Graduate Research Fellow}
\author{George Zhou}
\affiliation{Center for Astrophysics \textbar \ Harvard \& Smithsonian, 60 Garden Street, Cambridge, MA 02138, USA}
\affiliation{University of Southern Queensland, Centre for Astrophysics, Toowoomba, Queensland 4350, Australia}

\author[0000-0001-6588-9574]{Karen A.\ Collins}
\affiliation{Center for Astrophysics \textbar \ Harvard \& Smithsonian, 60 Garden Street, Cambridge, MA 02138, USA}
\author[0000-0001-9811-568X]{Adam L. Kraus}
\affiliation{Department of Astronomy, University of Texas at Austin, Austin, Texas 78712, USA}
\author[0000-0002-3481-9052]{Keivan G.\ Stassun}
\affiliation{Department of Physics and Astronomy, Vanderbilt University, 6301 Stevenson Center Ln., Nashville, TN 37235, USA}
\affiliation{Department of Physics, Fisk University, 1000 17th Avenue North, Nashville, TN 37208, USA}
\author{Zoe de Beurs}
\affiliation{Department of Physics and Kavli Institute for Astrophysics and Space Research, Massachusetts Institute of Technology, Cambridge, MA 02139, USA}
\affiliation{Department of Astronomy, University of Texas at Austin, Austin, Texas 78712, USA}
% ...



\begin{abstract}
Astronomers do not have a complete picture of the effects of wide-binary companions (semimajor axes greater than 100 AU) on the formation and evolution of exoplanets. We investigate these effects using new data from \Gaia\ EDR3 and the \TESS\ mission to characterize wide-binary systems with transiting exoplanets. We identify a sample of 67 systems of transiting exoplanet candidates (with well-determined, edge-on orbital inclinations) that reside in wide visual binary systems. We derive limits on orbital parameters for the wide-binary systems and measure the minimum difference in orbital inclination between the binary and planet orbits. We determine that there is statistically significant difference in the inclination distribution of wide-binary systems with transiting planets compared to a control sample, {with the probability that the two distributions are the same being 0.0037}. This implies that there is an overabundance of planets in binary systems whose orbits are aligned with those of the binary. The overabundance of aligned systems appears to primarily have semimajor axes less than 700 AU. We investigate some effects that could cause the alignment and conclude that a torque caused by a misaligned binary companion {on the protoplanetary disk} is the most promising explanation.

\end{abstract}

\keywords{planet-star interactions}

\section{Introduction}
\label{sec:intro}

Many stars in our galaxy reside in binary systems \citep{Rahavan2010, fischer1992, Frankowski2007}. These binary systems have semimajor axes ranging from less than 0.01 AU \citep{Dimitrov} to greater than 20,000 AU \citep{Latham1991,Esteban2019}. The extreme range in semimajor axes exhibited by binary systems makes it very challenging for any one formation mechanism to explain all observed systems; instead, there are likely multiple pathways by which binary stars may form. 

At close separations, binary stars with semimajor axes less than about 100 AU may form by disk fragmentation \citep{adams1989} and turbulent fragmentation at larger separations followed by migration \citep{Bate2018}. In disk fragmentation, instabilities in massive circumstellar disks collapse and form a second star orbiting in the plane of the disk. At larger separations, binary stars can form through turbulent fragmentation, where turbulence in the initial core leads to fragmentation of the core into an eventual wide-binary system \citep{Offner2010,Offner2016,Bate2018}. These binaries can in-turn migrate to smaller separations, thus close binaries form through a mixture of disk and turbulent fragmentation. Another viable mechanism for the formation of wide-binaries is core capture, in which initially unbounded stars form, for example, via the dissolution of open clusters \citep{Kouwenhoven2010} or from pre-stellar core capture \citep{Tokovinin2017}. 


Many binary stars are known to host exoplanets \citep{Mugrauer2019}. While some exoplanets orbit around \textit{both} stars in the binary (the so-called circumbinary system), e.g. \citealt{kepler16}), most exoplanets in binary systems orbit closely around just one of binary components (a circumstellar orbit). In wide-binary systems, it is believed that virtually all planets will be on circumstellar orbits. The effects of a wide-binary companion on a planetary system are debated. Theoretical work has shown that the dynamical influence of wide binary companions can eject planets and increase the eccentricity of planetary orbits \citep{Kaib2013,Correa2017,Akos2020}. Binary companions can affect planetary orbits via the Lidov-Kozai mechanism \citep{lidov, kozai,Vonzeipel1910}, potentially causing tidal migration of planets to tighter orbits. This mechanism could provide an explanation for the existence of hot Jupiters \citep{Wu2003,Fabrycky2007,Petrovich2015, Dawson2018, Li2020}, although this is not the only mechanism that can explain hot Jupiter orbits \citep{lin1996, Ngo2016, Batygin2016}, and at least some hot Jupiters could not have formed in this way \citep[e.g.][]{becker2015, Weiss2017, becker2017, canas2019, huang2020}. The presence of a torque from the binary companion could also misalign the protostellar disk \citep{Batygin2012, Lai2014,Hjorth2021}.

On the observational front, statistical analyses have shown that while wide binary companions with semimajor axes $\gtrsim$ 1000 AU do not seem to have a significant impact on planet occurrence \citep{deaconpanstarrs}, closer binary companions (of semimajor axes $\lesssim 100$ AU) seem to suppress planet occurrence \citep{Kraus2016,Ziegler2021}, possibly by disrupting the protoplanetary disk \citep{Duchene2010}. 

So far, most observational studies of binary companions to exoplanet hosts have focused on the effects of binary companions as a function of the projected separation, partly due to the difficulty of determining the true separation or orbital elements of binary systems. Traditionally, measuring visual binary star orbits requires repeated precise observations of the positions of the two stars over years, decades, or even centuries \citep{wds}. However, recently the extremely precise astrometry from ESA's \Gaia\ mission \citep{gaiamission} has made it possible to derive loose constraints on the orbital elements of visual binary stars using only the masses and instantaneous relative velocities of the two components \citep{Netwon2019, Pearce2020}. 




Meanwhile, the advent of exoplanet detecting space telescopes -- specifically the \Kepler\ mission \citep{koch} and the \textit{Transiting Exoplanet Survey Satellite} (\textit{TESS}, \citealt{ricker2015}) that use the transit method to detect exoplanets --  have resulted in an explosion in the numbers of planets known in visual binaries. Because the planets discovered by \Kepler\ and \TESS\ transit their host stars, we know that the planetary orbital planes are aligned to within a few degrees of our line of sight. 

In this paper, we take advantage of these new observations to study whether there is a tendency towards alignment in the orientation of the orbits of visual binary systems and the orbits of planets that reside in these systems. In particular, we measure the orbital inclination of a sample of visual binary stars in which one component is known to host a transiting exoplanet candidate. 

{It is important to note that we refer to alignment as the minimum alignment between the binary system orbit and exoplanet orbit, not the stellar rotation axis of the primary star and orbit of the exoplanet as is commonly measured using the Rossiter-McLaughlin effect. We make no assumptions on the orientation of the stellar rotation axis in our analysis.}

\begin{figure*}
    \centering
    \includegraphics[width=\textwidth]{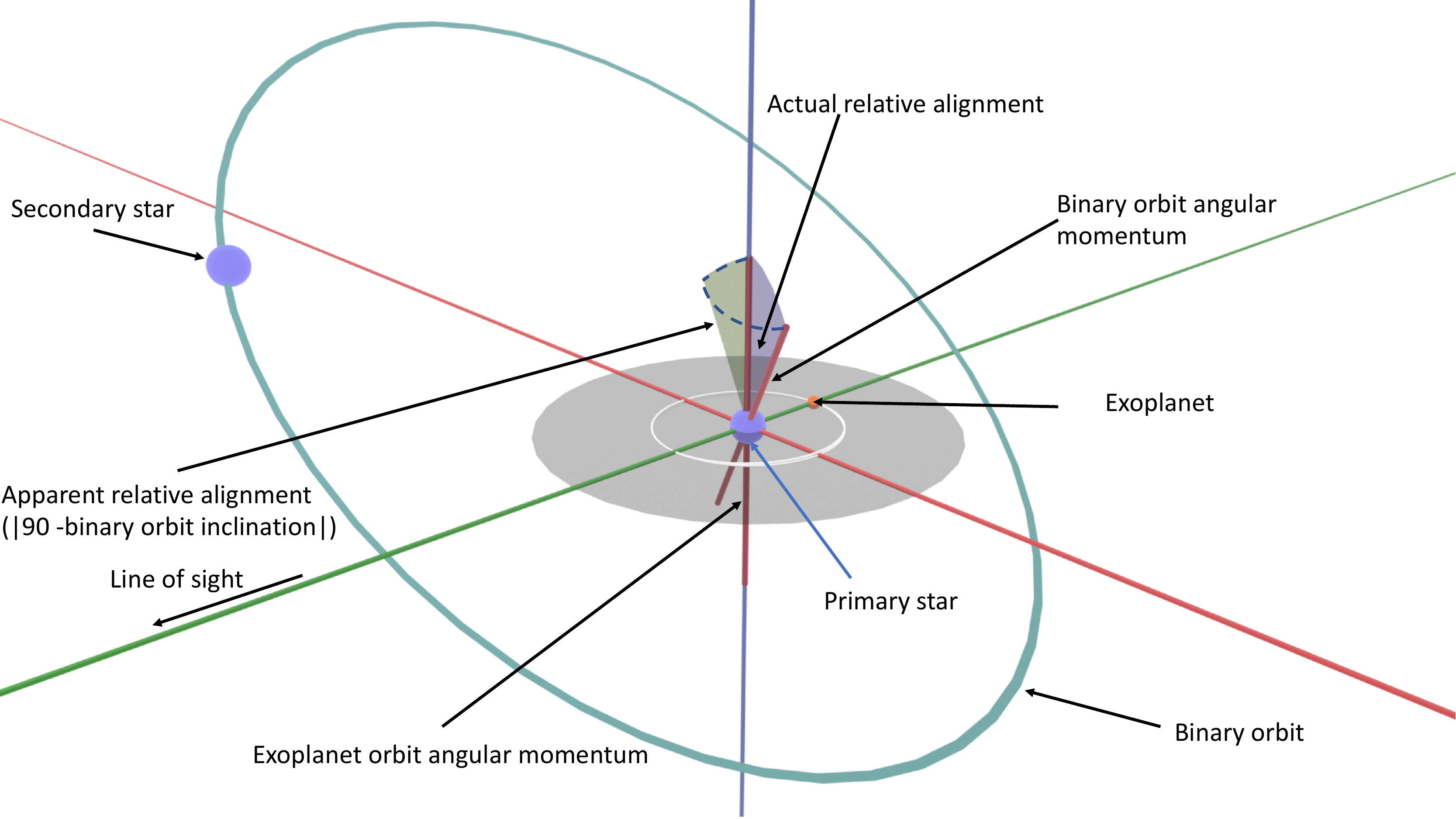}
    \caption{A diagram of the orbital configurations relevant to this paper. The diagram is centered on the primary star. Apparent relative alignment is calculated as $|90^\circ-i|$, where $i$ is inclination of the binary system (the transiting exoplanets will always have approximately 90 degrees inclination). {In the diagram, the green wedge is the inclination of the binary system.} The primary star is the star that hosts the exoplanet, while the binary companion is the companion star without detected exoplanets. The angular momentum of the star is the axis that the star rotates on. The exoplanet (orange) orbits at 90 degrees to the line of sight.}
    \label{fig:orbitalSchematic}
\end{figure*}

Because the orbital inclinations of the transiting planets must be close to 90$^\circ$, an overabundance of edge-on binary orbits implies a preferential alignment between the binary systems and their planets. The observed misalignment is really the minimum possible misalignment of the binary system. {If $\Omega_p$, $\Omega_b$, $i_p$, and $i_b$ are the longitude of ascending node} and inclination of the planet and binary respectively, then the misalignment $\Delta$ between the binary and planet can be expressed as
\begin{equation}
    \cos(\Delta)=\cos(i_p)\cos(i_b)+\sin(i_p)\sin(i_b)\cos(\Omega_p-\Omega_b)
\end{equation}
Since $i_p=90^\circ$,
\begin{equation}
    \cos(\Delta)=\cos(90-i_b)\cos(\Omega_p-\Omega_b)
\end{equation}
Thus the observed misalignment $|90-i_b|$ is only equivalent to the actual misalignment $\Delta$ if the longitude of ascending nodes of the binary system and planet happen to be the same and otherwise $\Delta$ is equivalent to the minimum misalignment between the binary system and planet. {A large observed relative inclination means a system is misaligned, while a system with small observed relative inclination could be aligned or misaligned depending on the relative (unknown) longitude of ascending nodes of the exoplanet and binary orbit. However, if many systems are observed, an overabundance of small relative inclinations has the physical interpretation that an overabundance of systems tend to be aligned since the longitude of ascending nodes of misaligned systems is expected to be distributed randomly and independently of relative inclination.} A diagram of the relevant parameters described in this paper is presented in Figure \ref{fig:orbitalSchematic}.

The paper is organized as follows. In Section \ref{sec:observations} we present the \Gaia\ EDR3, \TESS, and ground-based spectroscopic and photometric observations used in our study. In Section \ref{sec:Analysis} we describe the procedure we use to constrain the masses of the binary systems and subsequently model the orbits of the binary systems. In Section \ref{sec:Results}, we describe the statistical tests performed on the data and rule out possible biases. Section \ref{sec:dynamics} gives an analysis of two theoretical mechanisms that could possibly cause the observed alignment. In Section \ref{sec:discussion} we discuss two possible scenarios for the observed alignment and discuss future directions for our work. Finally in Section \ref{sec:conclusion} we summarize our results.

\section{Observations/Data}
\label{sec:observations}

\begin{figure}
    \centering
    \includegraphics[width=\columnwidth]{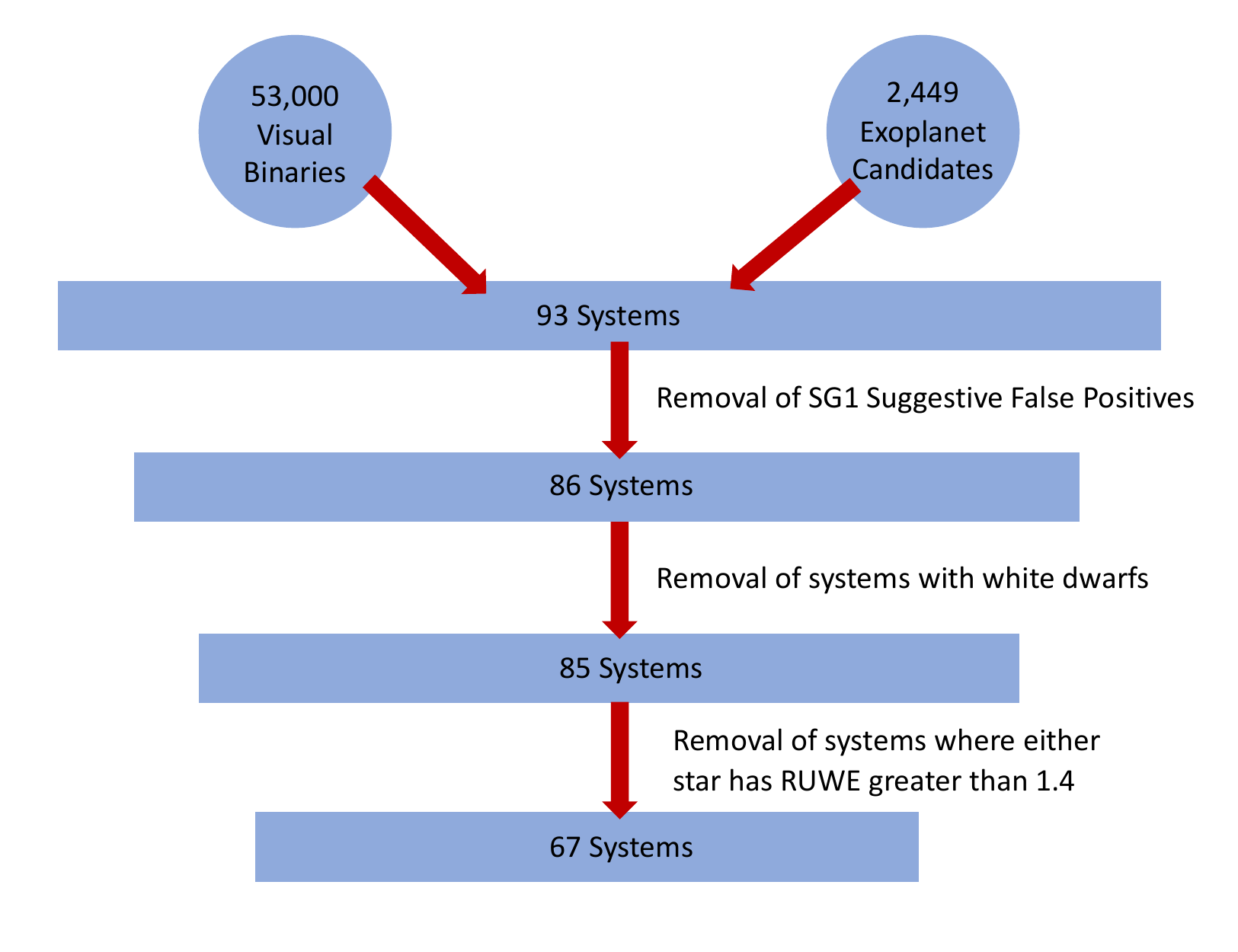}
    \caption{A hierarchy of the cuts performed on the sample of visual binaries with transiting exoplanet candidates. The same cuts were performed on the control sample. The specific cuts performed are described in detail in Section \ref{sec:observations}}
    \label{fig:cutHierarchy}
\end{figure}

To investigate whether there is a tendency toward alignment between the orbits of visual binary stars and their planetary systems, we need both a sample of likely transiting planet candidates and a constraint on the orbital inclinations of any visual binary companions to these planet host stars. For the former, we make use of planet candidates discovered by the \TESS\ mission and vetted with ground-based follow-up photometric and spectroscopic observations. For the latter, we use astrometric observations from Gaia, archival broadband photometry, and metallicity measurements from both new and archival spectra to determine the masses of the binary components using isochrone fitting. We also perform a variety of cuts on our sample of visual binaries with exoplanets and control sample. A diagram of the various cuts perform is shown in Figure \ref{fig:cutHierarchy}. We describe these inputs to our analysis and the cuts we performed in more detail in this section.


\subsection{TESS planet candidates}
\subsubsection{Identification with TESS}
\label{sec:Planets}
We start with the list of planet candidates reported by the \TESS\ mission (also known as TESS Objects of Interest, or TOIs). \TESS\ uses four 10-cm cameras to repeatedly image 96$^\circ$ by 24$^\circ$ regions of the sky for 28 days at a time. After the completion of each 28-day observation (called a sector), \TESS\ moves to a new field of view and repeats the process. Over the course of its two-year primary mission, \TESS\ observed approximately 70\% of the sky, and is continuing to observe in an extended mission.

The \TESS\ CCDs read out images of the sky every two seconds, but the data volume required to download each two-second image from orbit would be prohibitively large. Instead, \TESS\ co-adds the two-second images into longer observations before beaming the data back to Earth. Most of the sky is coadded to long-cadence Full Frame Images (FFIs) with exposure times of 30 minutes (in the primary mission) or 10 minutes (in the extended mission), while the pixels surrounding 20,000 pre-selected stars are coadded to two minutes and for the extended mission, 1000 of these targets are co-added to 20 seconds. 


Once the data have been received on Earth, they are analyzed as described by \citet{guerrero2021} to process the observations and identify planet candidates. We base our sample on the list of all \TESS\ planet candidates that had been reported online as of December 15th, 2020.  

\subsubsection{Sample of Visual Binaries with Planet Candidates}
We identify planet candidate hosting stars that also happen to reside in a visual binary system matching the underlying \Gaia\ DR2 ID of items in the TIC catalog to a catalog of visual binary stars identified in \Gaia\ data by \citet{elbadry2018}. This work reports approximately 53,000 visual binary systems within 200 pc of the Sun and with projected separations between 50 and 50,000 AU derived from Gaia DR2 astrometric observations. Although the catalog has binaries with separations as small as 50 AU, the vast majority of binaries in the catalog have much wider separations. At wider separations, it is more likely that the Gaia spacecraft will resolve the individual stars in the binary system. In total, after all cuts were performed, we identified a sample of 67 visual binary systems including a \TESS\ planet candidate host star with projected semimajor axes ranging from 61 to 34,700 AU and parallaxes ranging from 5 to 48 milliarcseconds.

\subsubsection{Follow-up Ground-based Time-series Photometry}

\begin{deluxetable*}{llDDc}
\label{table:facilities}
\tabletypesize{\scriptsize}
\tablecaption{Facilities used for SG1 Seeing-limited Photometric Follow-up Observations}
\tablehead{\colhead{Observatory/Telescope} & \colhead{Location} & \twocolhead{Aperture} & \twocolhead{Pixel scale} & \colhead{FOV}\\[-2mm]
 & & \twocolhead{(m)} & \twocolhead{(arcsec)} & \colhead{(arcmin$^2$)}
}
\decimals
\startdata
Acton Sky Portal (Private Observatory) & Acton, MA, USA & 0.36 & 0.69 & $17.3 \times 11.5$\\
Adams Observatory at Austin College & Sherman, Texas, USA & 0.61 & 0.38 & $26 \times 26$\\
Antarctic Search for Transiting ExoPlanets (ASTEP) & Concordia Station, Antarctica & 0.4 & 0.93 & $63 \times 63$\\
Chilean-Hungarian Automated Telescope (CHAT) & Las Campanas Observatory, Chile & 0.7 & 0.6 & $21 \times 21$ \\
Deep Sky West & Rowe, New Mexico, USA & 0.5 & 1.09 & $37 \times 37$\\
El Sauce Observatory (Evans Private Telescope) & Coquimbo, Chile & 0.36 & 1.47 & $19 \times 13$\\
Fred L. Whipple Observatory  (FLWO) & Amado, Arizona, USA & 1.2 & 0.672 & $23.1 \times 23.1$\\
George Mason University (GMU) & Fairfax, Virginia, USA & 0.8 & 0.35 & $23 \times 23$\\
Grand-Pra Observatory & Valais Sion, Switzerland & 0.4 & 0.73 & $12.9 \times 12.55$\\
Hazelwood Private Observatory & Churchill, Victoria, Australia & 0.32 & 0.55 & $20 \times 14$\\
Infrared Survey Facility (IRSF/SIRIUS) & South Africa &  1.4 &  0.45 &  $7.7 \times 7.7$\\
Las Cumbres Observatory Global Telescope (0.4m) & Spain, Australia & 0.4 &  0.571 &  $29.2 x 19.5$\\
Las Cumbres Observatory Global Telescope (1m) & Chile, South Africa, Australia, USA & 1.0 & 0.39 & $26 \times 26$ \\
Las Cumbres Observatory Global Telescope (2m/MuSCAT3) & Haleakala, Hawaii, USA & 2.0 &  0.27 &  $9.5 \times 9.5$\\
MEarth-South Observatory & La Serena, Chile & 0.4 & 0.84 & $29 \times 29$ \\
Mt. Kent Observatory (CDK700) & Toowoomba, Australia & 0.7 & 0.4 & $27 \times 27$\\
Mt. Stuart Observatory & Dunedin, New Zealand & 0.3175 & 0.88 & $44 \times 30$\\
Mt. Lemmon Observatory & Tucson, Arizona, USA & 0.61 & 0.39 & $26 \times 26$\\
Observatoire du Mont-M\'egantic (OMM) & Notre-Dame-des-Bois, Qu\'ebec, Canada & 1.6 & 0.47 & $7.95 \times 7.95$\\
Observatori Astron\`omic Albany\`a (OAA) & Albany\`a, Girona, Spain & 0.406 & 1.44 & $36 \times 36$\\
Okayama 188cm telescope (MuSCAT) & Okayama, Japan & 1.88 & 0.358 & $6.1 \times 6.1$\\
Perth Exoplanet Survey Telescope (PEST) & Perth, Australia & 0.3 & 1.2 & $31 \times 21$
\\
Kotizarovci observatory & Sarsoni, Croatia & 0.3 & 1.21 & $15 \times 10$\\
Private observatory of the Mount & Saint-Pierre-du-Mont, France & 0.20 & 0.69 & $38 \times 29$\\
Sierra Nevada Observatory & Granada, Andaluc\'ia, Spain & 1.5 & 0.232 & $7.92 \times 7.92$\\
Teide Observatory (MuSCAT2) & La Laguna, Spain & 1.52 & 0.44 & $7.4 \times 7.4 $\\
TRAPPIST-North & Oukaimeden Observatory, Morocco & 0.6 & 0.64 & $ 22 \times 22$ \\
Virtual Telescope Project & Ceccano, Italy & 0.43 & 1.2 & $16 \times 11$\\
Whitin Observatory at Wellesley College & Wellesley, MA USA & 0.7 & 0.67 & $23 \times 23$\\
\enddata
\end{deluxetable*}


We identified and removed additional false positive planet candidates using ground-based observations. The majority of these observations came from Sub-group 1 (SG1) of the TESS Follow-up Observing Program Working Group (TFOP WG), which performs seeing-limited time-series photometry of TESS Objects of Interest. The specific facilities used for follow-up observations are listed in Table~\ref{table:facilities}. SG1 observations have the primary purposes of ruling out the possibility of nearby eclipsing binaries (NEBs) as the source of the TESS detection and refining the parameters of transiting planet candidates. 

SG1 observations are used to classify TOIs into a variety of photometric dispositions, indicating whether a given candidate is a false positive, a plausible candidate, or a well-vetted likely planet. The dispositions used by SG1 are: 

\begin{itemize}
\item PC, or Planet Candidate, indicates that either no follow-up observations have been conducted, or that they are in-progress. 
\item PPC, or Promising Planet Candidate, indicates that follow-up observations have ruled out nearby eclipsing binary false positive scenarios on most stars in the field. 
\item CPC, or Cleared Planet Candidate, indicates that follow-up observations have ruled out nearby eclipsing binary false positive scenarios on \textit{all} stars in the field.
\item VPC, or Validated Planet Candidate, indicates that ground-based follow-up observations have detected the transit signal discovered by TESS, confirming that the signal is on-target and not a false alarm. 
\item KP, or Known Planet, indicates the candidate was previously identified and confirmed as a planet independently of TESS. 
\item LEPC, or Lost Ephemeris Planet Candidate, indicates that the uncertainty on predicted future transit times has grown large enough that ground-based photometric observations cannot efficiently screen for false positives. 
\item STPC, or Single Transit Planet Candidate, indicates that the orbital period of the planet is not known and therefore ground-based photometric observations cannot efficiently screen for false positives. 
\item NEB, or Nearby Eclipsing Binary, indicates the detection of a nearby eclipsing binary that is contaminating the TESS aperture.
\item PNEB indicates a Possible NEB
\item NPC, or Nearby Planet Candidate, indicates that the TESS detection was actually of a nearby star, but the TESS detection itself is not ruled out to be a false positive. However, the original TOI is retired as a false-positive in this case.
\item APC, or Ambiguous Planet Candidate, indicates that results are ambiguous, but suggest that confirming a planet candidate in the system would be difficult.
\item  BEB, or Blended Eclipsing Binary, and EB, Eclipsing Binary, indicate the presence of an eclipsing binary as the cause for the TESS detection.
\item  FA, or False Alarm, indicates an instrumental anomaly as the cause of the detection.
\end{itemize}

In this analysis, we consider any TOI with a photometric disposition of PNEB, NEB, NPC, APC, BEB, EB, and FA to be a false positive and remove them from our sample. After this false positive cut, there are 86 binary systems with exoplanets.


\subsection{Visual binaries from \Gaia}

\subsubsection{Control Sample of Visual Binaries without Planet Candidates}
We also identified a control sample of visual binary systems from the \citet{elbadry2018} catalog. \Kepler\ has taught us that most of these stars likely host planetary systems of their own \citep[e.g.][]{fressin, deaconpanstarrs}, but since they do not host any \textit{transiting} planets, their inclinations will be unknown. Therefore, performing our analysis on a control sample and comparing the results against the sample of binaries with planet candidates helps give us confidence that any features we see in the resulting distribution of inclination angles are astrophysical, and not due to selection effects. We specifically constructed our control sample to have nearly identical properties to the sample of binaries with planet candidates to make sure that our control sample incorporates any selection biases from the \TESS\ planet detection process. To achieve this goal, we defined a metric $\mathcal{M}$ to quantify the similarity between any two visual binary systems:  

\begin{multline}
      \mathcal{M} = \left(\frac{\Delta G_{1}}{4}\right)^2 + \left(\frac{\Delta G_{2}}{4}\right)^2 + \left(\frac{\Delta RP_{1}}{4}\right)^2+ \left(\frac{\Delta RP_{2}}{4}\right)^2 + \\ \left(\frac{\Delta BP_{1}}{4}\right)^2+
      \left(\frac{\Delta BP_{2}}{4}\right)^2+
      (\Delta \varpi)^2 + (\Delta s)^2
\end{multline}
 
\noindent where $s$ is the projected separation of the two stars in the binary system, $\varpi$ the system parallax, and $G$, $BP$ and $RP$ are the stars' apparent magnitudes in the three \Gaia\ passbands. Here, the $\Delta$ symbol represents the normalized fractional difference between the values for the two systems: a system with a transiting exoplanet and potential control sample system, and the subscripts 1 and 2 represent the primary and secondary star in each system. For instance, $\Delta G_1=\frac{(G_1-G_c)}{G_c}$, where $c$ represents the control sample. We arbitrarily divide all magnitude normalized differences by 4 so that not all weight is given to the magnitudes.

\begin{figure*}
    \centering
    \includegraphics[width=\textwidth]{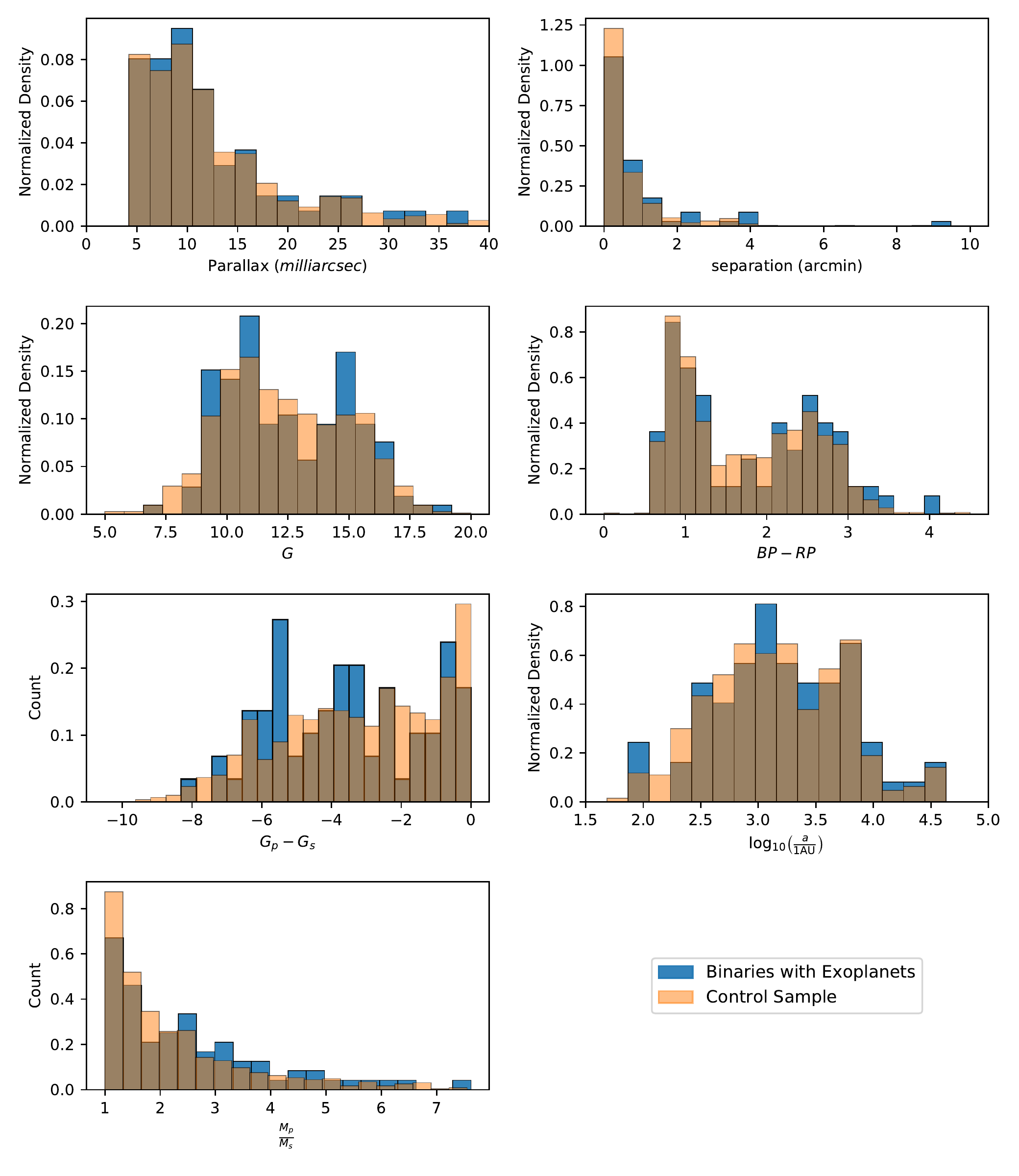}\\
    \caption{Histograms of properties of the binary systems. From left to right: Parallax in milliarcseconds, projected separation in arcminutes, apparent $G$ magnitude, $BP-RP$ color, $G_p-G_s$ where $p$ is the primary star (defined as the brighter star) and $s$ is the secondary star, $\log_{10}(a)$ where $a$ is the projected semimajor axis, and mass ratio of the primary and secondary star.}
    \label{fig:parameterDistributions}
\end{figure*}

For each of the visual binary systems with non-false positive exoplanets as of December 25th, 2020, we identified the 12 systems from the \citet{elbadry2018} catalog with the lowest $\mathcal{M}$ metric. Systems were not removed after each sampling procedure (i.e. they are allowed to be included twice), however due to the large number of systems present in the \citet{elbadry2018} catalog, the resulting control sample has no repeated systems. A subset of our planet candidate sample were also identified by \citet{elbadry2019} to have spectroscopic metallicity measurements from one of several large spectroscopic surveys (see Section \ref{sec:archivalSpec}). For these systems, we restricted our search for similar systems to those that also have an archival spectroscopic metallicity measurement from \citet{elbadry2019}. In total, we identify a control sample of 960 systems with very similar distributions of parameters to the input sample of binaries containing planet candidates. Figure \ref{fig:parameterDistributions} shows various properties of the sample with exoplanets and control sample.

\subsubsection{Astrometric Parameters}

Our analysis hinges on highly precise measurements of the positions, proper motions, and parallaxes of each star in the visual binary system. Originally we used parameters from \Gaia\ Data Release 2 (DR2, \citealt{gaiadr2, lindegren2018dr2solution}), which were based on 22 months of data. During the preparation of our manuscript, updated astrometric parameters based on 34 months of data became available in \Gaia\ Early Data Release 3 (EDR3, \citealt{gaiaedr3, lindegren2020}).  We performed our full analysis using data from both \Gaia\ DR2 and EDR3 and found consistent results between the two samples. We present the results from our analysis using the more precise \Gaia\ EDR3 data in the rest of this paper.




\subsubsection{Removing Incorrect Cross-matches, High RUWE solutions, and White Dwarfs}
We apply a variety of cuts to both the control sample and sample with exoplanets in order to ensure that only high quality astrometric parameters are preserved.

In the process of converting between Gaia DR2 and Gaia EDR3 IDs, a purely positional crossmatch can contaminate the sample due to proper motion movement from the Gaia DR2 epoch (2015.5) to the Gaia EDR3 epoch (2016) and the addition of new sources in EDR3. To ensure that there are no incorrectly crossmatched stars in our sample, we exclude 17 binary systems in the control sample for which $|G_{EDR3}-G_{DR2}|>0.05$.

RUWE (re-normalized unit weight error) can be used as an indicator of the quality of the Gaia astrometric solution for a star \citep{lindegren2018}. RUWE is the square root of the reduced $\chi^2$ divided by a correction function that eliminates dependence on $G$ magnitude and $BP-RP$ color. An RUWE of greater than 1.4 typically indicates a poor astrometric fit, so we eliminate any systems for which the RUWE for at least one of the stars is greater than 1.4. A high RUWE can indicate the presence of an unresolved companion \citep{Belokurov2020}. 

We also remove any binaries where either the host star or companion star is a white dwarf; it is more difficult to estimate masses for white dwarfs than for main sequence stars, and in these systems the binary orbit has been influenced by post-main-sequence mass loss. While these effects are very interesting in their own right, it is beyond the scope of this work to consider them.

\begin{figure}
    \centering
    \subfloat[Histogram of the exoplanets' radii in our sample. The median fractional uncertainty in radius  $\left(\displaystyle \frac{\sigma_R}{R}\right)$ is 0.16.]{\includegraphics[width=\columnwidth]{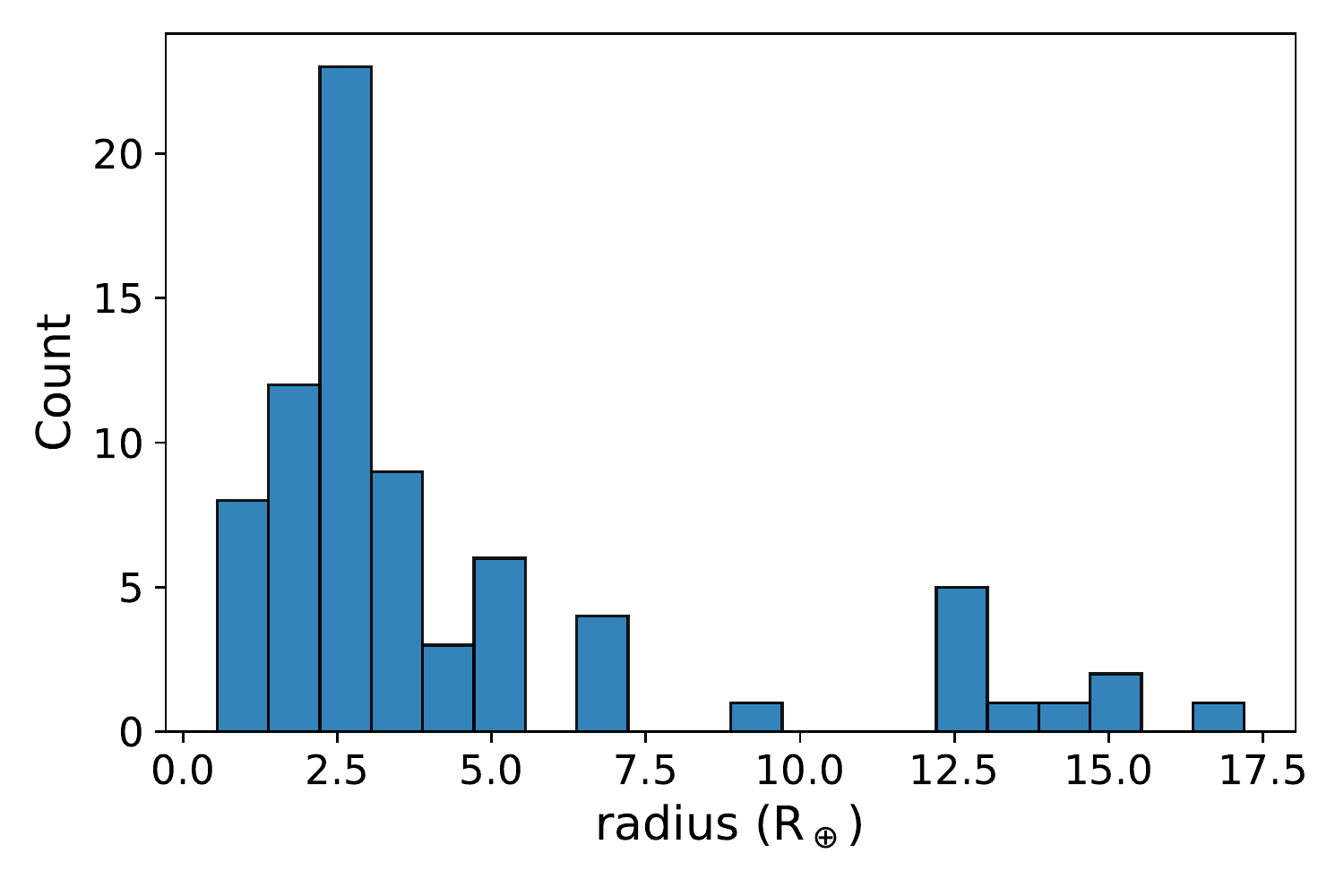}}\\
    \subfloat[Histogram of the exoplanets' periods in our sample. The error in the exoplanets' period is on average 0.004 days.]{\includegraphics[width=\columnwidth]{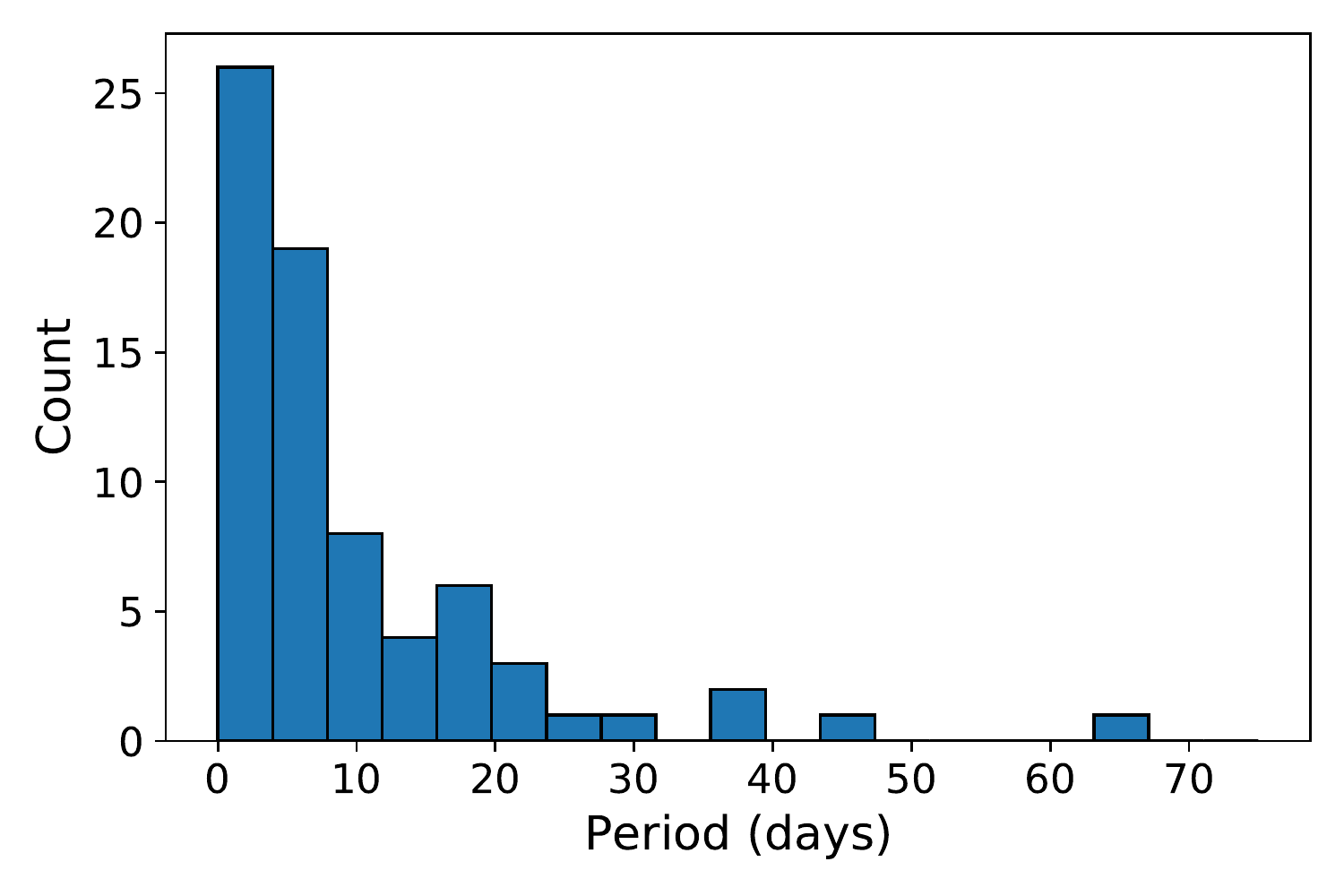}}\\
    \caption{Most of the planets in our sample are small (1-5 $R_\oplus$), and thus have relatively low false positive probabilities \citep[e.g.][]{mortonjohnson, guerrero2021} compared to giant planet candidates \citep{santerne2012}.}
    \label{fig:radiusComparison}
\end{figure}

After these cuts, there are 67 binary systems with exoplanets and 688 binary systems in the control sample. The distribution of the radii and period of exoplanets in our sample is shown in Figure \ref{fig:radiusComparison}. 


\subsubsection{Other work identifying visual binaries in \Gaia }

The \citet{elbadry2018} catalog we used in this study is not the only list of visual binary stars including planet candidates. Recently, \citet{Mugrauer2020} presented a sample of 193 binary companions of TESS exoplanets. Although \citet{Mugrauer2020} identify a significantly larger number of possible binary companions to TOIs, they do not identify visual binaries in non-planet-hosting stars with the same criteria that we could use to construct a control sample, so we cannot include these additional binaries in our analysis. \citet{Ziegler2020} uses speckle imaging with the Southern Astrophysical Research Observatory (SOAR) to search for binary companions to TOIs. They then compare their discovered companions to those discovered in \Gaia\ DR2. Many of their systems overlap with our sample.

During the final preparation of our manuscript, \citet{elbadry2021} reported an updated search for visual binaries using the more precise astrometric parameters from \Gaia\ EDR3, including a significant increase in the number of identified systems. In the future, we could perform the same analysis in this paper on their larger sample of visual binaries and potentially increase the statistical significance of our results. {We checked that most of the binary systems (92\%) in our sample lie in the sample of \citet{elbadry2021}, with the inclination distribution being virtually the same when excluding those systems not in \citet{elbadry2021}.}  

\subsection{Ground-based Spectroscopy}
Fitting binary orbits using only instantaneous positions and proper motions from \Gaia\ requires an estimate of the mass of each binary component, which in turn requires an estimate of each star's metallicity. To derive the metallicities of stars in our samples, we use both archival observations from large spectroscopic surveys and targeted follow-up observations of planet candidate host stars made by the \TESS\ Follow-up Observing Program (TFOP). Below, we describe the sources of our spectroscopic parameters and the procedure we used to determine the metallicities of the observed stars. Because the components of relatively wide binary stars are known to have nearly identical elemental abundances in most cases \citep{Hawkins2020}, we assume the metallicity of both stars in the binary are the same when we only have metallicity measurements for one of the pair.  

For stars that have more than one spectroscopic observation, we use an average of the metallicities derived from the separate spectroscopic observations and add the errors from each separate observation in quadrature.

\subsubsection{Archival Spectroscopy}
\label{sec:archivalSpec}

Many of the stars in our samples have archival spectra and published metallicity estimates. \citet{elbadry2019} cross-matched their sample of visual binary stars \citep{elbadry2018} with stars observed by large spectroscopic surveys and identify a subset of 8,507 binaries for which spectroscopic metallicities have been reported in the literature for at least one component. The archival metallicities they identify come from the following surveys or compilations: RAVE \citep{Steinmetz2003,Kunder2017}, LAMOST \citep{Zhao2012}, Hypatia \citep{Hinkel2014}, APOGEE \citep{Majewski2017}, and GALAH \citep{cotar2019,Buder2018}. Of the stars in our sample of binary stars with planet candidates, 16 stars have a metallicity from RAVE, 1 from LAMOST, 4 from the Hypatia catalog, 2 from APOGEE, and 2 from GALAH.

\subsubsection{LCO/NRES}

We obtained observations of 7 stars from our sample of binary stars with planet candidates using the Network of Robotic Echelle Spectrographs (NRES), a set of four identical optical echelle spectrographs connected to the Las Cumbres Observatory (LCO) global telescope network \citep{Brown2013,Eastman2014, NRES, NRES2}. Each spectrograph is fiber-fed by one of the 1m telescopes in the LCO network. NRES achieves a spectral resolution of $R \sim 53000$ over the wavelength range $390-860$ nm. We derive stellar parameters from NRES spectra using the SpecMatch algorithm \citep{Petigura2015,Petigura2017}, which compares the observed spectra to the synthetic spectra of \citet{Coelho2005}.
\subsubsection{TRES}
We observed 15 stars with Tillinghast Reflector Echelle Spectrograph (TRES), an optical echelle spectrograph with a wavelength range of $385-910$ nm. TRES is located on the 1.5m telescope at the Whipple Observatory on Mt. Hopkins in Arizona and has a resolution of $R=44,000$ \citep{gaborthesis, TRES}. The TRES metallicities are derived using the Stellar Parameter Classification tool (SPC). SPC cross-correlates the observed spectra of stars with a grid of around 51,000 model spectra. The peaks of the cross-correlation function (CCF) are then fitted with a polynomial over four stellar parameters ($T_{\text{eff}}$, log(g), [Fe/H], $v\sin(i)$) in an attempt to determine the location of the highest point between grid points. If multiple observations are available for a star, the mean metallicity is weighted by the corresponding highest peak of the CCF in the SPC results for each observation \citep{Buchave2012,Buchave2014,Buchave2015}. For all stars with $T_{\text{eff}}< 4500 K$ (indicating that the star is a cooler, dwarf-like star), a Yale-Yonsei isochrone is used as a prior for the star's  log(g), $T_{\text{eff}}$, and [Fe/H] \citep{Spada2013}; this additional step improves the reliability of spectroscopic parameters for cool stars.

\subsubsection{FIES}
We observed 6 stars with the FIbre-fed Echelle Spectrograph (FIES). FIES is located on the 2.56m Nordic Optical Telescope (NOT) in La Palma, Spain \citep{Telting2014}. FIES has three observation modes that offer different spectral resolution and throughput; our observations use the highest resolution (R=67000) 1.3 arcsecond fiber. {Metallicities are derived using SPC in a similar manner to the TRES metallicities.}


\subsubsection{CHIRON}
We observed 8 stars with the CTIO High Resolution spectrometer (CHIRON) located on the 1.5 meter SMARTS telescope at the Cerro Tololo Inter-American Observatory in Chile. CHIRON is a fiber-fed optical echelle spectrograph that achieves a resolution of $R \sim 79,000$ and a wavelength range of $415-880$ nm \citep{CHIRON,Tokovinin2013}. Metallicities are derived by interpolating the CHIRON spectra to a sample of around 10,000 TRES spectra with parameters derived by SPC using a gradient-boosting regressor. When multiple CHIRON observations are present for a star, the mean of the observations is used, with error added in quadrature.



\section{Analysis}
\label{sec:Analysis}
Here, we describe the calculations needed to constrain the inclination angle of each binary orbit in our sample. This involves two main steps: 1) estimating the mass of each star in the binary system, and 2) given these stellar masses and the \Gaia\ astrometric parameters, determining plausible orbital parameters using the LOFTI software package \citep{Pearce2020}. 

\subsection{Stellar Mass Determination}
\label{sec:massFitting}
\begin{figure}
    \centering
    \includegraphics[width=\columnwidth]{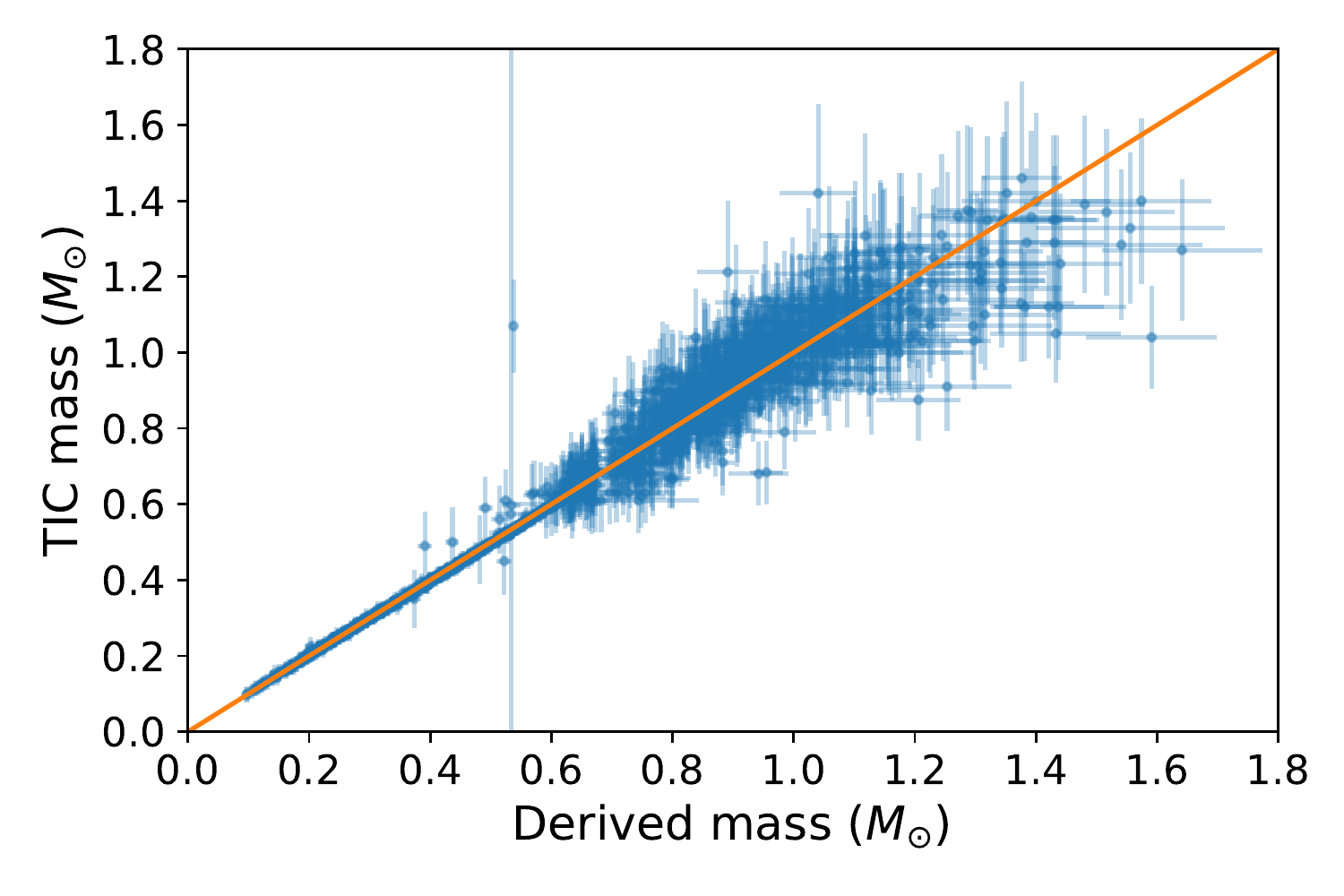}
    \caption{A comparison between the masses derived using isochrone fitting in this paper vs. the TIC derived masses for the control sample. The plotted error bars are one standard deviation. An orange line denotes where a 1:1 correspondence of masses would lie. We see good correspondence between our mass estimates and those in the TIC.}
    \label{fig:massComparison}
\end{figure}
We derive masses for the stars in our samples using the \verb|Isochrones| Python package \citep{Morton2015}. Given observable parameters like a star's apparent magnitudes in different bandpasses, its trigonometric parallax, and spectroscopic parameters, \verb|Isochrones| determines the star's most likely physical parameters and their uncertainties by comparing measured parameters to those predicted by stellar evolution and atmosphere models. \verb|Isochrones| supports several different suites of isochrone models; we use the MIST isochrones \citep{MIST, choi2016}. We input each star's parallax, metallicity (when available from archival spectroscopy or NRES, TRES, FIES, or CHIRON), and apparent magnitudes in the G, BP, and RP bandpasses from \Gaia\ and the J, H, and K bandpasses from the 2MASS survey \citep{twomass}. For uncertainties in bandpasses, we use the error floors of \citet{Exofast2}.

Not all of the stars in our samples have spectroscopic metallicity measurements. If a star doesn't have a metallicity value, the metallicity of its companion is used \citep[see][]{Hawkins2020}. If neither star in a binary pair has a spectroscopic metallicity measurement, we use the \verb|isochrones| default metallicity\footnote{
\url{https://github.com/timothydmorton/isochrones/blob/c134d271950fe63bd5e84ede4530585eba3f48a4/isochrones/priors.py\#L364}}, a prior based on a two-Gaussian fit to the distribution of metallicities reported by \citet{casagrande2011}.

Model isochrones are not as well calibrated for M-dwarf stars (which constitute around half of the stars in the sample of visual binaries) as they are for Sun-like stars \citep{Angus2019}. For M-dwarfs, we use the $M_K-M_\star$ relationship of \citet{Mann2019} using the accompanying \verb|M_-M_K-| python package\footnote{Available at \href{https://github.com/awmann/M\_-M\_K-}{https://github.com/awmann/M\_-M\_K-}} to estimate the stars' mass. $M_k$ magnitudes, when available, are taken from the \TESS\ Input Catalog (TIC) \citep{TIC}. Any stars in the range 4.5 $<$ $M_k$ $<$ 10.5 (the more conservative option given by \citealt{Mann2019}) are treated as M-dwarfs.

Rough mass estimates for these stars are also provided in the \TESS\ Input Catalog \citep{Stassun2019}. We compare our mass estimates to the TIC estimates, as demonstrated for the control sample in Figure \ref{fig:massComparison}, to ensure that there are no systematic discrepancies in estimation. The median uncertainty of the masses in our sample is 0.016 $M_\odot$ and in the TIC 0.082 $M_\odot$. We also compare our mass estimations to those derived from spectral energy distribution (SED) modeling \citep{stassun2018}, where stellar atmosphere model grids are interpolated in $T_{\rm eff}$, $\log{g}$, and [Fe/H] and combined with spectroscopic $\log{g}$ measurements. There is general agreement between the masses derived from both techniques, an independent check of the masses assigned to the stars.

\subsection{LOFTI modeling}
\begin{table}[]
    \begin{tabular}{|c|c|}
         \hline
         Parameter & Prior \\
         \hline
         Eccentricity ($e$) & Uniform($0,1$)\\
         Inclination ($i$) & Sin($0,180$)\\
         Argument of Periastron ($w$) & Uniform($0,360$)\\
         $t_\text{const}$ & Uniform($0,1$)\\
         Total Mass ($M_\text{tot}$) & Normal($M_0$,$M_\text{std}$)\\
         Distance & Normal($D_0$,$D_\text{std}$)\\
         \hline
    \end{tabular}
    \caption{{List of priors for orbital parameters that are randomly sampled in LOFTI. $t_\text{const}$ is a value used in the calculation of the epoch of periastron passage ($t_0$)}}.
    \label{table:priors}
\end{table}

\begin{table}[]
    \begin{tabular}{|c|c|}
         \hline
         Parameter & Formula \\
         \hline
         Period ($T$)& $\left(\displaystyle\frac{a^3}{M_\text{tot}}\right)$\\
         Epoch of Periastron Passage ($t_0$) & $\text{d}_\text{epoch}-t_\text{const}*T$\\
         Semimajor axis $a$ & Scale and Rotate\\
         Longirude of Ascending Node ($\Omega$) & Scale and Rotate\\
         \hline
    \end{tabular}
    \caption{{List of orbital parameters that are calculated either from sampled parameters or calculated using the scale and rotate method to match observed data. $d_\text{epoch}$ is the date of the epoch of the observations.}}
    \label{table:calcParams}
\end{table}

\label{sec:LOFTI}
\begin{figure}
    \centering
    \includegraphics[width=\columnwidth]{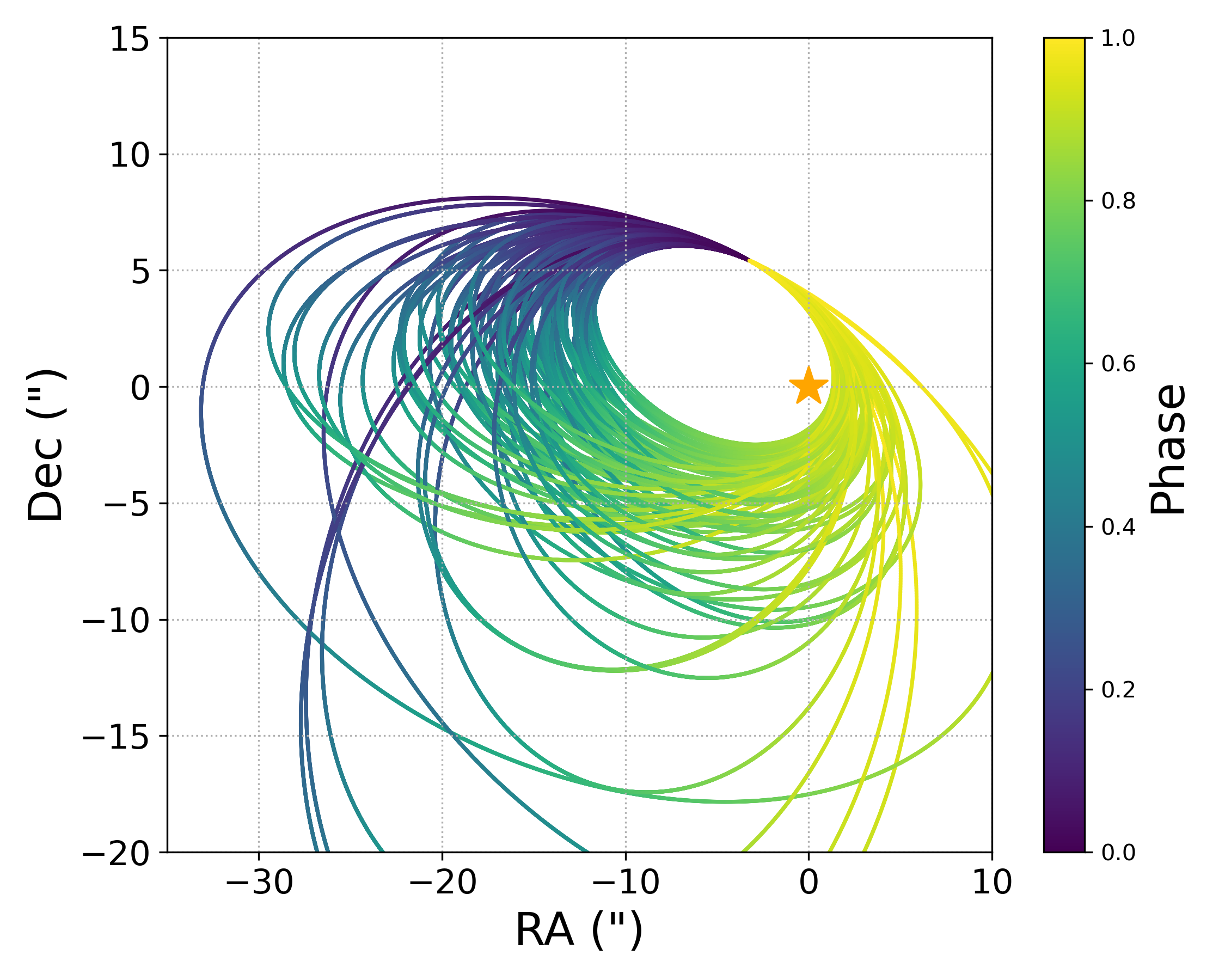}
    \caption{100 orbits from the posterior of a sample LOFTI fit.}
    \label{fig:LOFTI}
\end{figure}
Linear Orbits for The Impatient (LOFTI, \citealt{Pearce2020}) uses measurements of the relative positions of two stars (in both right ascension, $\Delta\alpha$, and declination, $\Delta\delta$), relative proper motions, relative radial velocity (if available), and stellar masses to constrain orbits of visual binaries. {In our analysis, we take all of these parameters from \Gaia. LOFTI calculates relative proper motion and position as primary minus companion, while relative radial velocity is calculated as companion minus primary.} LOFTI uses the Orbits for the Impatient (OFTI) sampling and rejection technique \citep{Blunt2017} applied to \Gaia\  astrometric parameters.

OFTI, on which LOFTI is based, optimizes orbit fitting via rejection sampling with the unique scale and rotate step. The algorithm randomly samples eccentricity ($e$), argument of periastron ($\omega$), mean anomaly from derived epoch of periastron ($t_o$), and cosine of inclination ($\cos(i)$) from uniform priors, then scales semimajor axis ($a$) and rotates longitude of the ascending node ($\Omega$) to match the observed separation in binaries, rejecting orbits if the $\chi^2$ probability (proportional to $e^{-\chi^2/2}$) is greater than a randomly chosen number in the range $[0,1]$ to form a posterior distribution of orbital parameters. OFTI is computationally advantageous over traditional MCMC methods for visual binary systems, where orbital parameters are often poorly constrained \citep{Blunt2017,Blunt2020}. {Table \ref{table:priors} is a list of priors used in the LOFTI algorithm on sampled parameters, while Table \ref{table:calcParams} shows parameters calculated either from sampled parameters or via the scale and rotate method.}

Orbital fitting using very short observational baselines (as done here) can lead to degeneracies in parameters. Particularly, for high eccentricity values, inclination is biased towards edge-on orbits \citep{ferrerchavez}. The inclusion of a control sample in our study should allow us to determine if there is a preferential alignment regardless of the presence of this bias.

We ran LOFTI on the binary systems with exoplanets and those in the control sample. We continued sampling the posterior distributions until we reached 100,000 accepted orbits. We performed this computationally intensive task on the Maverick 2 and Lonestar 5 supercomputer clusters at the Texas Advanced Computing Center.

\begin{figure*}
    \centering
    \includegraphics[width= \textwidth]{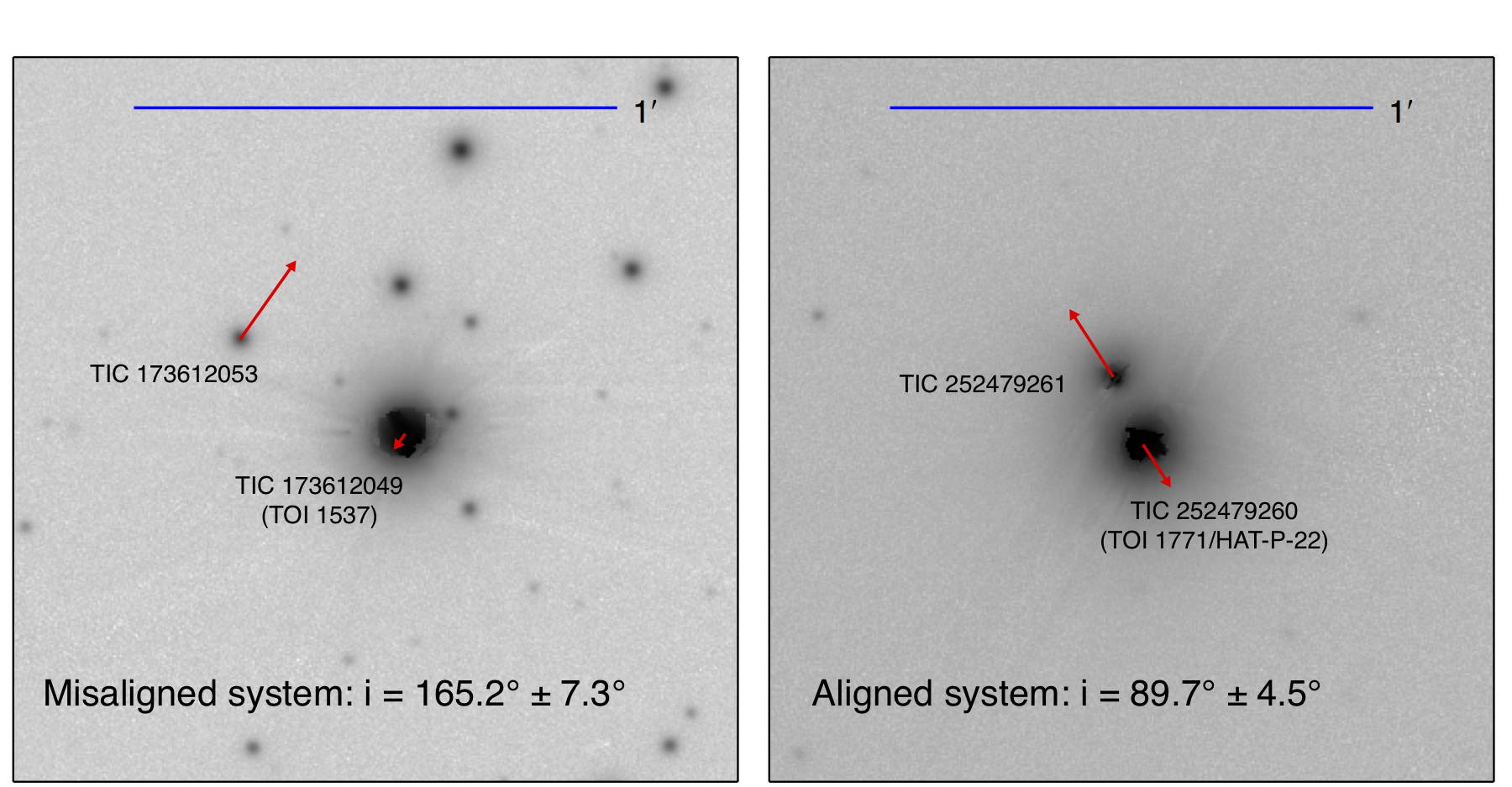}\\
    \caption{Images from the Pan-STARRS survey \citep{chambers2016} of selected binary systems from our sample of transiting exoplanet hosts. In each image, the exoplanet host and its binary companion are labeled (with the planet host labeled as a TOI) and red arrows show their relative proper motion after subtracting the velocity of the system's center of mass. The blue bar at the top of each image shows the scale. The two images are labeled with the binary inclination from our LOFTI fits. The image on the left is misaligned with the orbit of the transiting planet, while The image on the right shows an aligned orbit. {In general, only orbits with $i=90$ will have proper motion vectors that are parallel to the projected semimajor axis. All other angles between proper motion vectors and projected semimajor axes a priori allow all inclination values, although given a uniform prior on eccentricity, higher inclination values will be favored for nearly parallel semimajor axis and proper motion vectors}}
    \label{fig:images}
\end{figure*}

We show 100 orbits drawn from a sample LOFTI posterior in Figure \ref{fig:LOFTI} and a sample posterior distribution of parameters from LOFTI for one binary system {in Appendix \ref{cornerplot}}. We also show archival ground-based images of two characteristic systems, one aligned and one misaligned, annotated with astrometric information from \Gaia\ in Figure \ref{fig:images}.

\section{Results} \label{sec:Results}

\begin{figure}
    \centering
    \subfloat[Histogram of $|90-i|$ for the median $i$ of each system]{\includegraphics[width= \columnwidth]{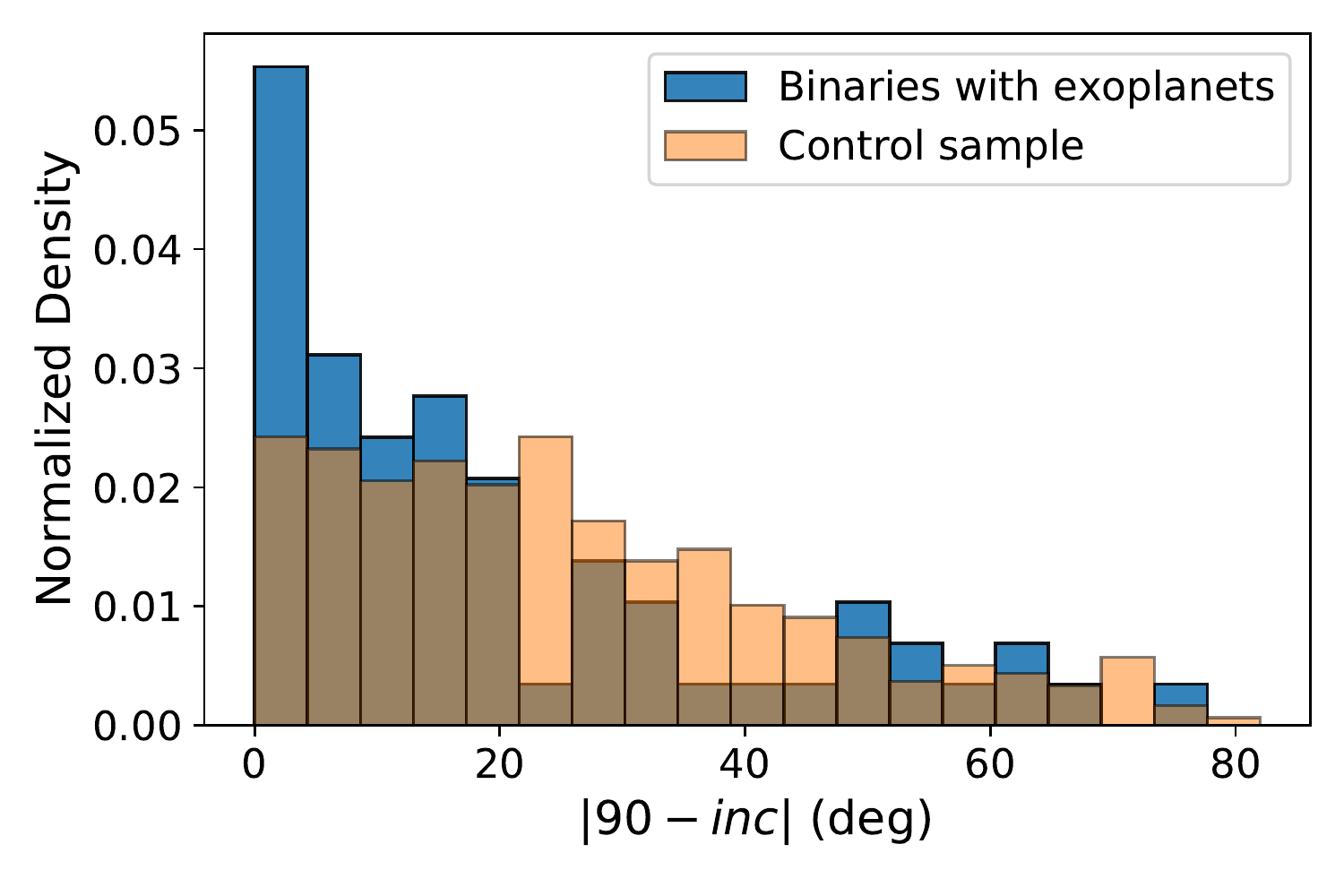}}\\
    \subfloat[Histogram of $\sin(|90-i|)$ for the median $i$ of each system. The dashed line indicates the expected isotropic distribution (uniform $\cos(i)$).]{\includegraphics[width= \columnwidth]{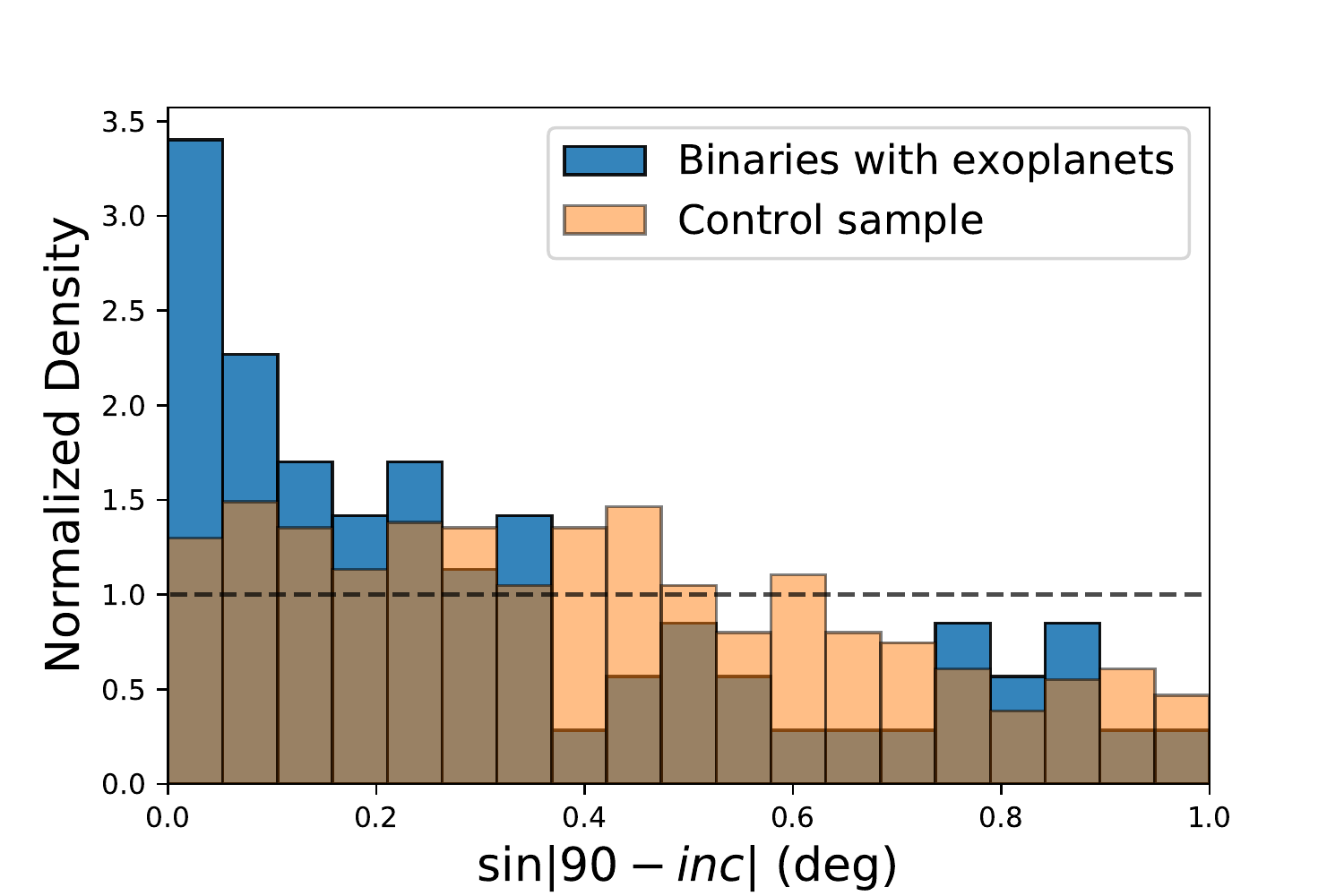}}\\
    \caption{A histogram of $|90-i|$ (a) and $\sin(|90-i|)$ (b). We see an overabundance of systems with $i\approx$ 90$^\circ$ in our exoplanet sample compared to our control sample, indicating that there is some preferential alignment between the orbital planes of close-in exoplanets and visual binaries. We also find that the inclinations of binaries in our control sample are significantly different from the expected purely random $\sin(i)$ distribution (the dashed line in (b)), likely due to either selection effects \citep{elbadry2018} or biases from inferring orbits from short arcs \citep{ferrerchavez}.}
    \label{fig:distribution}
\end{figure}

After obtaining orbital parameters for all of the visual binary systems in our samples, we calculated the difference in median inclination between the plane of each visual binary orbit and the orbit of the planetary system orbiting that same primary. Since the planets in our sample were all detected using the transit method by \TESS, they all have inclinations with respect to the plane of the sky very close to 90$^{\circ}$. The minimum difference in inclination between the binary star system and planet can be expressed by $\lvert 90^{\circ}-i \rvert$. We calculate this value for each system in both our sample of binaries containing transiting planet candidates and our control sample of binaries without detected exoplanets to account for selection effects in the \citet{elbadry2018} catalog or biases in our orbit fitting \citep{ferrerchavez}.

We show histograms of the median $\lvert 90^{\circ}-i \rvert$ for both our sample of planet candidates and our control sample in Figure \ref{fig:distribution}a. In Figure \ref{fig:distribution}b, we show the distribution of $\sin|90-i|$. If inclination is drawn from an isotropic distribution, $\sin|90-i|$ should be uniform. There is an apparent overdensity of visual binary systems hosting exoplanets with inclinations close to 90$^{\circ}$. {Although any given individual system with a binary inclination near 90$^{\circ}$ might not necessarily be aligned, the overall overdensity suggests a preferential alignment between exoplanet and wide binary orbit.}

\begin{figure}
    \centering
    \subfloat[A plot of semimajor axis vs. $\vert 90 - i \vert$ of the binary systems with exoplanet candidates. 90\% confidence intervals on both parameters are shown as error bars.]{\includegraphics[width=\columnwidth]{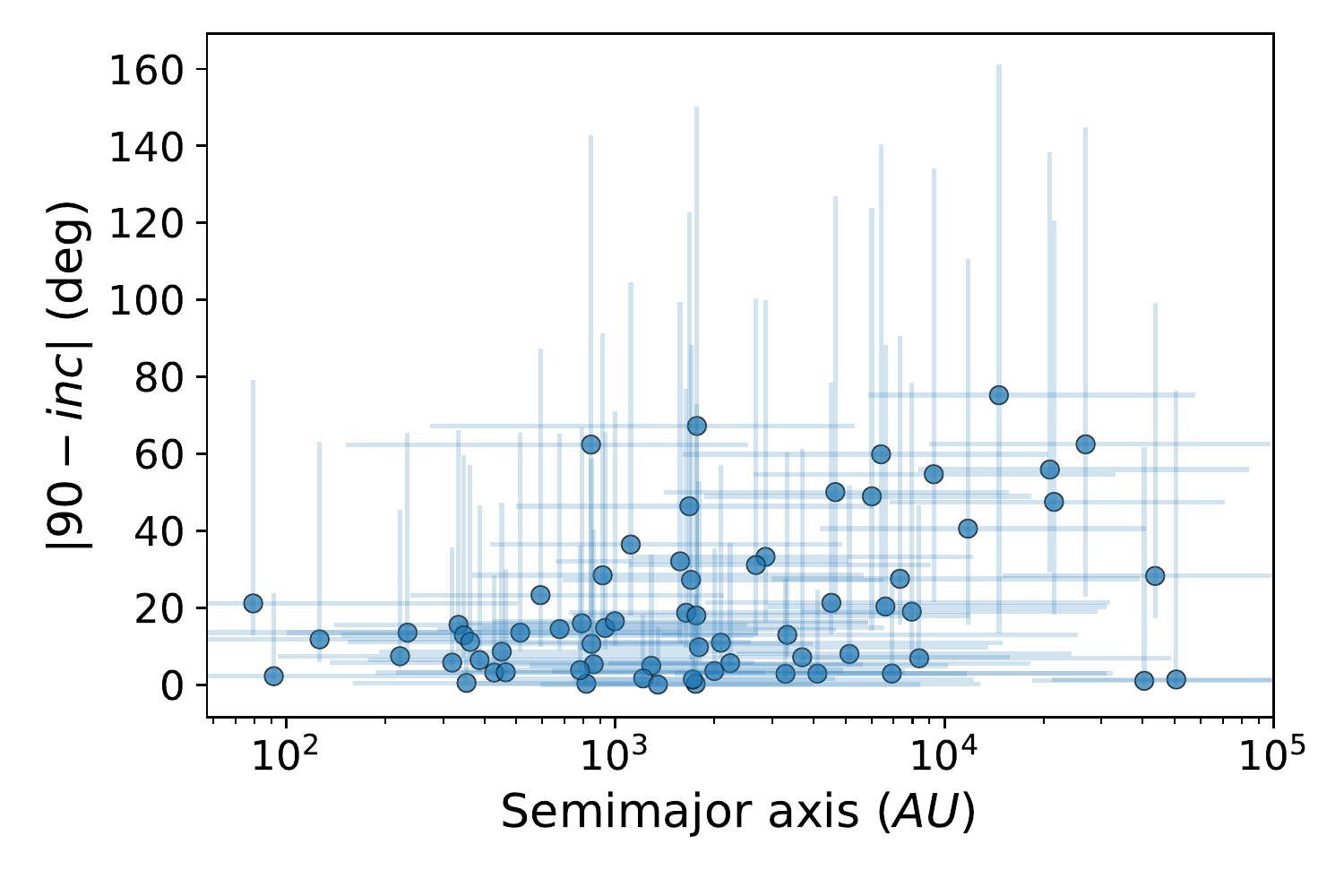}}\\
    \subfloat[A plot of semimajor axis vs. $\vert 90 - i \vert$ of the binary systems in the control sample. Error bars are not shown due to the large number of data points.]{\includegraphics[width= \columnwidth]{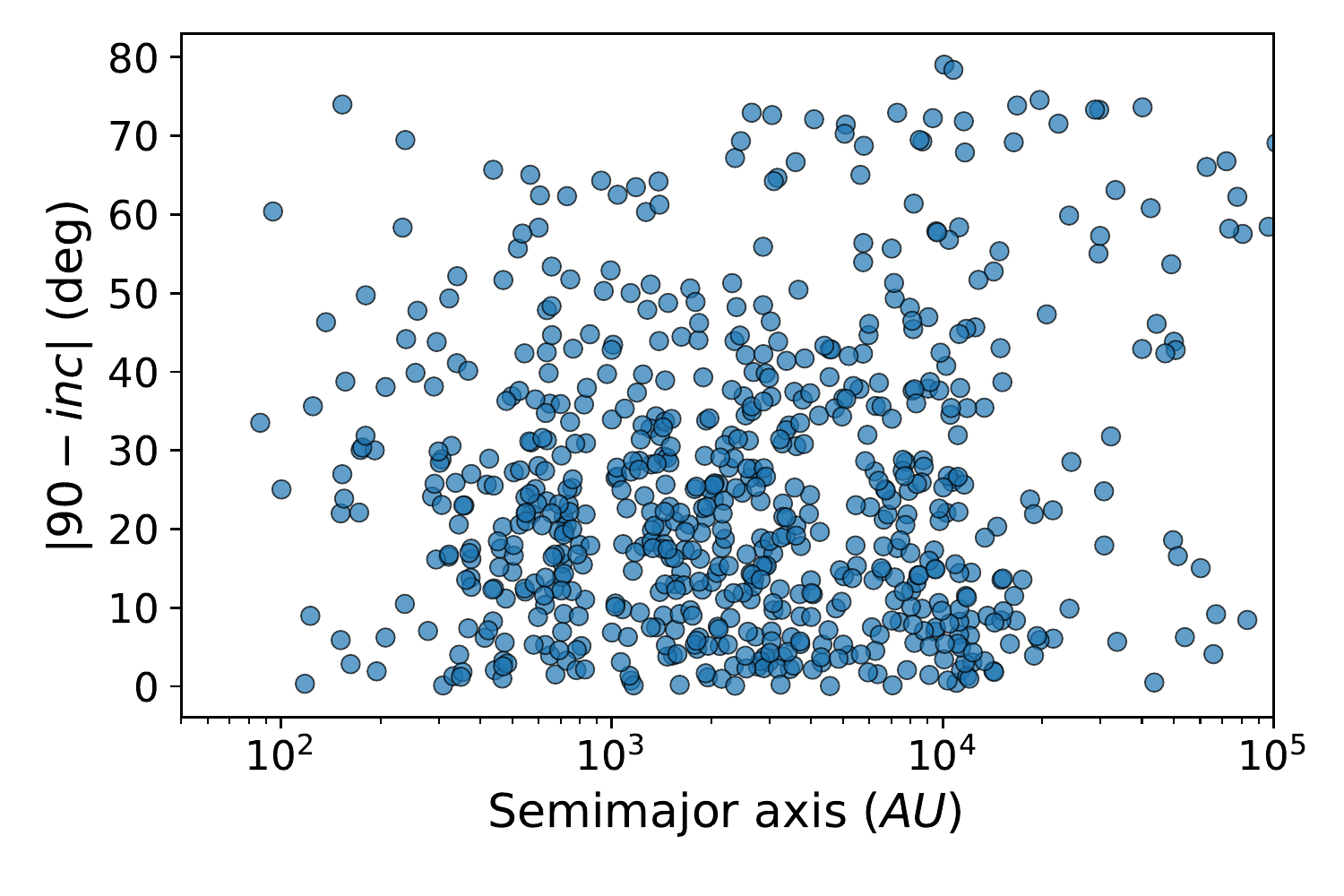}}\\
    \caption{A clear alignment of binary and exoplanet below approximately $ 700$ AU is seen in (a), the sample with exoplanet candidates. No such alignment below 700 AU is seen in (b), the control sample.}
    \label{fig:semimajorVsInclination}
\end{figure}
\begin{figure}
    \centering
    \includegraphics[width=\columnwidth]{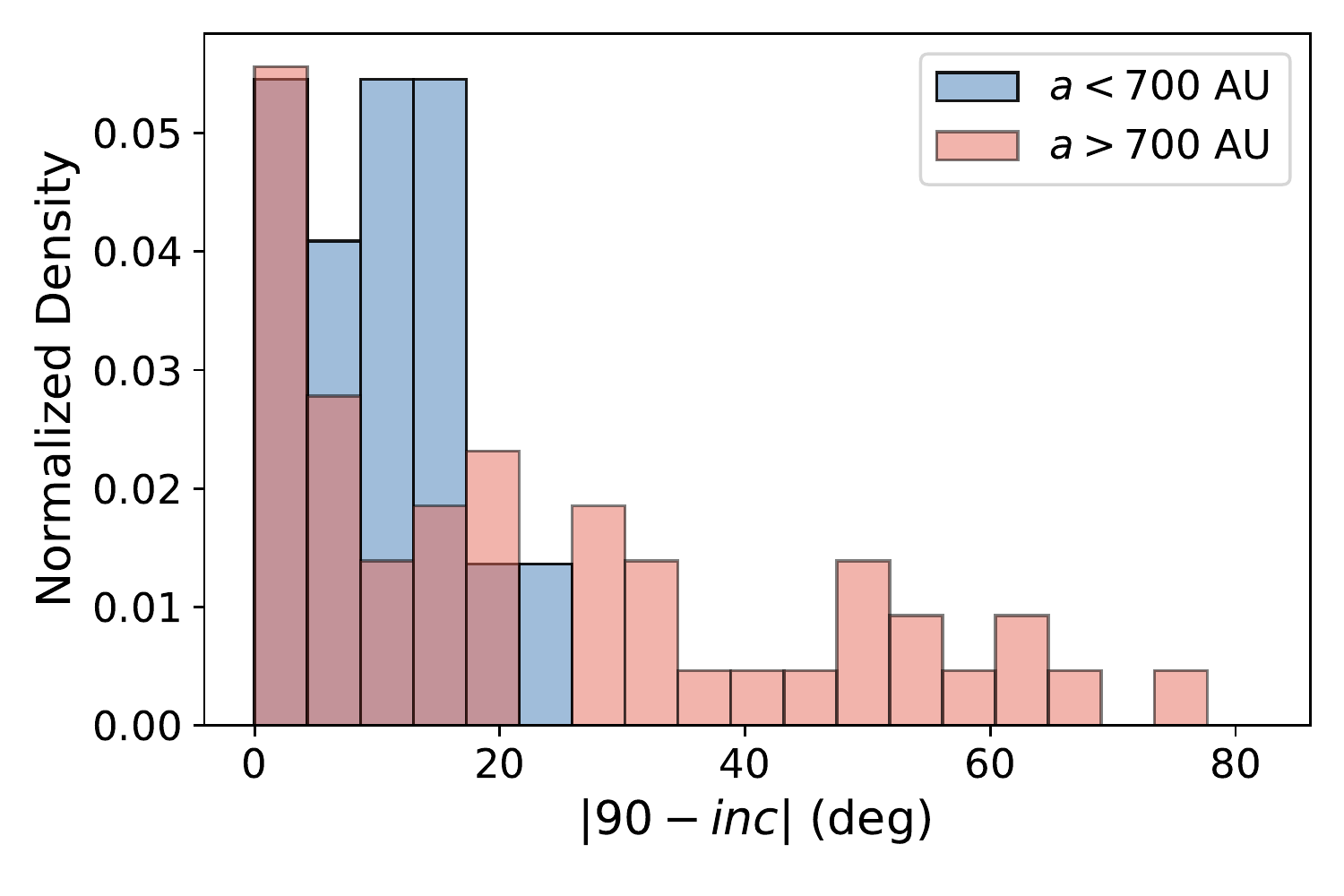}
    \caption{Distribution of $|90-i|$ (with median $i$) for binaries with exoplanets with semimajor axes ($a$) below and above 700 AU. The binaries with semimajor axes less than 700 AU are clustered within 20$^\circ$ of edge-on orbits, while the binaries with semimajor axes greater than 700 show a more uniform distribution of inclinations.  }
    \label{fig:700aucutoff}
\end{figure}

The overdensity of well-aligned binary orbits is most apparent at binary separations closer than approximately 700~AU. Figure \ref{fig:semimajorVsInclination} shows the minimum difference in inclination for each system as a function of the binary semimajor axis. Figure \ref{fig:700aucutoff} shows histograms of $\lvert 90^{\circ}-i \rvert$ for binary systems hosting exoplanets separated by semimajor axis. Evidently, planets are well aligned with the orbits of binary stars with semimajor axes less than about 700 AU, while any alignment (if present) between planets and binaries with semimajor axes greater than about 700 AU is much weaker. No such dependence in inclination on semimajor axis is seen in the control sample. As can be seen in Figure \ref{fig:semimajorVsInclination}, the 700 AU value is very approximate and the actual value could reasonably be much lower or higher due to the small sample sizes we use.

\subsection{Kolmogorov-Smirnov Test}
\label{sec:KStest}
\begin{figure}
    \centering
    \includegraphics[width=\columnwidth]{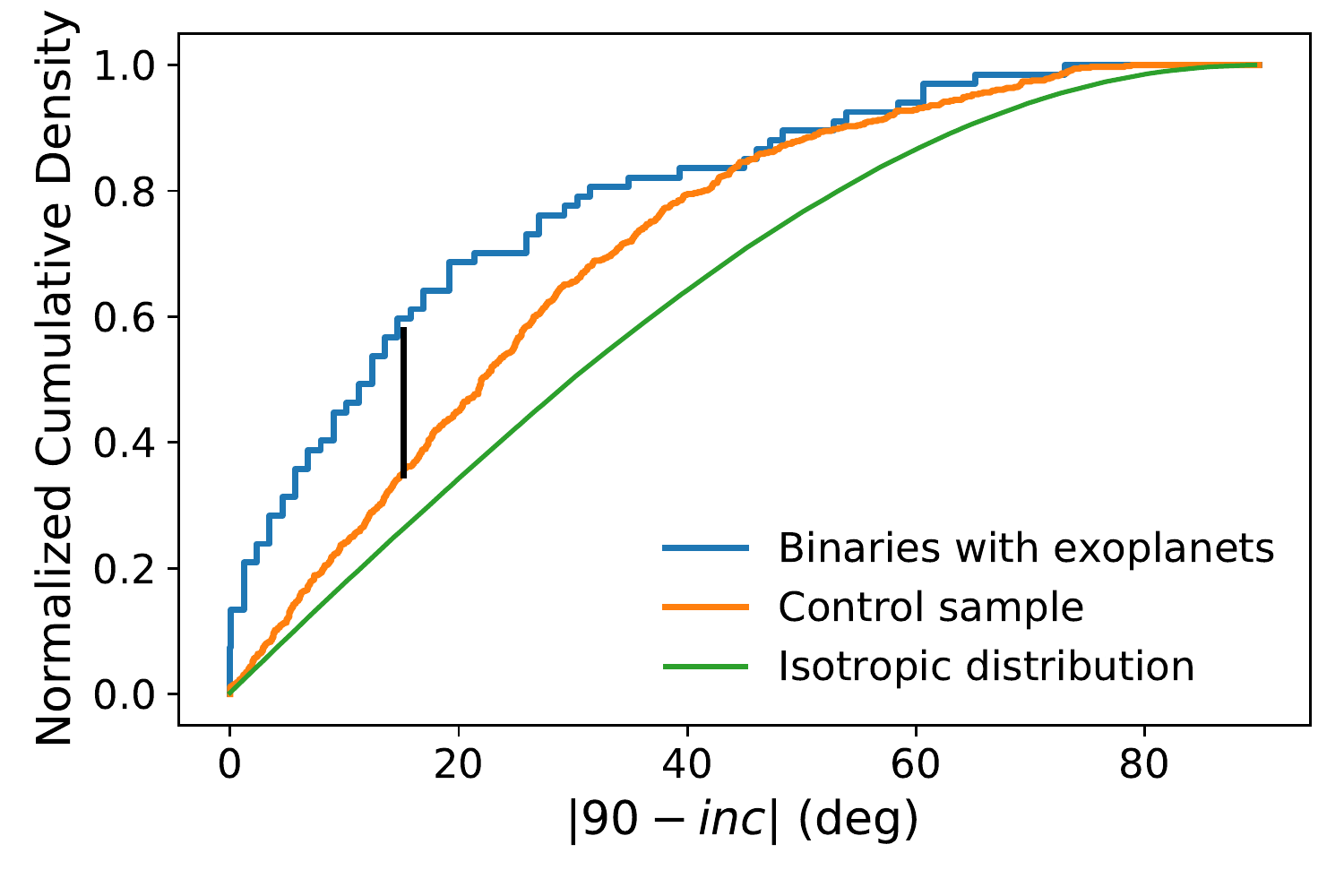}
    \caption{The two empirical cumulative distributions used in the K-S test. The black line shows the maximum distance between the two distributions which corresponds to a K-S value of 0.223 and a p-value of 0.0037.}
    \label{fig:KSDistribution}
\end{figure}

We assessed the statistical significance of the apparent alignment between visual binary orbits and their planetary systems using a Kolmogorov-Smirnov (K-S) test. The K-S test measures the maximum difference between the cumulative distributions of the two empirical distributions and then calculates a p-value representing the probability that the two empirical distributions were drawn from the same underlying distribution. A small K-S value indicates that the distributions being compared are not from the same underlying distribution. We perform the K-S test between the sample with exoplanets and control sample's $\lvert90-i \rvert$ values, where $i$ is the median value of the distribution of inclinations. The test rejects the null hypothesis that the two datasets are drawn from the same underlying distribution with  $p= 0.0037$. {To check that our significance value is not dependent on our somewhat arbitrary choice of RUWE cutoff we adjust our RUWE cutoff to 1.3, 1.2 and 1.1, finding that the p-values are respectively $p=0.016$, $p=0.051$ and $p=0.045$. Additionally, we use two other statistical tests specifically designed for comparing angular data: Wallraff's nonparametric test \citep{Wallraff1979} and Watson's nonparametric test \citep{watson1983statistics}. The Wallraff test yields a p-value of $0.031$, while the Watson test yields a p-value in the range $[0.001,0.01]$ (the test only gives a range of p-values).}

None of these tests account for error in the individual measurements of inclination. The K-S test is shown visually as the maximum difference in cumulative distributions in Figure \ref{fig:KSDistribution}.

Since most of the excess of aligned systems are found in binary systems with semimajor axes less than about $700$ AU, we also compare the distributions of $\lvert90-i \rvert$ for these close binaries. For systems with semimajor axes less than 700 AU compared to the control sample, the K-S test shows stronger evidence that the inclinations in the exoplanet sample and control sample are drawn from different distributions ($p=0.000250$). Finally, we investigate whether the distributions of inclination above and below 700 AU are significantly different from each other. For systems with semimajor axes less than 700 AU compared to those with semimajor axes greater than 700 AU, the K-S test still shows evidence for a difference between the distributions, but with lesser significance: $p=0.0172$.
\begin{figure}
    \centering
    \subfloat[The distribution of the 90\% confidence interval in $|90-i|$ for the sample with exoplanets vs. the control sample.]{\includegraphics[width=\columnwidth]{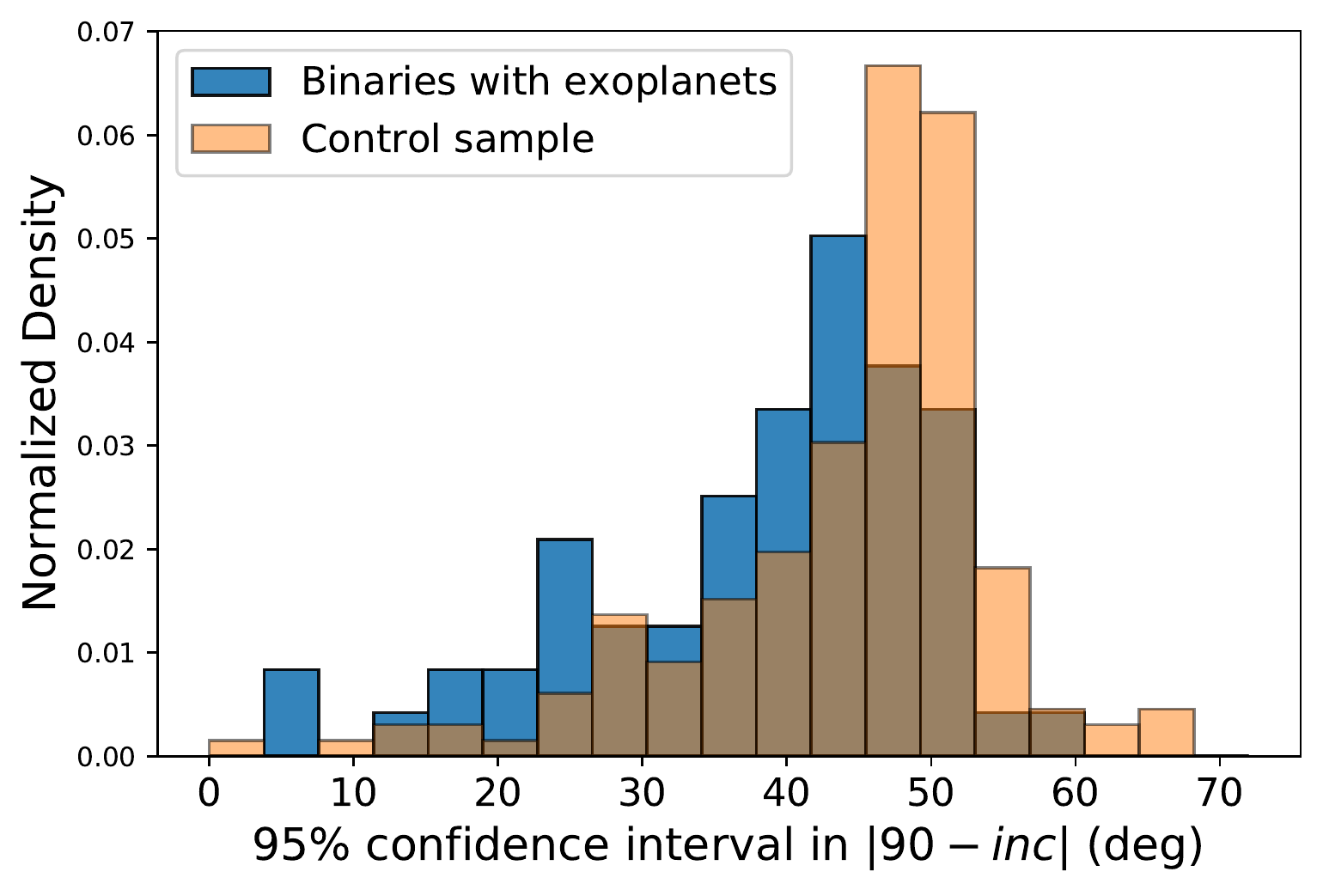}}\\
    \subfloat[A plot of $|90-i|$ vs. the error in $|90-i|$.]{\includegraphics[width=\columnwidth]{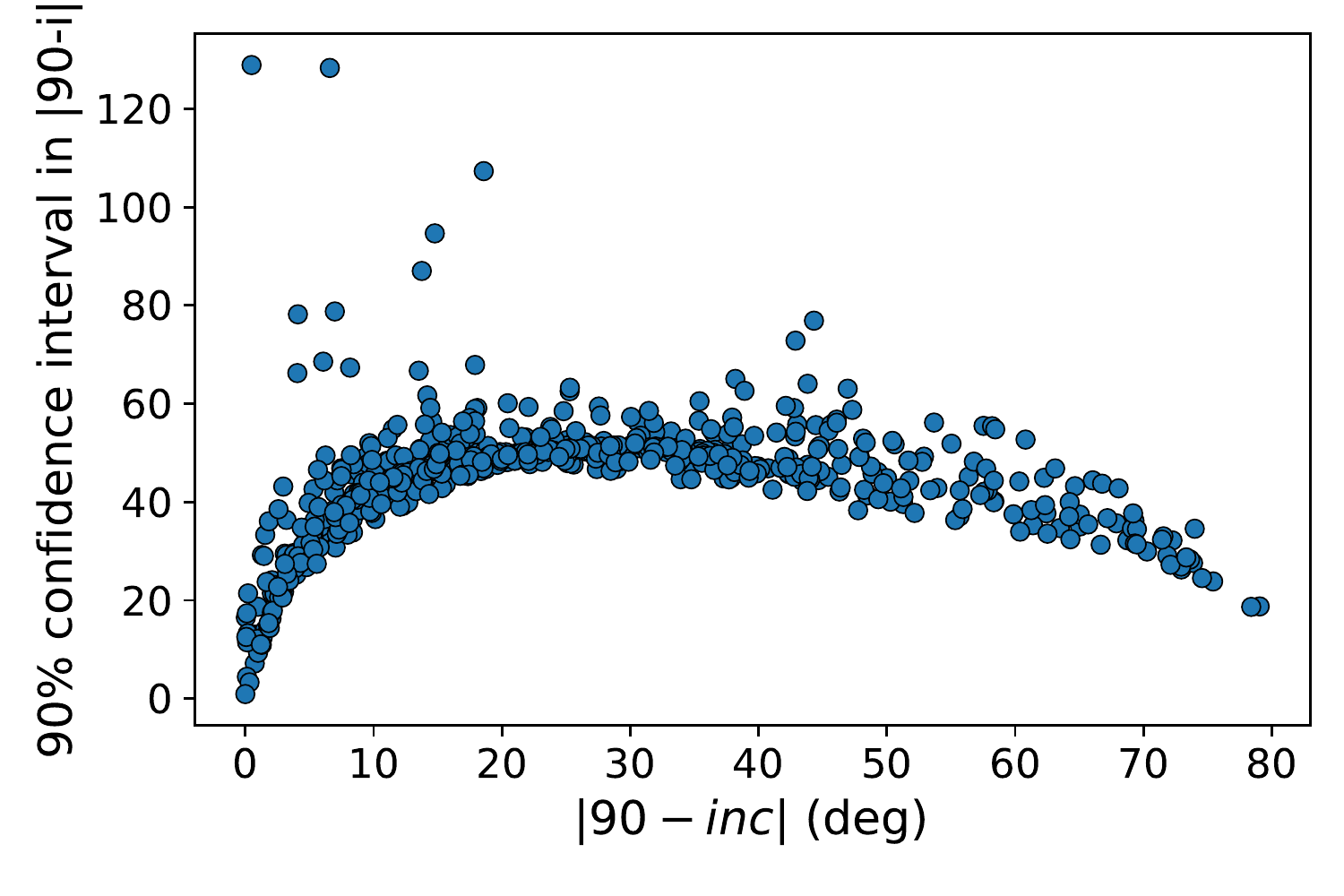}}\\
    \caption{A histogram (a) and scatterplot (b) showing the error in inclination's distribution and dependence on $90-i$ respectively. The noticeable difference in the errors of (a) is likely caused by more aligned systems, which are more prevalent in the sample with exoplanets, tending to have tighter constraints on inclination.}
    \label{fig:errorDistribution}
\end{figure}

The standard deviation in inclination is on average 14 degrees, and a large proportion of samples have a skewed distribution of error. Additionally, error varies with inclination. The distribution of errors in our sample is shown in Figure \ref{fig:errorDistribution}a, and the dependence of error on inclination in Figure \ref{fig:errorDistribution}b. To explore the effect of this large and often asymmetrical error on the results of the KS test, we perform a bootstrap procedure on the two distributions where we take a random sample from each binary system's LOFTI results, and compare the overall distributions of those random samples. We repeat this process 10,000 times to derive a bootstrap distribution. This process calculates the distribution of $p$-values that we would expect to measure if we ran the same experiment many times, with slightly larger uncertainties on the inclinations (by about a factor of $\sqrt{2}$). We found that when we performed this analysis comparing the full samples, 61.91\% of the bootstrap instances yielded a p-value less than 0.05, with the median p-value being $0.0300$. We run the same bootstrap test on the subset of systems with $a < 700$ AU. This test returns a median p-value of $0.0044$ with 87.35\% of the bootstrap instances being at least $p=0.05$.

We also simulated a K-S test of two $\sin(90-i)$ probability density functions with sample sizes similar to those of that in our study and various types of skewed error (e.g. skewed away from $i=90$ or towards $i=90$). All simulations returned uniform distributions of K-S p-values, indicating that any skewed error is likely not affecting the result of the K-S test.

For completeness, we tested whether the distribution of binary inclinations in the control sample is consistent with randomly oriented orbits. The expected probability density function of binary inclination angles for random orbital orientations is proportional to $\sin(i)$. When we compare the control sample to inclinations drawn from a $\sin(i)$ distribution with a K-S test, we find a significant difference between the two. We show a comparison of the control sample distribution and the expected isotropic distribution in Figure \ref{fig:distribution}b and Figure \ref{fig:KSDistribution}. This difference is likely caused by either subtle biases when fitting astrometric orbits in small-orbit arc fitting \citep{ferrerchavez} or selection effects in the \citet{elbadry2018} catalog. The control sample has slightly fewer face-on systems and more edge-on systems than we would expect from a perfectly random distribution of inclinations. Nevertheless, the difference between the distribution of inclinations in the control sample and the sample of binaries with exoplanets is still significant.
\subsubsection{Inclination-eccentricity degeneracy}
\begin{figure}
    \centering
    \includegraphics[width=\columnwidth]{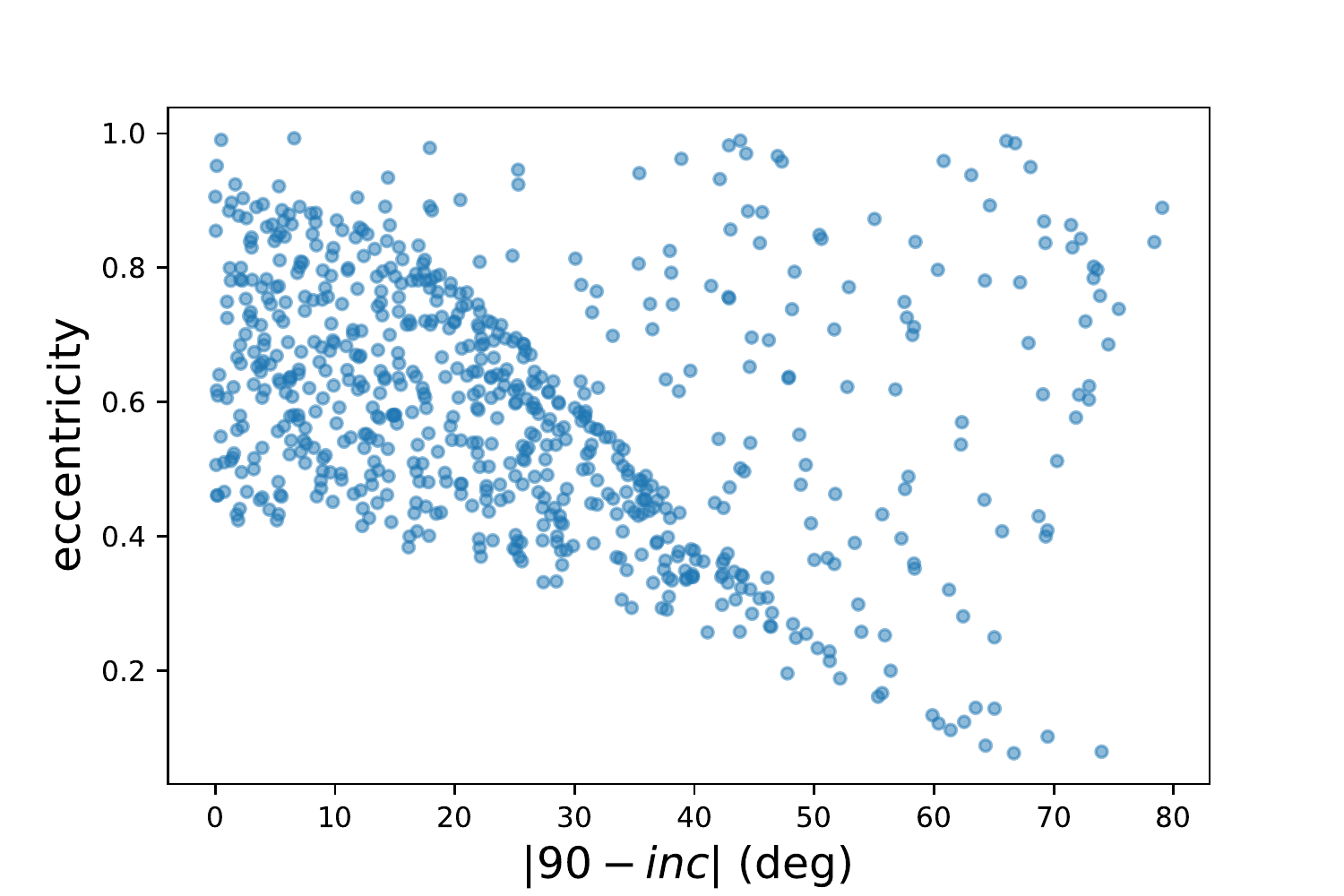}
    \caption{{The median inclination relative to 90 degrees versus median eccentricity for the control sample. A degeneracy exists between inclination and eccentricity, excluding any systems with inclination near 90 degrees and low eccentricity}}
    \label{fig:eccIncDegeneracy}
\end{figure}
{There is an apparent degeneracy between inclination and eccentricity, presumably related to the bias mentioned in \citet{ferrerchavez}. We plot inclination versus eccentricity for our systems in Figure \ref{fig:eccIncDegeneracy} where it is clear that LOFTI does not find any systems with low eccentricity and inclination near 90 degrees. This is not a feature of the OFTI method specifically: all orbits derived through short arc fitting will have a similar degeneracy between median inclination and eccentricity. Such a degeneracy is only between summary statistics of eccentricity and inclination. The full parameter space is covered amongst all individual samples.}

{We have no reason to suspect that the eccentricity of the binary orbit could somehow be preferentially modifying the orbit of the planet, so do not believe that this degeneracy can explain the overabundance of systems with inclination near 90 degrees. In fact, the difference in the distribution of eccentricity in the control sample versus the sample with exoplanets is not statistiscally significant ($p=0.119$).}

\subsection{Mixture model}
\label{sec:mixture}
We fit a mixture model to the orbital inclinations in our sample of binaries with planets in an attempt to determine what fraction of the binary systems with exoplanets can be explained by a random distribution versus what fraction is aligned. The probability density function of $i$ can be expressed as $\sin(i)$ (although selection effects slightly bias the distribution as shown in the previous section), while we represent the distribution of aligned binaries inclinations as a normal distribution. Our likelihood function is defined as: 
\begin{equation}
    p(y \vert \lambda,\sigma) = \prod_{n=1}^{N} (\lambda\cdot\mathcal{N}(y_n \vert \mu, \sigma) + (1-\lambda)\cdot\sin{y_n})
\end{equation}
Where $y$ are the binary orbital inclinations, $\mathcal{N}$ is a Normal distribution with mean $\mu$ and standard deviation $\sigma$, and $\lambda$ is the mixture parameter, or the fraction of systems in the aligned population. We impose the following priors: $\mu$ is fixed to be $\pi/2$ radians (perfectly aligned with the planet orbits), $\sigma$ is constrained with a half normal distribution with standard deviation $\frac{\pi}{2}$ radians and $\lambda$, the mixture parameter, has a uniform prior from 0 to 1. Specifically,
\begin{equation}
        \sigma \sim Normal(0,\frac{\pi}{2})
\end{equation}
\begin{equation}
    \sigma > 0
\end{equation}
\begin{equation}
    \lambda \sim Uniform(0, 1)
\end{equation}
\begin{equation}
    \mu = \pi/2
\end{equation}
We run the mixture model in the probabilistic programming language Stan \citep{pystan}. The resultant distribution of $\lambda$ is peaked near 50\%, with the 90\% confidence interval being [0.33,0.72]. Though the 90\% confidence interval is fairly wide, it does allow us to conclude that, assuming our Gaussian + $\sin(i)$ model for the distribution is a reasonable approximation, the process of alignment occurs in in between 26\% and 72\% of the systems in our sample with 90\% confidence.
\begin{figure}
    \includegraphics[width=\columnwidth]{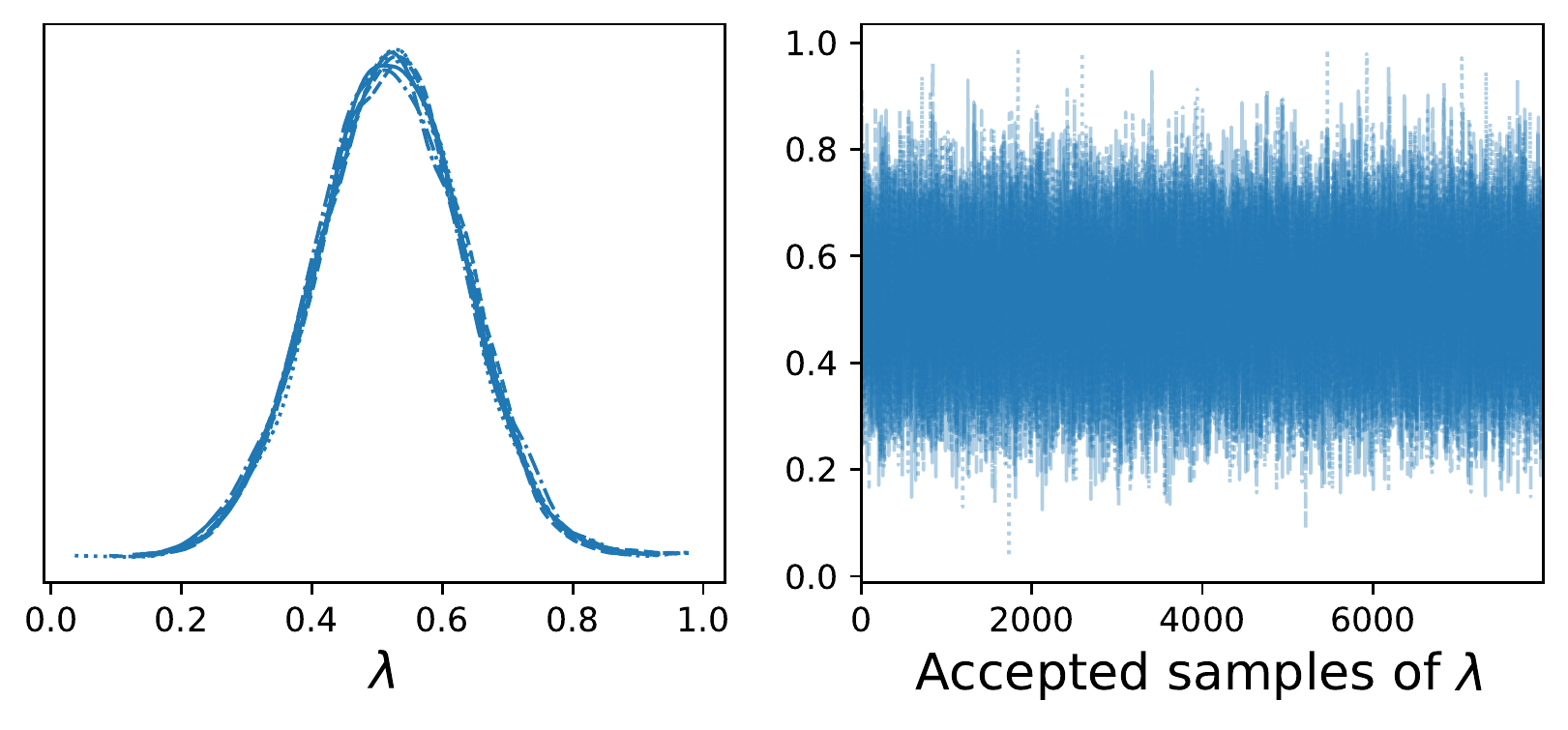}
    \caption{The posterior probability distribution of the mixture parameter $\lambda$ for 8 chains and the accepted samples of  $\lambda$. Our mixture model fit prefers a roughly 50\%/50\% ratio of systems in an aligned Gaussian population to those in a random $\sin(i)$ distribution.}
    \label{fig:my_label}
\end{figure}

\subsection{Assessing Possible Biases}
\label{sec:biases}
While we were careful to construct a control sample of binaries with nearly identical properties to the exoplaet sample, there are a few subtle differences that could plausibly introduce a difference in the distribution of inclinations. Here, we show that these differences do not significantly affect the distribution of inclination angles within the control sample, and therefore are highly unlikely to significantly contribute to the apparent alignment between planetary systems and wide binary orbits. 

\subsubsection{Metallicity measurements}
Some of the binary systems with exoplanets have metallicity measurements from our own TFOP follow-up spectroscopic observations (see Section \ref{sec:archivalSpec}). We do not have metallicity measurements for any binaries in the control sample aside from those with archival observations identified by \citet{elbadry2019}, so it is plausible that including the TFOP spectroscopy could introduce a small difference between the samples. To test this, we compare the inclination distributions of binary systems with spectroscopic metallicity measurements to those without using a K-S test.

\begin{figure}
    \centering
    \subfloat{\includegraphics[width= \columnwidth]{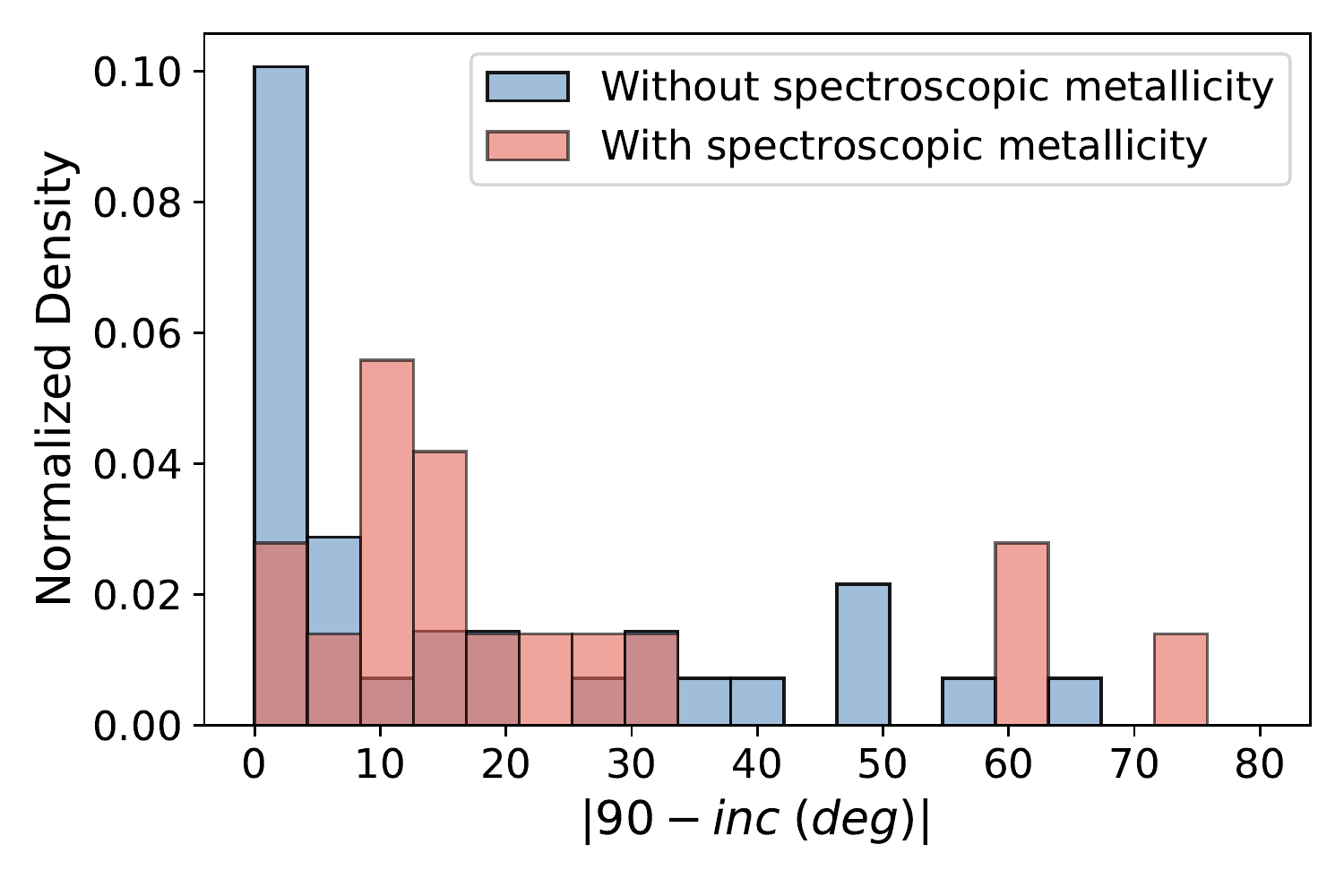}}\\
    \subfloat{\includegraphics[width= \columnwidth]{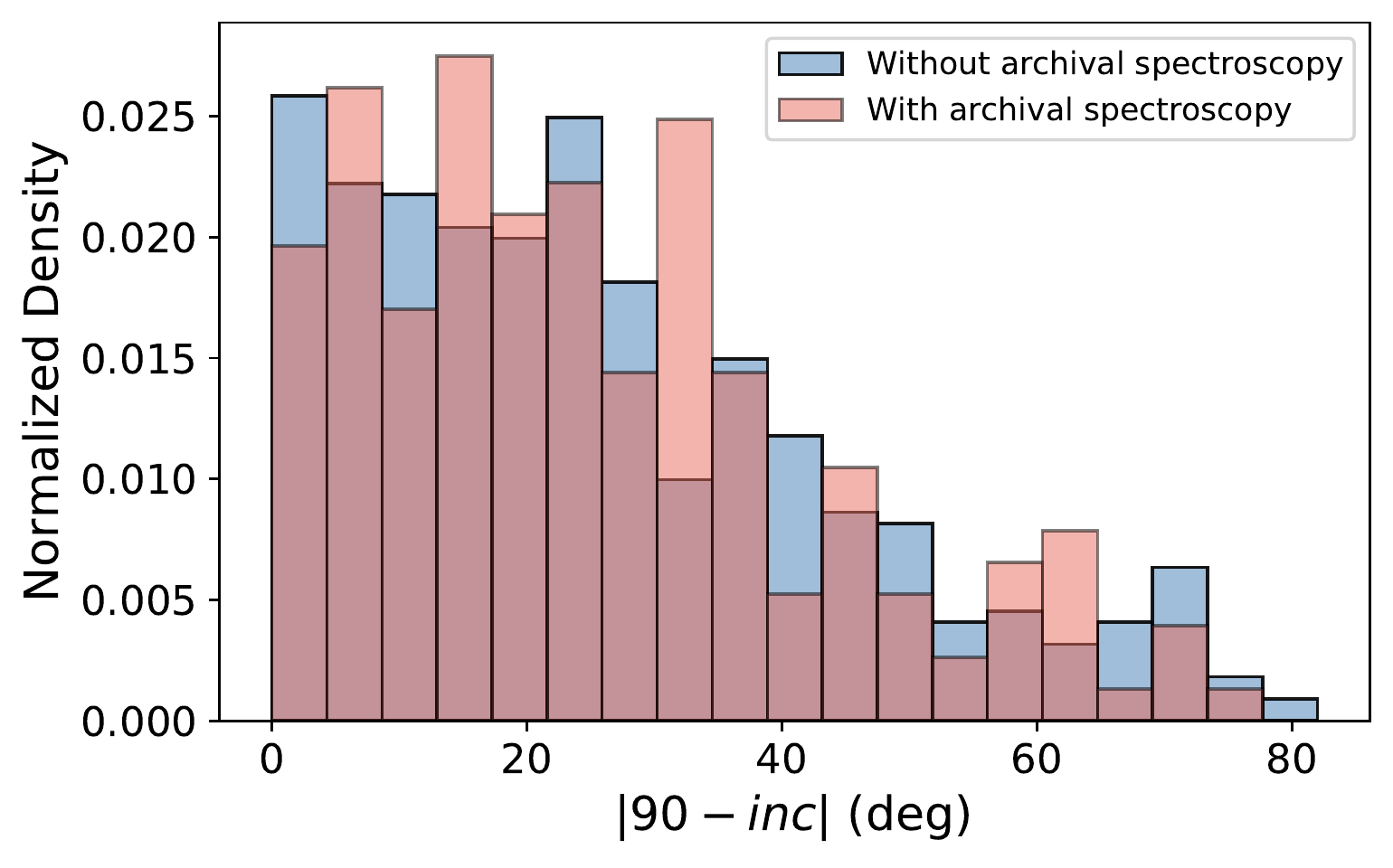}}\\
    \caption{A comparison of binaries with exoplanets with and without spectroscopic metallicity measurements from TFOP (a) and the binaries in the control sample with and without archival spectroscopy (which metallicity is derived from). Neither comparison shows a statistically significant difference, so differences in metallicity measurements are unlikely to explain the difference in the inclination distributions between the exoplanet and control sample. }
    \label{fig:metallicityComparison}
\end{figure}

First, we tested whether there is a significant difference between the inclination distribution of binaries with and without TFOP spectroscopy in our exoplanet sample. A K-S test reveals no statistically significant difference between these two subsamples ($p=0.32$). In fact, if anything, the addition of spectroscopy leads to less of an alignment as can be seen in Figure \ref{fig:metallicityComparison}a.

We also tested whether there is a significant difference between the inclination distribution of binaries with and without archival spectroscopic metallicity measurements in our control sample. Here, the K-S test also reveals no statistically significant difference between those samples in the control sample with and without archival spectroscopy (Section \ref{sec:archivalSpec}) as can be seen in Figure \ref{fig:metallicityComparison}b ($p=$0.699).

These tests show that evidently, the presence or absence of a spectroscopic metallicity measurement has a negligible impact on the resulting distribution of binary orbital inclinations. This effect is therefore highly unlikely to explain the apparent alignment of wide binary orbits with their planetary systems.

\subsubsection{Two-minute Cadence vs FFI light curves}

As described in Section \ref{sec:Planets}, \TESS\ observes 20,000 of the best planet-search target stars at two-minute cadence \citep{Stassun2019, Barclay2018} per sector, while the rest of the sky is observed at 30-minute cadence (and more recently, at 10-minute cadence in the extended mission). One of the selection criteria for two-minute-cadence observations is the extent to which nearby stars dilute the signal of the target star. Therefore, stars chosen for two-minute cadence observations were less likely to be nearby other bright stars, potentially including visual binary companions. This could lead to an increase in edge-on binaries in two-minute cadence observation since companions could be more likely to be very close to the primary star, and therefore less likely to be resolved in ground-based seeing-limited imaging and consequently less likely to be rejected due to the companion's flux dilution. Because transit searches in two-minute cadence observations are more sensitive to detecting planet candidates, this could plausibly introduce a bias in our results.

\begin{figure}

   \includegraphics[width= \columnwidth]{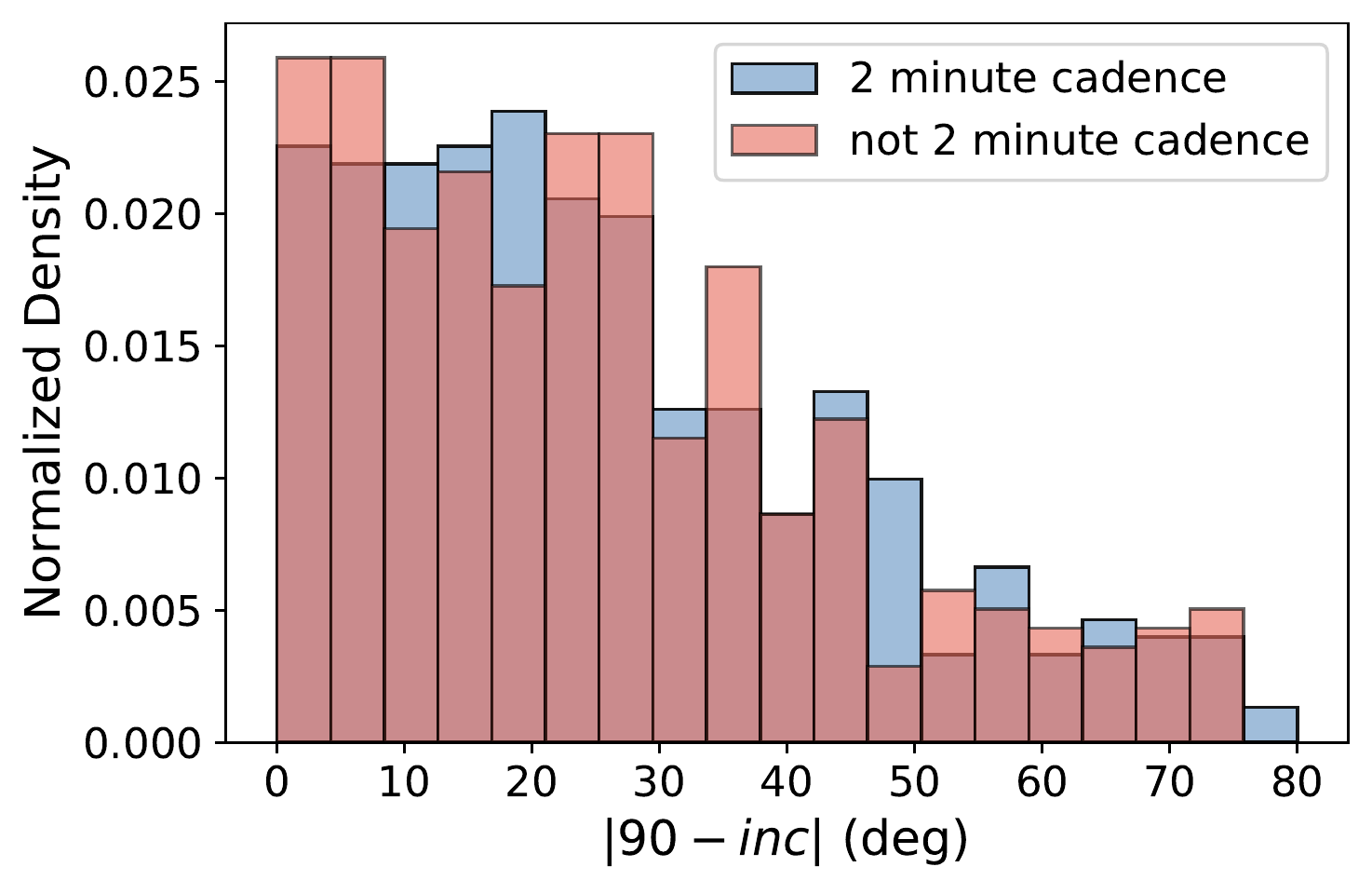}\\
 
    \caption{A comparison in the distribution of inclination in binary systems in the control sample with 2 minute cadence and those without. The two distributions are essentially indistinguishable, so selection biases from the \TESS\ two-minute target selection are unlikely to affect our results. }
    \label{fig:cadenceComparison}
\end{figure}

We tested whether the selection of two-minute targets significantly affects the inclination distribution of wide binary companions by comparing the distributions of stars in our control sample that were and were not observed in two-minute cadence mode. As demonstrated in Figure \ref{fig:cadenceComparison}, we find no significant difference in the distribution of inclinations for stars in the control sample observed in two-minute cadence. A K-S test fails to reject the null hypothesis that binary inclinations for stars observed at two-minute cadence and those observed in FFIs are drawn from the same population ($p=0.85$). The \TESS\ two-minute cadence target selection is therefore highly unlikely to be a source of bias.

\subsubsection{Additional Biases}
Visual binary stars that are not resolved by TESS introduce complex biases into transiting exoplanet detection that might plausibly affect the measured binary inclination distribution \citep{bouma2018}. In our study, 45 systems have separations less than 30 arcseconds, the approximate resolution of TESS. In this case, however, the difference in detectability of exoplanets in visual binary systems that are resolved in TESS vs. those that are not resolved should not cause a preferential alignment between exoplanet and binary system. A larger fraction of binary system that have face-on orbits with respect to the plane of the sky are expected to be resolved by TESS than those with edge-on orbits. Thus, if anything, the larger fraction of resolved misaligned systems would cause the opposite effect from what we observe.

It is known that because LOFTI is inferring orbits from such short orbital arcs,  inferred orbits can preferentially be biased towards higher inclinations (i.e. towards $i=90$, an alignment with the location of the planetary systems in our study) \citep{ferrerchavez}. However, there is no reason to suspect that this bias is affecting the sample with exoplanets anymore than the control sample.

\section{Dynamical Mechanisms to Explain the Existence of the Preferential Alignment}
\label{sec:dynamics}
In this section, we discuss two possible dynamical mechanisms that could explain a preferential alignment between visual binaries and the planets in those systems. 

\subsection{Lidov-Kozai Timescale}
\label{sec:lidovkozaisec}
The Lidov-Kozai mechanism induces oscillations in a pair of objects (in this case the planet and star) from the secular influence of a massive, far-away companion. Bodies undergoing the Lidov-Kozai effect experience oscillations in their inclination and eccentricity, trading off between the two quantities on a characteristic timescale set by the system parameters. The Lidov-Kozai mechanism has been invoked to explain a wide variety of astrophysical phenomena, including the existence of hot Jupiters \citep{Fabrycky2007, Naoz2012, Dawson2018, Li2020}. 

Since the planet mass is much smaller than the inner stellar mass, Lidov-Kozai effect will only drive substantial dynamical evolution when there is a significant mutual inclination ($\gtrsim 40^\circ$) between the orbit of the planet-star system and the orbit of the distant binary companion \citep{Naoz2013}. The fact that the Lidov-Kozai effect operates on systems with large mutual inclinations means that it could plausibly explain the alignment we observe between the orbits of wide binary stars and their close-in planetary systems. If, for example, the inclinations of wide binary orbits were randomly distributed with respect to the orbits of the planetary system, the Lidov-Kozai effect would only act to significantly alter the geometries of systems with large initial misalignments. If the Lidov-Kozai effect preferentially disrupts systems with large initial misalignments (either by driving the eccentricity high enough to cause a collision between the planet and the star, or by causing dynamical instabilities that lead to planetary collisions or ejections), the resultant observed distribution of relative inclinations could develop some non-uniformity, as seen in our sample.

\begin{figure}

   \includegraphics[width= \columnwidth]{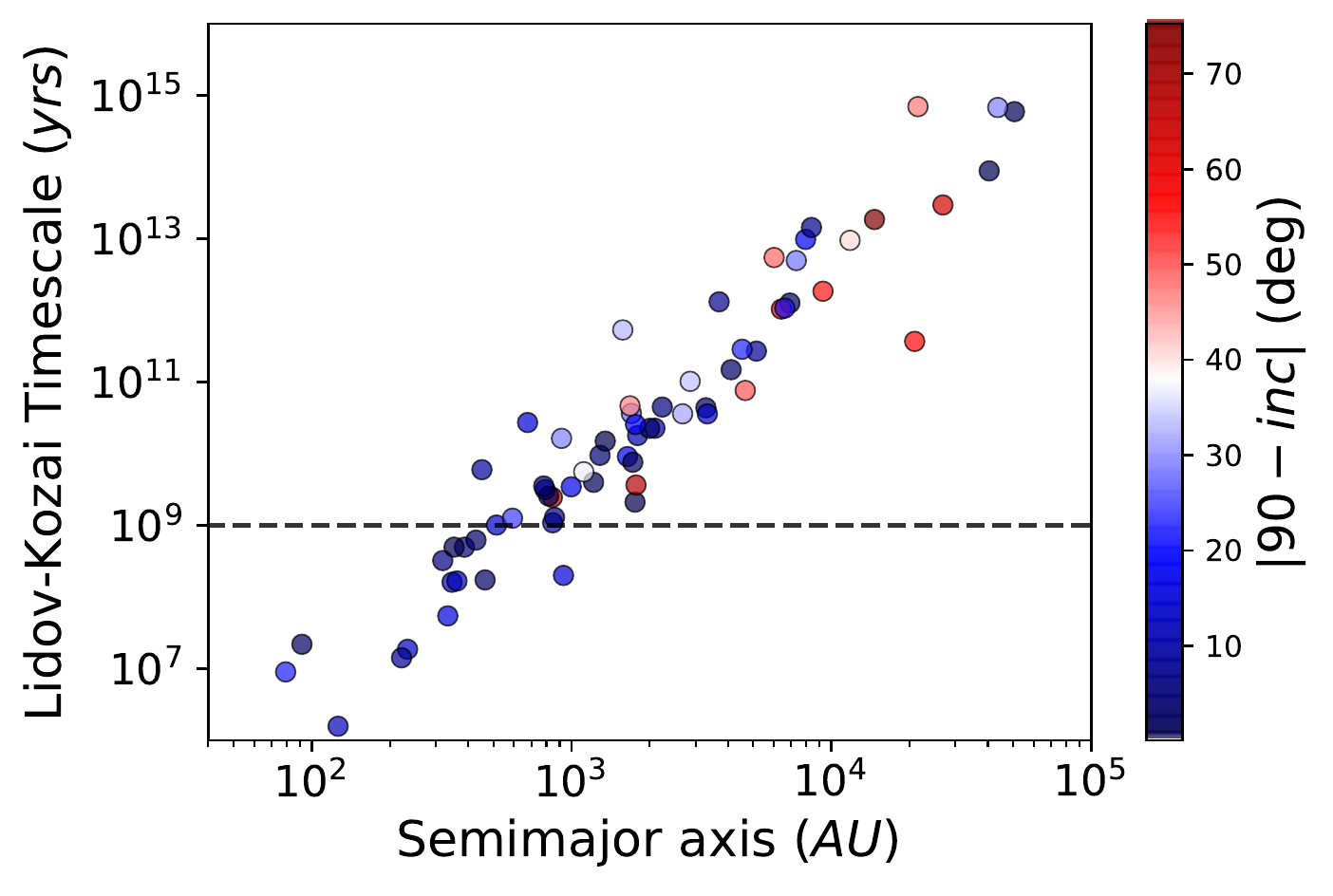}\\
 
    \caption{Lidov-Kozai timescale for binary systems with exoplanets plotted against semimajor axis of the binary companion. The colorbar represents $\vert 90^\circ-i \vert$ for each system. The dashed line represents the approximate ages of the stars ($\sim$ 1 Gyr). The wide binaries with semimajor axes less than about 700 AU are close enough that the Kozai-Lidov effect could operate on similar timescales, but other factors make this explanation for the alignment less likely.}
    \label{fig:lidovKozai}
\end{figure}

To assess whether the Lidov-Kozai effect could explain the alignment between wide binary orbits and planetary orbits we observe, we calculate the timescale for Lidov-Kozai oscillations to take place for each system, assuming a large enough primordial mutual inclination for the effect to take place. If the Lidov-Kozai timescale is significantly smaller than the total system age, it could have acted to sculpt the observed alignment.  The oscillations induced by the Lidov-Kozai timescale occur on a characteristic time-scale, $\tau_{\rm LK}$, which can be estimated as: 
\begin{equation}
    \tau_{\rm LK} \approx \frac{8}{15\pi} \left(1+\frac{m_{h}}{m_c}\right) \left(\frac{P^2_{c}}{P_{p}}\right)(1-e_p^2)^{(3/2)}
\end{equation}
where $m_{h}$ is the mass of the planet host star (i.e. the primary star), $m_c$ is the mass of the stellar companion, $P_c$ is the period of the binary companion, $P_p$ is the period of the planet, and $e_p$ is the eccentricity of the planet \citep{Holman1997, Antognini2015}. The period of the planet is very small compared to the mutual period of the stars. We calculate the Lidov-Kozai timescale for each system in our sample and show the results in Figure \ref{fig:lidovKozai}, as a function of the semimajor axis of the binary system. 

We find that in systems where the Lidov-Kozai timescale is significantly less than the total age of the system (the systems below the dashed line in Figure \ref{fig:lidovKozai}) the systems are preferentially aligned. This is a consequence of the fact that most systems with binary semimajor axes smaller than about 700 AU tend to have orbits aligned with their planetary systems in our sample. As a result, it is plausible that the Lidov-Kozai effect would operate rapidly enough in systems below 700 AU to contribute to the alignment we see. 

However, we suspect that the Lidov-Kozai effect is likely not a dominant effect causing the observed alignment between binary and planetary orbital planes. One reason is that for planets orbiting very close to their stars, even small apsidal precession rates (due to either general relativity, tides, or the stellar quadrupole moment) can completely suppress the Lidov-Kozai effect \citep{Sterne1939, MD99, Fabrycky2007}. Also, any sculpting by the Lidov-Kozai effect would tend to leave systems with mutual inclinations with up to 40 degree misalignments, while the overdensity of aligned systems in our sample is strongest within 10 degrees of alignment. Thus, it is likely that if the Lidov-Kozai effect is acting in systems, the effect is quite small and would need to be coupled with other effects to explain the observed alignment.

\subsection{Disk Timescale}
\label{sec:disksec}


Another process that could explain the preferential alignment of wide binary orbits with their close-in planetary systems is that the gravitational influence of the wide binary star could torque the protoplanetary disk into alignment early in the system's history. Consider, for example, a wide binary companion in an initially misaligned orbit around a star with a gas-rich protoplanetary disk.  The mechanism of \citet{Batygin2012} (see also \citealt{bate2000}) could be induced by the wide binary companion, causing the protoplanetary disk to precess in the reference frame of the binary star system, which would manifest as an oscillation of inclination in the reference frame of the host star's angular momentum vector. As the protoplanetary disk precesses, it can simultaneously dissipate energy and move towards its lowest energy state, an alignment with the binary star angular momentum and disk angular momentum vectors \citep{bate2000,Martin2017}. Any planets that subsequently form from the disk would tend to be in well aligned orbits with the wide binary companion.

The warping of a disk can lead to dissipation of energy in the disk. The effect of this warping on alignment of the disk has been studied in \citet{bate2000} and \citet{Zanazzi2018} in a debris (collision-less) disk. The timescale of alignment due to only warping of the debris disk is generally too long to act in the systems we observe according to \citet{Zanazzi2018}. However, it is well known that any gas disk, regardless of warping, will thermalize energy due to turbulence or viscosity \citep{Nelson2000,Dalessio2006}. As a disk radiates away this energy during precession, it will move to its lowest energy state, an alignment with the binary star system.



To explore the possibility that the disk is dissipating energy and moving towards an alignment via a precession, we use the approximate closed-form expression of \citet{Batygin2012} supplementary materials Equation 6 to estimate the timescale of one precession, and thus one cycle of oscillation in inclination. The timescale of precession is not the same as the timescale of alignment, but given that sufficiently fast precession is a necessary condition for alignment, this timescale can be used to somewhat estimate if alignment could take place in the systems in our sample. \citet{Batygin2012} gives
\begin{equation}
    \diff{\Omega_{disk}}{t} = - \frac{\int_{a_{in}}^{a_{out}}G \Sigma(r) M_{c} (\frac{r}{a_{c}})^2\Tilde{b}^{(1)}_{3/2}(0.95)\textrm{dr}}{4\int_{a_{in}}^{a_{out}} \Sigma(r) \sqrt{G M_{\star}r^3}\textrm{dr}} \cos(i)
    \label{eq:precessionEq}
\end{equation}
where $a_{c}$ and $M_{c}$ are the semimajor axis and mass of the companion star, respectively, $M_\star$ is the mass of the host star, $i$ is the inclination of the disk with respect to the binary orbit, $\Sigma(r)$ is the disk density profile, $a_{out}$ and $a_{in}$ are the inner and outer boundaries of the disk, and $\Tilde{b}^{(1)}_{3/2}$ is a Laplace coefficient with disk softening (see \citet{Batygin2012} supplementary materials Equation 4 for details). Since binaries with smaller separations tend to have smaller outer disk radii \citep{Manara2019}, we calculate all disk timescales for $a_{out}$ in the range $[a_{b}*0.05,a_{b}*0.25]$ if $a_b<400$ and otherwise $[20,100]$, where $a_b$ is the semimajor axis of the binary star system. \citet{Manara2019} calculates dust radii of disks, but the gas radii of disks (which is most responsive to torques) is expected to be larger than the dust radius, although the two tend to be proportional \citep{Ansdell2018}. In general, $\tau_\Omega$ has a very weak dependence upon the value of $a_{out}$ as long as $a_{out}$ is within a reasonable range of values. We use a density profile of the form
\begin{equation}
   \mathlarger{\Sigma}(r)=\mathlarger{\Sigma}_c \left (\frac{r}{R_c}\right)^\gamma \exp{\left [-\left (\frac{r}{R_c}\right)^{2-\gamma} \right ]}
\end{equation}
where $\gamma$ is 0.9 \citep{Andrew2010} and $R_c$ is 110 AU, the mean \citet{Andrew2010} value. Assuming the parameters of Equation \ref{eq:precessionEq} do not vary greatly over the precession cycle, the timescale of one complete precession cycle can be estimated as
\begin{equation}
    \tau_{\Omega} = \pi^2\left(\frac{4\int_{a_{in}}^{a_{out}} \Sigma(r) \sqrt{G M_{\star}r^3}\textrm{dr}}{\int_{a_{in}}^{a_{out}}G \Sigma(r) M_{c} (\frac{r}{a_{c}})^2\Tilde{b}^{(1)}_{3/2}(0.95)\textrm{dr}}\right)
\label{eq:timescalePrecession}
\end{equation}

\begin{figure}
    \includegraphics[width=\columnwidth]{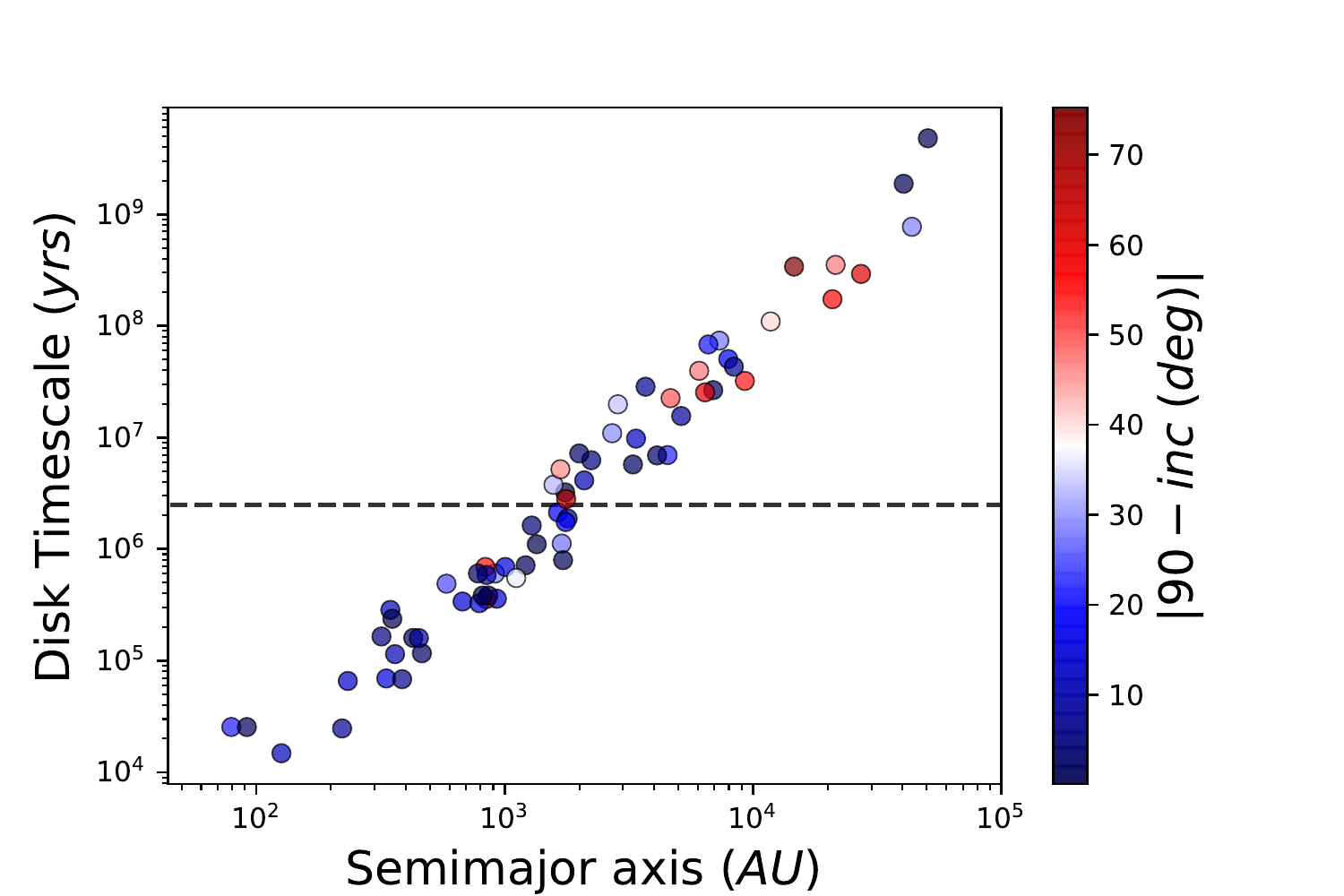}
    \caption{The disk precession timescale as simulated for the sample of binary star systems with exoplanets. The horizontal dashed line indicates the typical 2.5 Myr half-lifetime of protoplanetary disks \citep{Mamajek2009}. The well-aligned sample of binaries with semimajor axes shorter than about 700 AU typically have disk precession timescales short enough that this mechanism could explain the alignment.}
    \label{fig:diskFig}
\end{figure}

Here, Equation \ref{eq:timescalePrecession} is averaged over possible inclination values (the mean of $\cos^{-1}(i)$ for isotropic inclinations is $\pi/2$). We calculated the disk precession timescales for each system in our sample and show the resulting timescales in Figure \ref{fig:diskFig}. We find that systems with disk precession timescales less than a few Myr tend to be preferentially aligned, while systems with longer precession timescales show more large misalignments. It is noteworthy that a timescale of a few Myr happens to be the typical lifetime for protoplanetary disks \citep{Mamajek2009}, indicating that this mechanism could explain both the existence of an alignment and the greater level of alignment for binaries closer than about 700 AU.

\section{Discussion}
\label{sec:discussion}

Using data from the \TESS\ and \Gaia\ space missions, along with ground-based follow-up observations, we have shown that the orbits of wide binary stars with transiting exoplanets are significantly more likely to have an inclination near 90$^\circ$ than randomly selected binary systems. {Inclination is equal to the minimum possible alignment between binary orbit and exoplanet orbit, so} an excess of inclinations near 90$^\circ$ for transiting planet hosting systems, as compared to those without known transiting planets, suggests an underlying physical process that results in preferential alignment between outer stellar companions and inner transiting planets. Understanding the origin of such a preferential alignment will have implications for understanding the formation and evolution of both planets and binary stars. Here, we discuss three possible ways in which such an alignment may have come to be.

\subsection{Possibility 1: Primordial alignment}
\label{sec:possibility1}
One possibility is that the alignment between wide binary and close-in planetary orbits is primordial. That is, the planets and binary each formed in nearly their current aligned configuration. Binary star systems at relatively low separations can form through disk fragmentation and turbulent fragmentation. In disk fragmentation, the initial gravitationally unstable circumstellar disk fragments during the collapse of the system, forming a binary or higher multiplicity system \citep{adams1989,Bonnell1994,Sigalotti2018}. Disk fragmentation is expected to form binary systems where the stellar angular momenta are aligned with the binary orbit and the plane of the protoplanetary disk. Evidence for disk fragmentation includes aligned bipolar outflows \citep{Tobin2013} and the metallicity dependent binary fraction \citep{elbadry2018}. Recent observations have shown that disk fragmentation primarily takes place for binary stars with semimajor axes less than about 200 AU \citep{Tobin2016, elbadry2019}. More distant binaries likely formed via only turbulent fragmentation, which is not necessarily expected to produce binary stars with orbits in the same plane as any planetary system. Such systems can also migrate to smaller separations (i.e. below 200 AU). 

One way to explain our observation that there appears to be a preferential alignment between wide binaries and planets, and that the alignment is most prominent for binary stars with semimajor axes less than 700 AU, is that close-in binaries form aligned with the protoplanetary disk, while more distant binaries form via a different mechanism that leaves their orbits misaligned. In this case, presumably this would imply that the close-in binaries in our sample formed mainly via disk fragmentation, while the more distant binaries formed via turbulent fragmentation and, more rarely, capture of the binary companion. If this were true, it would  would necessitate that most binaries with semimajor axes less than 700 AU formed from disk fragmentation, which would be in tension with the apparent $\sim$200 AU cutoff for disk fragmentation \citep{Tobin2016,elbadry2018}, as well as the fact that some binaries below 200 AU should have formed via turbulent fragmentation and later migrated inwards.




\subsection{Possibility 2: Dynamically sculpted binary orbit misalignment}

If the binaries in our system tend to all form aligned (which is not supported by the observed occurrence of turbulent fragmentation at separations similar to those seen in our study), then the binary orbits of the systems above the threshold $\sim$700 AU could be sculpted by the dynamical influence of passing stars. \citet{Deacon2020} has shown that wide-binaries are rare in open clusters, suggesting that open clusters are a highly dynamical environment. Before complete disruption of a binary system, the system can be dynamically perturbed, leading to the larger fraction of misaligned systems we see above $\sim$700 AU. Such a dynamical sculpting need not take place in only clusters: \citet{Kaib2013} found that the orbits of binary systems can be altered by the galactic tide and passing stars even when systems are not in clusters.

However, as discussed in Section \ref{sec:possibility1}, observational evidence suggests that aligned systems form primarily below 200 AU and thus the cutoff is probably not primarily due to a dynamical sculpting of the binary orbits.

\subsection{Possibility 3: Dynamically sculpted alignment}
\label{sec:possibility2}
The final possibility we consider is that some dynamical process has sculpted the population of planets in wide binaries into orbital alignment after formation. Wide binary systems with separations similar to those probed in this work are expected to form primarily via turbulent fragmentation, where turbulence fractures the initial stellar core into multiple separate over-densities \citep{Offner2010,Offner2016,Bate2018}. In turbulent fragmentation, the two stars tend to have a distribution of angular momentum vectors that is less aligned on average than disk fragmentation \citep{Offner2016,Bate2018}. However, a larger portion of binaries that form via turbulent fragmentation are still expected to have aligned orbits than a completely isotropic distribution \citep{Bate2018}. Indirect evidence for turbulent fragmentation in wide-binary systems includes outflow orientations \citep{Lee2016} and direct imaging \citep{Fernandez2017}.

The initial misalignment of some binary systems that form via turbulent fragmentation will cause the disk to precess by the mechanism described in Section \ref{sec:disksec}, dissipate energy, and lead to an alignment between the disk and binary orbit. 
This scenario seems like it could explain both the observed alignment and the transition from predominantly aligned systems to more misaligned systems around $\sim$700 AU, which corresponds to a disk precession timescale similar to the lifetime of protoplanetary disks. 

\subsection{Future work}
We have presented evidence showing that the orbits of wide binary stars are preferentially aligned with the orbits of close-in planets in those systems. Although the alignment appears to be statistically significant, it is based on a relatively small sample of planetary systems. Confirming the conclusions of this work with a larger sample of binary stars hosting transiting exoplanets would be highly valuable.

Fortunately, it should be straightforward to increase the size of our sample. Recently, \citet{elbadry2021} released a sample of over a million wide binary stars from \textit{Gaia} EDR3, including 274 visual binaries found to have exoplanets by \TESS\ as of February 1st 2021. With this new sample of binaries, it will be possible to triple the size of our current sample.  In addition, with improved \textit{Gaia} astrometry from future data releases, such as more systems with radial velocity and an astrometric acceleration term, the inclinations of systems can be better constrained. If our hypothesis that most of the binaries form via turbulent fragmentation and were dynamically sculpted into aligned systems is true, then the statistical significance of the difference in inclination distribution of binaries above and below 700 AU should increase, and a clearer dividing timescale between aligned and randomly distributed systems should emerge.

A larger sample of binaries will also make it possible to subdivide the sample and look for correlations with planetary parameters. For example, initially misaligned wide binary companions could help trigger the formation of hot Jupiters via Lidov-Kozai oscillations, so it would be interesting to see whether the inclinations of wide binary companions in hot Jupiter systems are different from those in systems with small planets. Another important characteristic to investigate is the transiting planet system multiplicity. Compact multiplanet systems planets are susceptible to inclination oscillations from exterior companions that can change the planet's mutual inclinations enough to prevent them all from transiting \citep{beckeradams}. 

Another valuable avenue may be to make similar measurements for different samples of planetary systems. For example, ground-based high-resolution imaging observations of binaries too close to be resolved by \Gaia\ can help determine whether binaries with semimajor axes less than those probed in this work (a $\lesssim$ 100 AU) also show preferential alignment. It may also be possible to increase the sample of particularly wide binaries (a $\gtrsim$ 1000 AU) by including planet candidates orbiting the more distant stars targeted by \Kepler\ and K2 that reside in binary systems that can be resolved by \Gaia.

Finally, we note that as follow-up of \TESS\ planets continues, the purity of the \TESS\ planet candidate sample will increase and systematics in TESS planet detection will become better understood. The ongoing TFOP observation campaign will continue identifying false positives among the \TESS\ planet candidate sample and increase the likelihood that any given candidate in the surviving sample is indeed a planet candidate. Identifying these false positives will remove noise from the distribution of wide binary inclinations. As the systematics in TESS planet detection become better understood, it can be determined with a higher degree of confidence whether the observed alignment is affected by a bias in TESS' detection method.

\section{Conclusion}
\label{sec:conclusion}
Given the high frequency of wide-binary systems, understanding the evolution of protoplanetary disks and later, planets, is important to having a holistic understanding of planet formation and evolution. Various dynamical effects have been proposed for wide-binary systems (e.g. \citealt{Wu2003,Batygin2012}), but observational studies of planets in wide-binary systems are sparse, limiting the confirmation of these dynamical effects.

We conducted a study of planets in wide-binary systems and demonstrated that the orbits of wide binaries and planets residing in the binaries are aligned ($p=0.0037$). We first gathered a sample of wide-binary systems with exoplanets along with a control sample with matching properties (Section \ref{sec:observations}). We then derived stellar masses (Section \ref{sec:massFitting}) and estimated probable orbits for the wide binary systems (Section \ref{sec:LOFTI}). {We found that there was a statistically significant overabundance of systems around $i=90$, suggesting that planets and the wide-binary systems they reside in tend to be aligned (Section \ref{sec:KStest})}. We found that the alignment appears to occur primarily in systems with binary semimajor axes less than $\sim$700 AU.

We then presented a mixture model to attempt to derive the amount of wide binaries that are aligned (Section \ref{sec:mixture}). Although the results were relatively uninformative, it is likely that about half of the systems in our sample are aligned (with at least 25\% alignment at 97.5\% confidence). We found that no biases we could identify were causing the alignment (Section \ref{sec:biases}).

We derived Lidov-Kozai timescales for the binary systems with planets (Section \ref{sec:lidovkozaisec}). Additionally, we estimated the disk precession timescale \citep{Batygin2012} for the systems (Section \ref{sec:disksec}). We observed that in the case of the disk precession mechanism, almost all misaligned systems have disk precession timescales greater than the typical age of protoplanetary disks.

Finally, we discussed possible mechanisms for the observed alignment (Section \ref{sec:discussion}). The binary systems that form via disk fragmentation are expected to be aligned initially, while the binary systems that form via turbulent fragmentation and capture should have a more or less isotropic distribution of alignments. In order to explain an alignment for the binary systems that form via turbulent fragmentation, we proposed that the disk mechanism and, to a lesser extent, the Lidov-Kozai effect, can lead to preferential alignment of these systems early in their lifetime (Section \ref{sec:possibility2}).

As more exoplanets are discovered by TESS, the effects observed in this study are worth revisiting. Particularly, with a larger sample, in the future we could conclusively detect the alignment, probe the effect of parameters like planet mass or multiplicity, and more conclusively determine whether the alignment is stronger for binary systems with semimajor axes less than $\sim$700 AU.

\section{Acknowledgements}

We thank the anonymous referee and Eric Feigelson (the AAS Journals statistics editor) for valuable comments that strengthened our manuscript. We acknowledge the members of Dave Latham’s Coffee Club for their helpful feedback and insights. 
We thank Coco Zhang, Konstantin Batygin and Darryl Seligman for useful conversations. 

The authors acknowledge the Texas Advanced Computing Center (TACC) at The University of Texas at Austin for providing high performance computing resources that have contributed to the research results reported within this paper. URL: \url{http://www.tacc.utexas.edu} 

This paper includes data collected by the TESS mission, which are publicly available from the Mikulski Archive for Space Telescopes (MAST). Funding for the TESS mission is provided by NASA's Science Mission directorate.

This work has made use of data from the European Space Agency (ESA) mission
{\it Gaia} (\url{https://www.cosmos.esa.int/gaia}), processed by the {\it Gaia}
Data Processing and Analysis Consortium (DPAC,
\url{https://www.cosmos.esa.int/web/gaia/dpac/consortium}). Funding for the DPAC
has been provided by national institutions, in particular the institutions
participating in the {\it Gaia} Multilateral Agreement.

This work makes use of observations from the LCOGT network. Part of the LCOGT telescope time was granted by NOIRLab through the Mid-Scale Innovations Program (MSIP). MSIP is funded by NSF.

This article is based on observations made with the MuSCAT2 instrument, developed by ABC, at Telescopio Carlos Sánchez operated on the island of Tenerife by the IAC in the Spanish Observatorio del Teide.

This paper is based on observations made with the MuSCAT3 instrument, developed by the Astrobiology Center and under financial supports by JSPS KAKENHI (JP18H05439) and JST PRESTO (JPMJPR1775), at Faulkes Telescope North on Maui, HI, operated by the Las Cumbres Observatory.

The IRSF project is a collaboration between Nagoya University and the South African Astronomical Observatory (SAAO) supported by the Grants-in-Aid for Scientific Research on Priority Areas (A) (Nos. 10147207 and 10147214) and Optical \& Near-Infrared Astronomy Inter-University Cooperation Program, from the Ministry of Education, Culture, Sports, Science and Technology (MEXT) of Japan and the National Research Foundation (NRF) of South Africa.

This work is partly supported by JSPS KAKENHI Grant Number JP18H05439, and JST PRESTO Grant Number JPMJPR1775, and a University Research Support Grant from the National Astronomical Observatory of Japan (NAOJ).

This work is partly supported by Grant-in-Aid for JSPS Fellows, Grant Number JP20J21872.

This work is partly supported by JSPS KAKENHI Grant Number JP17H04574.

This work is partly supported by JSPS KAKENHI Grant Number JP20K14518, and by Astrobiology Center SATELLITE Research project AB022006

M.T. is supported by MEXT/JSPS KAKENHI grant Nos. 18H05442, 15H02063, and 22000005.

This work is partly supported by JSPS KAKENHI Grant Number JP21K13955.

This work is partly supported by JSPS KAKENHI Grant Number 20K14521.

C. R-L acknowledges financial support from the State Agency for Research of the Spanish MCIU through the Center of Excellence Severo Ochoa award for the Instituto de Astrofísica de Andalucía (SEV-2017-0709)

M. R. acknowledges support from the Universidad Cat\'olica de lo Sant\'isima Concepci\'on grant DI-FIAI 03/2021.

P.J.A. acknowledges support from grant AYA2016-79425-C3-3-P of the Spanish Ministory of Economony and Competitiveness (MINECO) and the Centre of Excellence ``Severo Ochoa'' award to the Instituto de Astrof\'isica de Andaluc\'ia (SEV-2017-0709)

This paper is based on observations made with the T150 telescope at the Sierra Nevada observatory (Granada, Spain), operated by the Instituto de Astrofísica de Andalucía (IAA - CSIC).

The research leading to these results has received funding from the ARC grant for Concerted Research Actions, financed by the Wallonia-Brussels Federation. TRAPPIST
is funded by the Belgian Fund for Scientific Research (Fond National de la Recherche Scientifique, FNRS) under the grant FRFC
2.5.594.09.F. TRAPPIST-North is a project funded by the University of Liège (Belgium), in collaboration with Cadi Ayyad University of Marrakech (Morocco).

D. D. acknowledges support from the TESS Guest Investigator Program grant 80NSSC19K1727 and NASA Exoplanet Research Program grant 18-2XRP18\_2-0136.

MG and EJ are F.R.S.-FNRS Senior Research Associates.

K.K.M. acknowledges support from the New York Community Trust’s Fund for Astrophysical Research.

This work has been carried out within the framework of the NCCR PlanetS supported by the Swiss National Science Foundation.

This work makes use of observations from the ASTEP telescope. ASTEP benefited from the support of the French and Italian polar agencies IPEV and PNRA in the framework of the Concordia station program and from Idex UCAJEDI (ANR-15-IDEX-01).

\section{Data Availability}
We present all supplementary data tables in the arxiv source code. Individual LOFTI posteriors are available upon request.



\textit{Facilities:} Gaia, TESS, FLWO:1.5m (TRES), CTIO:1.5m (CHIRON), UKST (RAVE), LCOGT (NRES), FIES, LAMOST, APOGEE, GALAH, ASTEP \citep{Guillot2015}, MuSCAT \citep{muscat1}, MuSCAT2 \citep{muscat2}, MuSCAT3 \citep{muscat3}, SIRIUS \citep{SIRIUS}

\textit{Software:} LOFTI \citep{Pearce2020}, Astropy \citep{astropy}, Pystan \citep{pystan}, Isochrones \citep{Morton2015}, TOPCAT \citep{Taylor2006}, AstroImageJ \citep{Collins:2017}, TAPIR \citep{Jensen:2013}

\appendix
\section{Full Corner Plot of Derived LOFTI Parameters}
\label{cornerplot}
\begin{figure*}
    \centering
    \includegraphics[width=\textwidth]{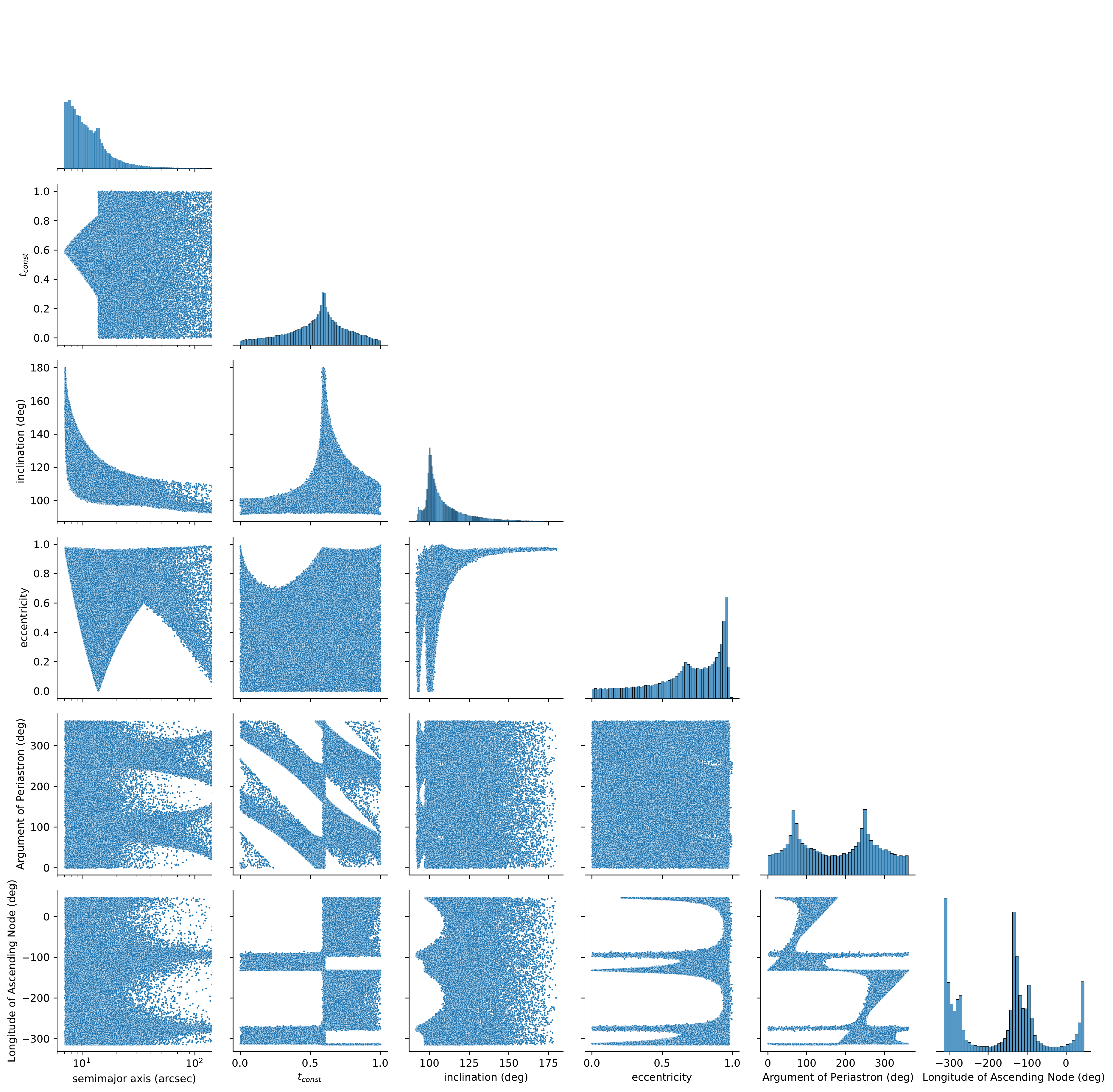}\\
    \caption{A corner plot of a sample full orbital fit. Note that Period and Epoch of Periastron Passage ($t_0$) are calculated according to Table \ref{table:priors}}
    \label{fig:cornerplotfull}
\end{figure*}


\begin{thebibliography}{}
    \expandafter\ifx\csname natexlab\endcsname\relax\def\natexlab#1{#1}\fi
    \providecommand{\url}[1]{\href{#1}{#1}}
    \providecommand{\dodoi}[1]{doi:~\href{http://doi.org/#1}{\nolinkurl{#1}}}
    \providecommand{\doeprint}[1]{\href{http://ascl.net/#1}{\nolinkurl{http://ascl.net/#1}}}
    \providecommand{\doarXiv}[1]{\href{https://arxiv.org/abs/#1}{\nolinkurl{https://arxiv.org/abs/#1}}}
    
    \bibitem[{{Adams} {et~al.}(1989){Adams}, {Ruden}, \& {Shu}}]{adams1989}
    {Adams}, F.~C., {Ruden}, S.~P., \& {Shu}, F.~H. 1989, \apj, 347, 959,
      \dodoi{10.1086/168187}
    
    \bibitem[{{Andrews} {et~al.}(2010){Andrews}, {Wilner}, {Hughes}, {Qi}, \&
      {Dullemond}}]{Andrew2010}
    {Andrews}, S.~M., {Wilner}, D.~J., {Hughes}, A.~M., {Qi}, C., \& {Dullemond},
      C.~P. 2010, \apj, 723, 1241, \dodoi{10.1088/0004-637X/723/2/1241}
    
    \bibitem[{{Angus} {et~al.}(2019){Angus}, {Morton}, {Foreman-Mackey}, {van
      Saders}, {Curtis}, {Kane}, {Bedell}, {Kiman}, {Hogg}, \&
      {Brewer}}]{Angus2019}
    {Angus}, R., {Morton}, T.~D., {Foreman-Mackey}, D., {et~al.} 2019, \aj, 158,
      173, \dodoi{10.3847/1538-3881/ab3c53}
    
    \bibitem[{{Ansdell} {et~al.}(2018){Ansdell}, {Williams}, {Trapman}, {van
      Terwisga}, {Facchini}, {Manara}, {van der Marel}, {Miotello}, {Tazzari},
      {Hogerheijde}, {Guidi}, {Testi}, \& {van Dishoeck}}]{Ansdell2018}
    {Ansdell}, M., {Williams}, J.~P., {Trapman}, L., {et~al.} 2018, \apj, 859, 21,
      \dodoi{10.3847/1538-4357/aab890}
    
    \bibitem[{{Antognini}(2015)}]{Antognini2015}
    {Antognini}, J.~M.~O. 2015, \mnras, 452, 3610, \dodoi{10.1093/mnras/stv1552}
    
    \bibitem[{{Barclay} {et~al.}(2018){Barclay}, {Pepper}, \&
      {Quintana}}]{Barclay2018}
    {Barclay}, T., {Pepper}, J., \& {Quintana}, E.~V. 2018, \apjs, 239, 2,
      \dodoi{10.3847/1538-4365/aae3e9}
    
    \bibitem[{{Bate}(2018)}]{Bate2018}
    {Bate}, M.~R. 2018, \mnras, 475, 5618, \dodoi{10.1093/mnras/sty169}
    
    \bibitem[{{Bate} {et~al.}(2000){Bate}, {Bonnell}, {Clarke}, {Lubow}, {Ogilvie},
      {Pringle}, \& {Tout}}]{bate2000}
    {Bate}, M.~R., {Bonnell}, I.~A., {Clarke}, C.~J., {et~al.} 2000, \mnras, 317,
      773, \dodoi{10.1046/j.1365-8711.2000.03648.x}
    
    \bibitem[{{Batygin}(2012)}]{Batygin2012}
    {Batygin}, K. 2012, \nat, 491, 418, \dodoi{10.1038/nature11560}
    
    \bibitem[{{Batygin} {et~al.}(2016){Batygin}, {Bodenheimer}, \&
      {Laughlin}}]{Batygin2016}
    {Batygin}, K., {Bodenheimer}, P.~H., \& {Laughlin}, G.~P. 2016, \apj, 829, 114,
      \dodoi{10.3847/0004-637X/829/2/114}
    
    \bibitem[{{Bazs{\'o}} \& {Pilat-Lohinger}(2020)}]{Akos2020}
    {Bazs{\'o}}, {\'A}., \& {Pilat-Lohinger}, E. 2020, \aj, 160, 2,
      \dodoi{10.3847/1538-3881/ab9104}
    
    \bibitem[{{Becker} \& {Adams}(2017)}]{beckeradams}
    {Becker}, J.~C., \& {Adams}, F.~C. 2017, \mnras, 468, 549,
      \dodoi{10.1093/mnras/stx461}
    
    \bibitem[{{Becker} {et~al.}(2017){Becker}, {Vanderburg}, {Adams}, {Khain}, \&
      {Bryan}}]{becker2017}
    {Becker}, J.~C., {Vanderburg}, A., {Adams}, F.~C., {Khain}, T., \& {Bryan}, M.
      2017, \aj, 154, 230, \dodoi{10.3847/1538-3881/aa9176}
    
    \bibitem[{{Becker} {et~al.}(2015){Becker}, {Vanderburg}, {Adams}, {Rappaport},
      \& {Schwengeler}}]{becker2015}
    {Becker}, J.~C., {Vanderburg}, A., {Adams}, F.~C., {Rappaport}, S.~A., \&
      {Schwengeler}, H.~M. 2015, \apjl, 812, L18,
      \dodoi{10.1088/2041-8205/812/2/L18}
    
    \bibitem[{{Belokurov} {et~al.}(2020){Belokurov}, {Penoyre}, {Oh}, {Iorio},
      {Hodgkin}, {Evans}, {Everall}, {Koposov}, {Tout}, {Izzard}, {Clarke}, \&
      {Brown}}]{Belokurov2020}
    {Belokurov}, V., {Penoyre}, Z., {Oh}, S., {et~al.} 2020, \mnras, 496, 1922,
      \dodoi{10.1093/mnras/staa1522}
    
    \bibitem[{{Blunt} {et~al.}(2017){Blunt}, {Nielsen}, {De Rosa}, {Konopacky},
      {Ryan}, {Wang}, {Pueyo}, {Rameau}, {Marois}, {Marchis}, {Macintosh},
      {Graham}, {Duch{\^e}ne}, \& {Schneider}}]{Blunt2017}
    {Blunt}, S., {Nielsen}, E.~L., {De Rosa}, R.~J., {et~al.} 2017, \aj, 153, 229,
      \dodoi{10.3847/1538-3881/aa6930}
    
    \bibitem[{{Blunt} {et~al.}(2020){Blunt}, {Wang}, {Angelo}, {Ngo}, {Cody}, {De
      Rosa}, {Graham}, {Hirsch}, {Nagpal}, {Nielsen}, {Pearce}, {Rice}, \&
      {Tejada}}]{Blunt2020}
    {Blunt}, S., {Wang}, J.~J., {Angelo}, I., {et~al.} 2020, \aj, 159, 89,
      \dodoi{10.3847/1538-3881/ab6663}
    
    \bibitem[{{Bonnell} \& {Bate}(1994)}]{Bonnell1994}
    {Bonnell}, I.~A., \& {Bate}, M.~R. 1994, \mnras, 271, 999,
      \dodoi{10.1093/mnras/271.4.999}
    
    \bibitem[{{Bouma} {et~al.}(2018){Bouma}, {Masuda}, \& {Winn}}]{bouma2018}
    {Bouma}, L.~G., {Masuda}, K., \& {Winn}, J.~N. 2018, \aj, 155, 244,
      \dodoi{10.3847/1538-3881/aabfb8}
    
    \bibitem[{{Brown} {et~al.}(2013){Brown}, {Baliber}, {Bianco}, {Bowman},
      {Burleson}, {Conway}, {Crellin}, {Depagne}, {De Vera}, {Dilday}, {Dragomir},
      {Dubberley}, {Eastman}, {Elphick}, {Falarski}, {Foale}, {Ford}, {Fulton},
      {Garza}, {Gomez}, {Graham}, {Greene}, {Haldeman}, {Hawkins}, {Haworth},
      {Haynes}, {Hidas}, {Hjelstrom}, {Howell}, {Hygelund}, {Lister}, {Lobdill},
      {Martinez}, {Mullins}, {Norbury}, {Parrent}, {Paulson}, {Petry}, {Pickles},
      {Posner}, {Rosing}, {Ross}, {Sand}, {Saunders}, {Shobbrook}, {Shporer},
      {Street}, {Thomas}, {Tsapras}, {Tufts}, {Valenti}, {Vander Horst}, {Walker},
      {White}, \& {Willis}}]{Brown2013}
    {Brown}, T.~M., {Baliber}, N., {Bianco}, F.~B., {et~al.} 2013, \pasp, 125,
      1031, \dodoi{10.1086/673168}
    
    \bibitem[{{Buchhave} \& {Latham}(2015)}]{Buchave2015}
    {Buchhave}, L.~A., \& {Latham}, D.~W. 2015, \apj, 808, 187,
      \dodoi{10.1088/0004-637X/808/2/187}
    
    \bibitem[{{Buchhave} {et~al.}(2012){Buchhave}, {Latham}, {Johansen},
      {Bizzarro}, {Torres}, {Rowe}, {Batalha}, {Borucki}, {Brugamyer}, {Caldwell},
      {Bryson}, {Ciardi}, {Cochran}, {Endl}, {Esquerdo}, {Ford}, {Geary},
      {Gilliland}, {Hansen}, {Isaacson}, {Laird}, {Lucas}, {Marcy}, {Morse},
      {Robertson}, {Shporer}, {Stefanik}, {Still}, \& {Quinn}}]{Buchave2012}
    {Buchhave}, L.~A., {Latham}, D.~W., {Johansen}, A., {et~al.} 2012, \nat, 486,
      375, \dodoi{10.1038/nature11121}
    
    \bibitem[{{Buchhave} {et~al.}(2014){Buchhave}, {Bizzarro}, {Latham},
      {Sasselov}, {Cochran}, {Endl}, {Isaacson}, {Juncher}, \&
      {Marcy}}]{Buchave2014}
    {Buchhave}, L.~A., {Bizzarro}, M., {Latham}, D.~W., {et~al.} 2014, \nat, 509,
      593, \dodoi{10.1038/nature13254}
    
    \bibitem[{{Buder} {et~al.}(2018){Buder}, {Asplund}, {Duong}, {Kos}, {Lind},
      {Ness}, {Sharma}, {Bland-Hawthorn}, {Casey}, {de Silva}, {D'Orazi},
      {Freeman}, {Lewis}, {Lin}, {Martell}, {Schlesinger}, {Simpson}, {Zucker},
      {Zwitter}, {Amarsi}, {Anguiano}, {Carollo}, {Casagrande}, {{\v{C}}otar},
      {Cottrell}, {da Costa}, {Gao}, {Hayden}, {Horner}, {Ireland}, {Kafle},
      {Munari}, {Nataf}, {Nordlander}, {Stello}, {Ting}, {Traven}, {Watson},
      {Wittenmyer}, {Wyse}, {Yong}, {Zinn}, {{\v{Z}}erjal}, \& {Galah
      Collaboration}}]{Buder2018}
    {Buder}, S., {Asplund}, M., {Duong}, L., {et~al.} 2018, \mnras, 478, 4513,
      \dodoi{10.1093/mnras/sty1281}
    
    \bibitem[{{Ca{\~n}as} {et~al.}(2019){Ca{\~n}as}, {Wang}, {Mahadevan}, {Bender},
      {De Lee}, {Fleming}, {Garc{\'\i}a-Hern{\'a}ndez}, {Hearty}, {Majewski},
      {Roman-Lopes}, {Schneider}, \& {Stassun}}]{canas2019}
    {Ca{\~n}as}, C.~I., {Wang}, S., {Mahadevan}, S., {et~al.} 2019, \apjl, 870,
      L17, \dodoi{10.3847/2041-8213/aafa1e}
    
    \bibitem[{{Casagrande} {et~al.}(2011){Casagrande}, {Sch{\"o}nrich}, {Asplund},
      {Cassisi}, {Ram{\'\i}rez}, {Mel{\'e}ndez}, {Bensby}, \&
      {Feltzing}}]{casagrande2011}
    {Casagrande}, L., {Sch{\"o}nrich}, R., {Asplund}, M., {et~al.} 2011, \aap, 530,
      A138, \dodoi{10.1051/0004-6361/201016276}
    
    \bibitem[{{Chambers} {et~al.}(2016){Chambers}, {Magnier}, {Metcalfe},
      {Flewelling}, {Huber}, {Waters}, {Denneau}, {Draper}, {Farrow}, {Finkbeiner},
      {Holmberg}, {Koppenhoefer}, {Price}, {Rest}, {Saglia}, {Schlafly}, {Smartt},
      {Sweeney}, {Wainscoat}, {Burgett}, {Chastel}, {Grav}, {Heasley}, {Hodapp},
      {Jedicke}, {Kaiser}, {Kudritzki}, {Luppino}, {Lupton}, {Monet}, {Morgan},
      {Onaka}, {Shiao}, {Stubbs}, {Tonry}, {White}, {Ba{\~n}ados}, {Bell},
      {Bender}, {Bernard}, {Boegner}, {Boffi}, {Botticella}, {Calamida},
      {Casertano}, {Chen}, {Chen}, {Cole}, {Deacon}, {Frenk}, {Fitzsimmons},
      {Gezari}, {Gibbs}, {Goessl}, {Goggia}, {Gourgue}, {Goldman}, {Grant},
      {Grebel}, {Hambly}, {Hasinger}, {Heavens}, {Heckman}, {Henderson}, {Henning},
      {Holman}, {Hopp}, {Ip}, {Isani}, {Jackson}, {Keyes}, {Koekemoer}, {Kotak},
      {Le}, {Liska}, {Long}, {Lucey}, {Liu}, {Martin}, {Masci}, {McLean}, {Mindel},
      {Misra}, {Morganson}, {Murphy}, {Obaika}, {Narayan}, {Nieto-Santisteban},
      {Norberg}, {Peacock}, {Pier}, {Postman}, {Primak}, {Rae}, {Rai}, {Riess},
      {Riffeser}, {Rix}, {R{\"o}ser}, {Russel}, {Rutz}, {Schilbach}, {Schultz},
      {Scolnic}, {Strolger}, {Szalay}, {Seitz}, {Small}, {Smith}, {Soderblom},
      {Taylor}, {Thomson}, {Taylor}, {Thakar}, {Thiel}, {Thilker}, {Unger},
      {Urata}, {Valenti}, {Wagner}, {Walder}, {Walter}, {Watters}, {Werner},
      {Wood-Vasey}, \& {Wyse}}]{chambers2016}
    {Chambers}, K.~C., {Magnier}, E.~A., {Metcalfe}, N., {et~al.} 2016, arXiv
      e-prints, arXiv:1612.05560.
    \newblock \doarXiv{1612.05560}
    
    \bibitem[{{Choi} {et~al.}(2016){Choi}, {Dotter}, {Conroy}, {Cantiello},
      {Paxton}, \& {Johnson}}]{choi2016}
    {Choi}, J., {Dotter}, A., {Conroy}, C., {et~al.} 2016, \apj, 823, 102,
      \dodoi{10.3847/0004-637X/823/2/102}
    
    \bibitem[{{Coelho} {et~al.}(2005){Coelho}, {Barbuy}, {Mel{\'e}ndez},
      {Schiavon}, \& {Castilho}}]{Coelho2005}
    {Coelho}, P., {Barbuy}, B., {Mel{\'e}ndez}, J., {Schiavon}, R.~P., \&
      {Castilho}, B.~V. 2005, \aap, 443, 735, \dodoi{10.1051/0004-6361:20053511}
    
    \bibitem[{{Collins} {et~al.}(2017){Collins}, {Kielkopf}, {Stassun}, \&
      {Hessman}}]{Collins:2017}
    {Collins}, K.~A., {Kielkopf}, J.~F., {Stassun}, K.~G., \& {Hessman}, F.~V.
      2017, \aj, 153, 77, \dodoi{10.3847/1538-3881/153/2/77}
    
    \bibitem[{{Correa-Otto} \& {Gil-Hutton}(2017)}]{Correa2017}
    {Correa-Otto}, J.~A., \& {Gil-Hutton}, R.~A. 2017, \aap, 608, A116,
      \dodoi{10.1051/0004-6361/201731229}
    
    \bibitem[{{D'Alessio} {et~al.}(2006){D'Alessio}, {Calvet}, {Hartmann},
      {Franco-Hern{\'a}ndez}, \& {Serv{\'\i}n}}]{Dalessio2006}
    {D'Alessio}, P., {Calvet}, N., {Hartmann}, L., {Franco-Hern{\'a}ndez}, R., \&
      {Serv{\'\i}n}, H. 2006, \apj, 638, 314, \dodoi{10.1086/498861}
    
    \bibitem[{{Dawson} \& {Johnson}(2018)}]{Dawson2018}
    {Dawson}, R.~I., \& {Johnson}, J.~A. 2018, \araa, 56, 175,
      \dodoi{10.1146/annurev-astro-081817-051853}
    
    \bibitem[{{Deacon} \& {Kraus}(2020)}]{Deacon2020}
    {Deacon}, N.~R., \& {Kraus}, A.~L. 2020, \mnras, 496, 5176,
      \dodoi{10.1093/mnras/staa1877}
    
    \bibitem[{{Deacon} {et~al.}(2016){Deacon}, {Kraus}, {Mann}, {Magnier},
      {Chambers}, {Wainscoat}, {Tonry}, {Kaiser}, {Waters}, {Flewelling}, {Hodapp},
      \& {Burgett}}]{deaconpanstarrs}
    {Deacon}, N.~R., {Kraus}, A.~L., {Mann}, A.~W., {et~al.} 2016, \mnras, 455,
      4212, \dodoi{10.1093/mnras/stv2132}
    
    \bibitem[{{Dimitrov} \& {Kjurkchieva}(2010)}]{Dimitrov}
    {Dimitrov}, D.~P., \& {Kjurkchieva}, D.~P. 2010, \mnras, 406, 2559,
      \dodoi{10.1111/j.1365-2966.2010.16843.x}
    
    \bibitem[{{Dotter}(2016)}]{MIST}
    {Dotter}, A. 2016, \apjs, 222, 8, \dodoi{10.3847/0067-0049/222/1/8}
    
    \bibitem[{{Doyle} {et~al.}(2011){Doyle}, {Carter}, {Fabrycky}, {Slawson},
      {Howell}, {Winn}, {Orosz}, {P{\v{r}}sa}, {Welsh}, {Quinn}, {Latham},
      {Torres}, {Buchhave}, {Marcy}, {Fortney}, {Shporer}, {Ford}, {Lissauer},
      {Ragozzine}, {Rucker}, {Batalha}, {Jenkins}, {Borucki}, {Koch}, {Middour},
      {Hall}, {McCauliff}, {Fanelli}, {Quintana}, {Holman}, {Caldwell}, {Still},
      {Stefanik}, {Brown}, {Esquerdo}, {Tang}, {Furesz}, {Geary}, {Berlind},
      {Calkins}, {Short}, {Steffen}, {Sasselov}, {Dunham}, {Cochran}, {Boss},
      {Haas}, {Buzasi}, \& {Fischer}}]{kepler16}
    {Doyle}, L.~R., {Carter}, J.~A., {Fabrycky}, D.~C., {et~al.} 2011, Science,
      333, 1602, \dodoi{10.1126/science.1210923}
    
    \bibitem[{{Duch{\^e}ne}(2010)}]{Duchene2010}
    {Duch{\^e}ne}, G. 2010, \apjl, 709, L114, \dodoi{10.1088/2041-8205/709/2/L114}
    
    \bibitem[{{Eastman}(2017)}]{Exofast2}
    {Eastman}, J. 2017, {EXOFASTv2: Generalized publication-quality exoplanet
      modeling code}.
    \newblock \doeprint{1710.003}
    
    \bibitem[{Eastman {et~al.}(2014)Eastman, Brown, Hygelund, van Eyken, Tufts, \&
      Barnes}]{Eastman2014}
    Eastman, J.~D., Brown, T.~M., Hygelund, J., {et~al.} 2014, in Ground-based and
      Airborne Instrumentation for Astronomy V, ed. S.~K. Ramsay, I.~S. McLean, \&
      H.~Takami, Vol. 9147, International Society for Optics and Photonics (SPIE),
      436 -- 450, \dodoi{10.1117/12.2054699}
    
    \bibitem[{{El-Badry} \& {Rix}(2018)}]{elbadry2018}
    {El-Badry}, K., \& {Rix}, H.-W. 2018, \mnras, 480, 4884,
      \dodoi{10.1093/mnras/sty2186}
    
    \bibitem[{{El-Badry} \& {Rix}(2019)}]{elbadry2019}
    ---. 2019, \mnras, 482, L139, \dodoi{10.1093/mnrasl/sly206}
    
    \bibitem[{{El-Badry} {et~al.}(2021){El-Badry}, {Rix}, \&
      {Heintz}}]{elbadry2021}
    {El-Badry}, K., {Rix}, H.-W., \& {Heintz}, T.~M. 2021, arXiv e-prints,
      arXiv:2101.05282.
    \newblock \doarXiv{2101.05282}
    
    \bibitem[{{Fabrycky} \& {Tremaine}(2007)}]{Fabrycky2007}
    {Fabrycky}, D., \& {Tremaine}, S. 2007, \apj, 669, 1298, \dodoi{10.1086/521702}
    
    \bibitem[{{Fern{\'a}ndez-L{\'o}pez} {et~al.}(2017){Fern{\'a}ndez-L{\'o}pez},
      {Zapata}, \& {Gabbasov}}]{Fernandez2017}
    {Fern{\'a}ndez-L{\'o}pez}, M., {Zapata}, L.~A., \& {Gabbasov}, R. 2017, \apj,
      845, 10, \dodoi{10.3847/1538-4357/aa7d51}
    
    \bibitem[{{Ferrer-Ch{\'a}vez} {et~al.}(2021){Ferrer-Ch{\'a}vez}, {Wang}, \&
      {Blunt}}]{ferrerchavez}
    {Ferrer-Ch{\'a}vez}, R., {Wang}, J.~J., \& {Blunt}, S. 2021, arXiv e-prints,
      arXiv:2103.12877.
    \newblock \doarXiv{2103.12877}
    
    \bibitem[{F\H{u}r\'esz(2008)}]{gaborthesis}
    F\H{u}r\'esz, G. 2008, PhD thesis, University of Szeged, Hungary,
      \url{www.sao.arizona.edu/html/FLWO/60/TRES/GABORthesis.pdf}
    
    \bibitem[{{Fischer} \& {Marcy}(1992)}]{fischer1992}
    {Fischer}, D.~A., \& {Marcy}, G.~W. 1992, \apj, 396, 178,
      \dodoi{10.1086/171708}
    
    \bibitem[{{Frankowski} {et~al.}(2007){Frankowski}, {Jancart}, \&
      {Jorissen}}]{Frankowski2007}
    {Frankowski}, A., {Jancart}, S., \& {Jorissen}, A. 2007, \aap, 464, 377,
      \dodoi{10.1051/0004-6361:20065526}
    
    \bibitem[{{Fressin} {et~al.}(2013){Fressin}, {Torres}, {Charbonneau}, {Bryson},
      {Christiansen}, {Dressing}, {Jenkins}, {Walkowicz}, \& {Batalha}}]{fressin}
    {Fressin}, F., {Torres}, G., {Charbonneau}, D., {et~al.} 2013, \apj, 766, 81,
      \dodoi{10.1088/0004-637X/766/2/81}
    
    \bibitem[{{Gaia Collaboration} {et~al.}(2016){Gaia Collaboration}, {Prusti},
      {de Bruijne}, {Brown}, {Vallenari}, {Babusiaux}, {Bailer-Jones}, {Bastian},
      {Biermann}, {Evans}, \& et~al.}]{gaiamission}
    {Gaia Collaboration}, {Prusti}, T., {de Bruijne}, J.~H.~J., {et~al.} 2016,
      \aap, 595, A1, \dodoi{10.1051/0004-6361/201629272}
    
    \bibitem[{{Gaia Collaboration} {et~al.}(2018){Gaia Collaboration}, {Brown},
      {Vallenari}, {Prusti}, {de Bruijne}, {Babusiaux}, {Bailer-Jones}, {Biermann},
      {Evans}, {Eyer}, \& et~al.}]{gaiadr2}
    {Gaia Collaboration}, {Brown}, A.~G.~A., {Vallenari}, A., {et~al.} 2018, \aap,
      616, A1, \dodoi{10.1051/0004-6361/201833051}
    
    \bibitem[{{Gaia Collaboration} {et~al.}(2021){Gaia Collaboration}, {Brown},
      {Vallenari}, {Prusti}, {de Bruijne}, {Babusiaux}, {Biermann}, {Creevey},
      {Evans}, {Eyer}, {Hutton}, {Jansen}, {Jordi}, {Klioner}, {Lammers},
      {Lindegren}, {Luri}, {Mignard}, {Panem}, {Pourbaix}, {Randich}, {Sartoretti},
      {Soubiran}, {Walton}, {Arenou}, {Bailer-Jones}, {Bastian}, {Cropper},
      {Drimmel}, {Katz}, {Lattanzi}, {van Leeuwen}, {Bakker}, {Cacciari},
      {Casta{\~n}eda}, {De Angeli}, {Ducourant}, {Fabricius}, {Fouesneau},
      {Fr{\'e}mat}, {Guerra}, {Guerrier}, {Guiraud}, {Jean-Antoine Piccolo},
      {Masana}, {Messineo}, {Mowlavi}, {Nicolas}, {Nienartowicz}, {Pailler},
      {Panuzzo}, {Riclet}, {Roux}, {Seabroke}, {Sordo}, {Tanga}, {Th{\'e}venin},
      {Gracia-Abril}, {Portell}, {Teyssier}, {Altmann}, {Andrae}, {Bellas-Velidis},
      {Benson}, {Berthier}, {Blomme}, {Brugaletta}, {Burgess}, {Busso}, {Carry},
      {Cellino}, {Cheek}, {Clementini}, {Damerdji}, {Davidson}, {Delchambre},
      {Dell'Oro}, {Fern{\'a}ndez-Hern{\'a}ndez}, {Galluccio}, {Garc{\'\i}a-Lario},
      {Garcia-Reinaldos}, {Gonz{\'a}lez-N{\'u}{\~n}ez}, {Gosset}, {Haigron},
      {Halbwachs}, {Hambly}, {Harrison}, {Hatzidimitriou}, {Heiter},
      {Hern{\'a}ndez}, {Hestroffer}, {Hodgkin}, {Holl}, {Jan{\ss}en}, {Jevardat de
      Fombelle}, {Jordan}, {Krone-Martins}, {Lanzafame}, {L{\"o}ffler}, {Lorca},
      {Manteiga}, {Marchal}, {Marrese}, {Moitinho}, {Mora}, {Muinonen}, {Osborne},
      {Pancino}, {Pauwels}, {Petit}, {Recio-Blanco}, {Richards}, {Riello},
      {Rimoldini}, {Robin}, {Roegiers}, {Rybizki}, {Sarro}, {Siopis}, {Smith},
      {Sozzetti}, {Ulla}, {Utrilla}, {van Leeuwen}, {van Reeven}, {Abbas}, {Abreu
      Aramburu}, {Accart}, {Aerts}, {Aguado}, {Ajaj}, {Altavilla}, {{\'A}lvarez},
      {{\'A}lvarez Cid-Fuentes}, {Alves}, {Anderson}, {Anglada Varela}, {Antoja},
      {Audard}, {Baines}, {Baker}, {Balaguer-N{\'u}{\~n}ez}, {Balbinot}, {Balog},
      {Barache}, {Barbato}, {Barros}, {Barstow}, {Bartolom{\'e}}, {Bassilana},
      {Bauchet}, {Baudesson-Stella}, {Becciani}, {Bellazzini}, {Bernet}, {Bertone},
      {Bianchi}, {Blanco-Cuaresma}, {Boch}, {Bombrun}, {Bossini}, {Bouquillon},
      {Bragaglia}, {Bramante}, {Breedt}, {Bressan}, {Brouillet}, {Bucciarelli},
      {Burlacu}, {Busonero}, {Butkevich}, {Buzzi}, {Caffau}, {Cancelliere},
      {C{\'a}novas}, {Cantat-Gaudin}, {Carballo}, {Carlucci}, {Carnerero},
      {Carrasco}, {Casamiquela}, {Castellani}, {Castro-Ginard}, {Castro Sampol},
      {Chaoul}, {Charlot}, {Chemin}, {Chiavassa}, {Cioni}, {Comoretto}, {Cooper},
      {Cornez}, {Cowell}, {Crifo}, {Crosta}, {Crowley}, {Dafonte}, {Dapergolas},
      {David}, {David}, {de Laverny}, {De Luise}, {De March}, {De Ridder}, {de
      Souza}, {de Teodoro}, {de Torres}, {del Peloso}, {del Pozo}, {Delbo},
      {Delgado}, {Delgado}, {Delisle}, {Di Matteo}, {Diakite}, {Diener},
      {Distefano}, {Dolding}, {Eappachen}, {Edvardsson}, {Enke}, {Esquej}, {Fabre},
      {Fabrizio}, {Faigler}, {Fedorets}, {Fernique}, {Fienga}, {Figueras},
      {Fouron}, {Fragkoudi}, {Fraile}, {Franke}, {Gai}, {Garabato},
      {Garcia-Gutierrez}, {Garc{\'\i}a-Torres}, {Garofalo}, {Gavras}, {Gerlach},
      {Geyer}, {Giacobbe}, {Gilmore}, {Girona}, {Giuffrida}, {Gomel}, {Gomez},
      {Gonzalez-Santamaria}, {Gonz{\'a}lez-Vidal}, {Granvik},
      {Guti{\'e}rrez-S{\'a}nchez}, {Guy}, {Hauser}, {Haywood}, {Helmi}, {Hidalgo},
      {Hilger}, {H{\l}adczuk}, {Hobbs}, {Holland}, {Huckle}, {Jasniewicz},
      {Jonker}, {Juaristi Campillo}, {Julbe}, {Karbevska}, {Kervella}, {Khanna},
      {Kochoska}, {Kontizas}, {Kordopatis}, {Korn}, {Kostrzewa-Rutkowska},
      {Kruszy{\'n}ska}, {Lambert}, {Lanza}, {Lasne}, {Le Campion}, {Le Fustec},
      {Lebreton}, {Lebzelter}, {Leccia}, {Leclerc}, {Lecoeur-Taibi}, {Liao},
      {Licata}, {Lindstr{\o}m}, {Lister}, {Livanou}, {Lobel}, {Madrero Pardo},
      {Managau}, {Mann}, {Marchant}, {Marconi}, {Marcos Santos}, {Marinoni},
      {Marocco}, {Marshall}, {Martin Polo}, {Mart{\'\i}n-Fleitas}, {Masip},
      {Massari}, {Mastrobuono-Battisti}, {Mazeh}, {McMillan}, {Messina},
      {Michalik}, {Millar}, {Mints}, {Molina}, {Molinaro}, {Moln{\'a}r},
      {Montegriffo}, {Mor}, {Morbidelli}, {Morel}, {Morris}, {Mulone}, {Munoz},
      {Muraveva}, {Murphy}, {Musella}, {Noval}, {Ord{\'e}novic}, {Orr{\`u}},
      {Osinde}, {Pagani}, {Pagano}, {Palaversa}, {Palicio}, {Panahi}, {Pawlak},
      {Pe{\~n}alosa Esteller}, {Penttil{\"a}}, {Piersimoni}, {Pineau}, {Plachy},
      {Plum}, {Poggio}, {Poretti}, {Poujoulet}, {Pr{\v{s}}a}, {Pulone}, {Racero},
      {Ragaini}, {Rainer}, {Raiteri}, {Rambaux}, {Ramos}, {Ramos-Lerate}, {Re
      Fiorentin}, {Regibo}, {Reyl{\'e}}, {Ripepi}, {Riva}, {Rixon}, {Robichon},
      {Robin}, {Roelens}, {Rohrbasser}, {Romero-G{\'o}mez}, {Rowell}, {Royer},
      {Rybicki}, {Sadowski}, {Sagrist{\`a} Sell{\'e}s}, {Sahlmann}, {Salgado},
      {Salguero}, {Samaras}, {Sanchez Gimenez}, {Sanna}, {Santove{\~n}a},
      {Sarasso}, {Schultheis}, {Sciacca}, {Segol}, {Segovia}, {S{\'e}gransan},
      {Semeux}, {Shahaf}, {Siddiqui}, {Siebert}, {Siltala}, {Slezak}, {Smart},
      {Solano}, {Solitro}, {Souami}, {Souchay}, {Spagna}, {Spoto}, {Steele},
      {Steidelm{\"u}ller}, {Stephenson}, {S{\"u}veges}, {Szabados}, {Szegedi-Elek},
      {Taris}, {Tauran}, {Taylor}, {Teixeira}, {Thuillot}, {Tonello}, {Torra},
      {Torra}, {Turon}, {Unger}, {Vaillant}, {van Dillen}, {Vanel}, {Vecchiato},
      {Viala}, {Vicente}, {Voutsinas}, {Weiler}, {Wevers}, {Wyrzykowski}, {Yoldas},
      {Yvard}, {Zhao}, {Zorec}, {Zucker}, {Zurbach}, \& {Zwitter}}]{gaiaedr3}
    ---. 2021, \aap, 649, A1, \dodoi{10.1051/0004-6361/202039657}
    
    \bibitem[{{Guerrero} {et~al.}(2021){Guerrero}, {Seager}, {Huang}, {Vanderburg},
      {Garcia Soto}, {Mireles}, {Hesse}, {Fong}, {Glidden}, {Shporer}, {Latham},
      {Collins}, {Quinn}, {Burt}, {Dragomir}, {Crossfield}, {Vanderspek},
      {Fausnaugh}, {Burke}, {Ricker}, {Daylan}, {Essack}, {G{\"u}nther}, {Osborn},
      {Pepper}, {Rowden}, {Sha}, {Villanueva}, {Yahalomi}, {Yu}, {Ballard},
      {Batalha}, {Berardo}, {Chontos}, {Dittmann}, {Esquerdo}, {Mikal-Evans},
      {Jayaraman}, {Krishnamurthy}, {Louie}, {Mehrle}, {Niraula}, {Rackham},
      {Rodriguez}, {Rowden}, {Sousa-Silva}, {Watanabe}, {Wong}, {Zhan},
      {Zivanovic}, {Christiansen}, {Ciardi}, {Swain}, {Lund}, {Mullally},
      {Fleming}, {Rodriguez}, {Boyd}, {Quintana}, {Barclay}, {Col{\'o}n},
      {Rinehart}, {Schlieder}, {Clampin}, {Jenkins}, {Twicken}, {Caldwell},
      {Coughlin}, {Henze}, {Lissauer}, {Morris}, {Rose}, {Smith}, {Tenenbaum},
      {Ting}, {Wohler}, {Bakos}, {Bean}, {Berta-Thompson}, {Bieryla}, {Bouma},
      {Buchhave}, {Butler}, {Charbonneau}, {Doty}, {Ge}, {Holman}, {Howard},
      {Kaltenegger}, {Kjeldsen}, {Kreidberg}, {Lin}, {Minsky}, {Narita}, {Paegert},
      {P{\'a}l}, {Palle}, {Sasselov}, {Spencer}, {Sozzetti}, {Stassun}, {Torres},
      {Udry}, \& {Winn}}]{guerrero2021}
    {Guerrero}, N.~M., {Seager}, S., {Huang}, C.~X., {et~al.} 2021, arXiv e-prints,
      arXiv:2103.12538.
    \newblock \doarXiv{2103.12538}
    
    \bibitem[{{Guillot} {et~al.}(2015){Guillot}, {Abe}, {Agabi}, {Rivet}, {Daban},
      {M{\'e}karnia}, {Aristidi}, {Schmider}, {Crouzet}, {Gon{\c{c}}alves},
      {Gouvret}, {Ottogalli}, {Faradji}, {Blanc}, {Bondoux}, \&
      {Valbousquet}}]{Guillot2015}
    {Guillot}, T., {Abe}, L., {Agabi}, A., {et~al.} 2015, Astronomische
      Nachrichten, 336, 638, \dodoi{10.1002/asna.201512174}
    
    \bibitem[{{Hawkins} {et~al.}(2020){Hawkins}, {Lucey}, {Ting}, {Ji}, {Katzberg},
      {Thompson}, {El-Badry}, {Teske}, {Nelson}, \& {Carrillo}}]{Hawkins2020}
    {Hawkins}, K., {Lucey}, M., {Ting}, Y.-S., {et~al.} 2020, \mnras, 492, 1164,
      \dodoi{10.1093/mnras/stz3132}
    
    \bibitem[{{Hinkel} {et~al.}(2014){Hinkel}, {Timmes}, {Young}, {Pagano}, \&
      {Turnbull}}]{Hinkel2014}
    {Hinkel}, N.~R., {Timmes}, F.~X., {Young}, P.~A., {Pagano}, M.~D., \&
      {Turnbull}, M.~C. 2014, \aj, 148, 54, \dodoi{10.1088/0004-6256/148/3/54}
    
    \bibitem[{{Hjorth} {et~al.}(2021){Hjorth}, {Albrecht}, {Hirano}, {Winn},
      {Dawson}, {Zanazzi}, {Knudstrup}, \& {Sato}}]{Hjorth2021}
    {Hjorth}, M., {Albrecht}, S., {Hirano}, T., {et~al.} 2021, Proceedings of the
      National Academy of Science, 118, 2017418118, \dodoi{10.1073/pnas.2017418118}
    
    \bibitem[{{Holman} {et~al.}(1997){Holman}, {Touma}, \& {Tremaine}}]{Holman1997}
    {Holman}, M., {Touma}, J., \& {Tremaine}, S. 1997, \nat, 386, 254,
      \dodoi{10.1038/386254a0}
    
    \bibitem[{{Huang} {et~al.}(2020){Huang}, {Quinn}, {Vanderburg}, {Becker},
      {Rodriguez}, {Pozuelos}, {Gandolfi}, {Zhou}, {Mann}, {Collins}, {Crossfield},
      {Barkaoui}, {Collins}, {Fridlund}, {Gillon}, {Gonzales}, {G{\"u}nther},
      {Henry}, {Howell}, {James}, {Jao}, {Jehin}, {Jensen}, {Kane}, {Lissauer},
      {Matthews}, {Matson}, {Paredes}, {Schlieder}, {Stassun}, {Shporer}, {Sha},
      {Tan}, {Georgieva}, {Mathur}, {Palle}, {Persson}, {Van Eylen}, {Ricker},
      {Vanderspek}, {Latham}, {Winn}, {Seager}, {Jenkins}, {Burke}, {Goeke},
      {Rinehart}, {Rose}, {Ting}, {Torres}, \& {Wong}}]{huang2020}
    {Huang}, C.~X., {Quinn}, S.~N., {Vanderburg}, A., {et~al.} 2020, \apjl, 892,
      L7, \dodoi{10.3847/2041-8213/ab7302}
    
    \bibitem[{{Jensen}(2013)}]{Jensen:2013}
    {Jensen}, E. 2013, {Tapir: A web interface for transit/eclipse observability},
      Astrophysics Source Code Library.
    \newblock \doeprint{1306.007}
    
    \bibitem[{{Jim{\'e}nez-Esteban} {et~al.}(2019){Jim{\'e}nez-Esteban}, {Solano},
      \& {Rodrigo}}]{Esteban2019}
    {Jim{\'e}nez-Esteban}, F.~M., {Solano}, E., \& {Rodrigo}, C. 2019, \aj, 157,
      78, \dodoi{10.3847/1538-3881/aafacc}
    
    \bibitem[{{Kaib} {et~al.}(2013){Kaib}, {Raymond}, \& {Duncan}}]{Kaib2013}
    {Kaib}, N.~A., {Raymond}, S.~N., \& {Duncan}, M. 2013, \nat, 493, 381,
      \dodoi{10.1038/nature11780}
    
    \bibitem[{{Koch} {et~al.}(2010){Koch}, {Borucki}, {Basri}, {Batalha}, {Brown},
      {Caldwell}, {Christensen-Dalsgaard}, {Cochran}, {DeVore}, {Dunham},
      {Gautier}, {Geary}, {Gilliland}, {Gould}, {Jenkins}, {Kondo}, {Latham},
      {Lissauer}, {Marcy}, {Monet}, {Sasselov}, {Boss}, {Brownlee}, {Caldwell},
      {Dupree}, {Howell}, {Kjeldsen}, {Meibom}, {Morrison}, {Owen}, {Reitsema},
      {Tarter}, {Bryson}, {Dotson}, {Gazis}, {Haas}, {Kolodziejczak}, {Rowe}, {Van
      Cleve}, {Allen}, {Chandrasekaran}, {Clarke}, {Li}, {Quintana}, {Tenenbaum},
      {Twicken}, \& {Wu}}]{koch}
    {Koch}, D.~G., {Borucki}, W.~J., {Basri}, G., {et~al.} 2010, \apjl, 713, L79,
      \dodoi{10.1088/2041-8205/713/2/L79}
    
    \bibitem[{{Kouwenhoven} {et~al.}(2010){Kouwenhoven}, {Goodwin}, {Parker},
      {Davies}, {Malmberg}, \& {Kroupa}}]{Kouwenhoven2010}
    {Kouwenhoven}, M.~B.~N., {Goodwin}, S.~P., {Parker}, R.~J., {et~al.} 2010,
      \mnras, 404, 1835, \dodoi{10.1111/j.1365-2966.2010.16399.x}
    
    \bibitem[{{Kozai}(1962)}]{kozai}
    {Kozai}, Y. 1962, \aj, 67, 591, \dodoi{10.1086/108790}
    
    \bibitem[{{Kraus} {et~al.}(2016){Kraus}, {Ireland}, {Huber}, {Mann}, \&
      {Dupuy}}]{Kraus2016}
    {Kraus}, A.~L., {Ireland}, M.~J., {Huber}, D., {Mann}, A.~W., \& {Dupuy}, T.~J.
      2016, \aj, 152, 8, \dodoi{10.3847/0004-6256/152/1/8}
    
    \bibitem[{{Kunder} {et~al.}(2017){Kunder}, {Kordopatis}, {Steinmetz},
      {Zwitter}, {McMillan}, {Casagrande}, {Enke}, {Wojno}, {Valentini},
      {Chiappini}, {Matijevi{\v{c}}}, {Siviero}, {de Laverny}, {Recio-Blanco},
      {Bijaoui}, {Wyse}, {Binney}, {Grebel}, {Helmi}, {Jofre}, {Antoja}, {Gilmore},
      {Siebert}, {Famaey}, {Bienaym{\'e}}, {Gibson}, {Freeman}, {Navarro},
      {Munari}, {Seabroke}, {Anguiano}, {{\v{Z}}erjal}, {Minchev}, {Reid},
      {Bland-Hawthorn}, {Kos}, {Sharma}, {Watson}, {Parker}, {Scholz}, {Burton},
      {Cass}, {Hartley}, {Fiegert}, {Stupar}, {Ritter}, {Hawkins}, {Gerhard},
      {Chaplin}, {Davies}, {Elsworth}, {Lund}, {Miglio}, \& {Mosser}}]{Kunder2017}
    {Kunder}, A., {Kordopatis}, G., {Steinmetz}, M., {et~al.} 2017, \aj, 153, 75,
      \dodoi{10.3847/1538-3881/153/2/75}
    
    \bibitem[{{Lai}(2014)}]{Lai2014}
    {Lai}, D. 2014, \mnras, 440, 3532, \dodoi{10.1093/mnras/stu485}
    
    \bibitem[{{Latham} {et~al.}(1991){Latham}, {Mazeh}, {Davis}, {Stefanik}, \&
      {Abt}}]{Latham1991}
    {Latham}, D.~W., {Mazeh}, T., {Davis}, R.~J., {Stefanik}, R.~P., \& {Abt},
      H.~A. 1991, \aj, 101, 625, \dodoi{10.1086/115711}
    
    \bibitem[{{Lee} {et~al.}(2016){Lee}, {Dunham}, {Myers}, {Arce}, {Bourke},
      {Goodman}, {J{\o}rgensen}, {Kristensen}, {Offner}, {Pineda}, {Tobin}, \&
      {Vorobyov}}]{Lee2016}
    {Lee}, K.~I., {Dunham}, M.~M., {Myers}, P.~C., {et~al.} 2016, \apjl, 820, L2,
      \dodoi{10.3847/2041-8205/820/1/L2}
    
    \bibitem[{{Li} {et~al.}(2020){Li}, {Mustill}, \& {Davies}}]{Li2020}
    {Li}, D., {Mustill}, A.~J., \& {Davies}, M.~B. 2020, arXiv e-prints,
      arXiv:2008.08842.
    \newblock \doarXiv{2008.08842}
    
    \bibitem[{{Lidov}(1962)}]{lidov}
    {Lidov}, M.~L. 1962, \planss, 9, 719, \dodoi{10.1016/0032-0633(62)90129-0}
    
    \bibitem[{{Lin} {et~al.}(1996){Lin}, {Bodenheimer}, \& {Richardson}}]{lin1996}
    {Lin}, D.~N.~C., {Bodenheimer}, P., \& {Richardson}, D.~C. 1996, \nat, 380,
      606, \dodoi{10.1038/380606a0}
    
    \bibitem[{Lindegren(2018)}]{lindegren2018}
    Lindegren, L. 2018.
    \newblock \url{http://www.rssd.esa.int/doc_fetch.php?id=3757412}
    
    \bibitem[{{Lindegren} {et~al.}(2018){Lindegren}, {Hern{\'a}ndez}, {Bombrun},
      {Klioner}, {Bastian}, {Ramos-Lerate}, {de Torres}, {Steidelm{\"u}ller},
      {Stephenson}, {Hobbs}, {Lammers}, {Biermann}, {Geyer}, {Hilger}, {Michalik},
      {Stampa}, {McMillan}, {Casta{\~n}eda}, {Clotet}, {Comoretto}, {Davidson},
      {Fabricius}, {Gracia}, {Hambly}, {Hutton}, {Mora}, {Portell}, {van Leeuwen},
      {Abbas}, {Abreu}, {Altmann}, {Andrei}, {Anglada}, {Balaguer-N{\'u}{\~n}ez},
      {Barache}, {Becciani}, {Bertone}, {Bianchi}, {Bouquillon}, {Bourda},
      {Br{\"u}semeister}, {Bucciarelli}, {Busonero}, {Buzzi}, {Cancelliere},
      {Carlucci}, {Charlot}, {Cheek}, {Crosta}, {Crowley}, {de Bruijne}, {de
      Felice}, {Drimmel}, {Esquej}, {Fienga}, {Fraile}, {Gai}, {Garralda},
      {Gonz{\'a}lez-Vidal}, {Guerra}, {Hauser}, {Hofmann}, {Holl}, {Jordan},
      {Lattanzi}, {Lenhardt}, {Liao}, {Licata}, {Lister}, {L{\"o}ffler},
      {Marchant}, {Martin-Fleitas}, {Messineo}, {Mignard}, {Morbidelli}, {Poggio},
      {Riva}, {Rowell}, {Salguero}, {Sarasso}, {Sciacca}, {Siddiqui}, {Smart},
      {Spagna}, {Steele}, {Taris}, {Torra}, {van Elteren}, {van Reeven}, \&
      {Vecchiato}}]{lindegren2018dr2solution}
    {Lindegren}, L., {Hern{\'a}ndez}, J., {Bombrun}, A., {et~al.} 2018, \aap, 616,
      A2, \dodoi{10.1051/0004-6361/201832727}
    
    \bibitem[{{Lindegren} {et~al.}(2021){Lindegren}, {Klioner}, {Hern{\'a}ndez},
      {Bombrun}, {Ramos-Lerate}, {Steidelm{\"u}ller}, {Bastian}, {Biermann}, {de
      Torres}, {Gerlach}, {Geyer}, {Hilger}, {Hobbs}, {Lammers}, {McMillan},
      {Stephenson}, {Casta{\~n}eda}, {Davidson}, {Fabricius}, {Gracia-Abril},
      {Portell}, {Rowell}, {Teyssier}, {Torra}, {Bartolom{\'e}}, {Clotet},
      {Garralda}, {Gonz{\'a}lez-Vidal}, {Torra}, {Abbas}, {Altmann}, {Anglada
      Varela}, {Balaguer-N{\'u}{\~n}ez}, {Balog}, {Barache}, {Becciani}, {Bernet},
      {Bertone}, {Bianchi}, {Bouquillon}, {Brown}, {Bucciarelli}, {Busonero},
      {Butkevich}, {Buzzi}, {Cancelliere}, {Carlucci}, {Charlot}, {Cioni},
      {Crosta}, {Crowley}, {del Peloso}, {del Pozo}, {Drimmel}, {Esquej}, {Fienga},
      {Fraile}, {Gai}, {Garcia-Reinaldos}, {Guerra}, {Hambly}, {Hauser},
      {Jan{\ss}en}, {Jordan}, {Kostrzewa-Rutkowska}, {Lattanzi}, {Liao}, {Licata},
      {Lister}, {L{\"o}ffler}, {Marchant}, {Masip}, {Mignard}, {Mints}, {Molina},
      {Mora}, {Morbidelli}, {Murphy}, {Pagani}, {Panuzzo}, {Pe{\~n}alosa Esteller},
      {Poggio}, {Re Fiorentin}, {Riva}, {Sagrist{\`a} Sell{\'e}s}, {Sanchez
      Gimenez}, {Sarasso}, {Sciacca}, {Siddiqui}, {Smart}, {Souami}, {Spagna},
      {Steele}, {Taris}, {Utrilla}, {van Reeven}, \& {Vecchiato}}]{lindegren2020}
    {Lindegren}, L., {Klioner}, S.~A., {Hern{\'a}ndez}, J., {et~al.} 2021, \aap,
      649, A2, \dodoi{10.1051/0004-6361/202039709}
    
    \bibitem[{{Majewski} {et~al.}(2016){Majewski}, {APOGEE Team}, \& {APOGEE-2
      Team}}]{Majewski2017}
    {Majewski}, S.~R., {APOGEE Team}, \& {APOGEE-2 Team}. 2016, Astronomische
      Nachrichten, 337, 863, \dodoi{10.1002/asna.201612387}
    
    \bibitem[{{Mamajek}(2009)}]{Mamajek2009}
    {Mamajek}, E.~E. 2009, in American Institute of Physics Conference Series, Vol.
      1158, Exoplanets and Disks: Their Formation and Diversity, ed. T.~{Usuda},
      M.~{Tamura}, \& M.~{Ishii}, 3--10, \dodoi{10.1063/1.3215910}
    
    \bibitem[{{Manara} {et~al.}(2019){Manara}, {Tazzari}, {Long}, {Herczeg},
      {Lodato}, {Rota}, {Cazzoletti}, {van der Plas}, {Pinilla}, {Dipierro},
      {Edwards}, {Harsono}, {Johnstone}, {Liu}, {Menard}, {Nisini}, {Ragusa},
      {Boehler}, \& {Cabrit}}]{Manara2019}
    {Manara}, C.~F., {Tazzari}, M., {Long}, F., {et~al.} 2019, \aap, 628, A95,
      \dodoi{10.1051/0004-6361/201935964}
    
    \bibitem[{{Mann} {et~al.}(2019){Mann}, {Dupuy}, {Kraus}, {Gaidos}, {Ansdell},
      {Ireland}, {Rizzuto}, {Hung}, {Dittmann}, {Factor}, {Feiden}, {Martinez},
      {Ru{\'\i}z-Rodr{\'\i}guez}, \& {Thao}}]{Mann2019}
    {Mann}, A.~W., {Dupuy}, T., {Kraus}, A.~L., {et~al.} 2019, \apj, 871, 63,
      \dodoi{10.3847/1538-4357/aaf3bc}
    
    \bibitem[{{Martin} \& {Lubow}(2017)}]{Martin2017}
    {Martin}, R.~G., \& {Lubow}, S.~H. 2017, \apjl, 835, L28,
      \dodoi{10.3847/2041-8213/835/2/L28}
    
    \bibitem[{{Mason} {et~al.}(2001){Mason}, {Wycoff}, {Hartkopf}, {Douglass}, \&
      {Worley}}]{wds}
    {Mason}, B.~D., {Wycoff}, G.~L., {Hartkopf}, W.~I., {Douglass}, G.~G., \&
      {Worley}, C.~E. 2001, \aj, 122, 3466, \dodoi{10.1086/323920}
    
    \bibitem[{{Mink}(2011)}]{TRES}
    {Mink}, D.~J. 2011, in Astronomical Society of the Pacific Conference Series,
      Vol. 442, Astronomical Data Analysis Software and Systems XX, ed. I.~N.
      {Evans}, A.~{Accomazzi}, D.~J. {Mink}, \& A.~H. {Rots}, 305
    
    \bibitem[{{Morton}(2015)}]{Morton2015}
    {Morton}, T.~D. 2015, {isochrones: Stellar model grid package}.
    \newblock \doeprint{1503.010}
    
    \bibitem[{{Morton} \& {Johnson}(2011)}]{mortonjohnson}
    {Morton}, T.~D., \& {Johnson}, J.~A. 2011, \apj, 738, 170,
      \dodoi{10.1088/0004-637X/738/2/170}
    
    \bibitem[{{Mugrauer}(2019)}]{Mugrauer2019}
    {Mugrauer}, M. 2019, \mnras, 490, 5088, \dodoi{10.1093/mnras/stz2673}
    
    \bibitem[{{Mugrauer} \& {Michel}(2020)}]{Mugrauer2020}
    {Mugrauer}, M., \& {Michel}, K.~U. 2020, arXiv e-prints, arXiv:2009.12234.
    \newblock \doarXiv{2009.12234}
    
    \bibitem[{{Murray} \& {Dermott}(1999)}]{MD99}
    {Murray}, C.~D., \& {Dermott}, S.~F. 1999, {Solar system dynamics}
    
    \bibitem[{{Nagayama} {et~al.}(2003){Nagayama}, {Nagashima}, {Nakajima},
      {Nagata}, {Sato}, {Nakaya}, {Yamamuro}, {Sugitani}, \& {Tamura}}]{SIRIUS}
    {Nagayama}, T., {Nagashima}, C., {Nakajima}, Y., {et~al.} 2003, in Society of
      Photo-Optical Instrumentation Engineers (SPIE) Conference Series, Vol. 4841,
      Instrument Design and Performance for Optical/Infrared Ground-based
      Telescopes, ed. M.~{Iye} \& A.~F.~M. {Moorwood}, 459--464,
      \dodoi{10.1117/12.460770}
    
    \bibitem[{{Naoz} {et~al.}(2013){Naoz}, {Farr}, {Lithwick}, {Rasio}, \&
      {Teyssandier}}]{Naoz2013}
    {Naoz}, S., {Farr}, W.~M., {Lithwick}, Y., {Rasio}, F.~A., \& {Teyssandier}, J.
      2013, \mnras, 431, 2155, \dodoi{10.1093/mnras/stt302}
    
    \bibitem[{{Naoz} {et~al.}(2012){Naoz}, {Farr}, \& {Rasio}}]{Naoz2012}
    {Naoz}, S., {Farr}, W.~M., \& {Rasio}, F.~A. 2012, \apjl, 754, L36,
      \dodoi{10.1088/2041-8205/754/2/L36}
    
    \bibitem[{{Narita} {et~al.}(2015){Narita}, {Fukui}, {Kusakabe}, {Onitsuka},
      {Ryu}, {Yanagisawa}, {Izumiura}, {Tamura}, \& {Yamamuro}}]{muscat1}
    {Narita}, N., {Fukui}, A., {Kusakabe}, N., {et~al.} 2015, Journal of
      Astronomical Telescopes, Instruments, and Systems, 1, 045001,
      \dodoi{10.1117/1.JATIS.1.4.045001}
    
    \bibitem[{{Narita} {et~al.}(2019){Narita}, {Fukui}, {Kusakabe}, {Watanabe},
      {Palle}, {Parviainen}, {Monta{\~n}{\'e}s-Rodr{\'\i}guez}, {Murgas},
      {Monelli}, {Aguiar}, {Perez Prieto}, {Oscoz}, {de Leon}, {Mori}, {Tamura},
      {Yamamuro}, {B{\'e}jar}, {Crouzet}, {Hidalgo}, {Klagyivik}, {Luque}, \&
      {Nishiumi}}]{muscat2}
    ---. 2019, Journal of Astronomical Telescopes, Instruments, and Systems, 5,
      015001, \dodoi{10.1117/1.JATIS.5.1.015001}
    
    \bibitem[{{Narita} {et~al.}(2020){Narita}, {Fukui}, {Yamamuro}, {Harbeck},
      {Bowman}, {Elphick}, {Nation}, {Armstrong}, {Han}, {Abe}, {Ikoma}, {Isogai},
      {Kawauchi}, {Kurita}, {Kusakabe}, {de Leon}, {Livingston}, {Mori},
      {Nishiumi}, {Tamura}, {Watanabe}, {Volgenau}, {Heinrich-Josties}, {Foale},
      {Daily}, {McCully}, {Kirby}, {Smith}, {Haworth}, {Conway},
      {Storrie-Lombardi}, {Rosing}, {Chatelain}, {Bachelet}, {Johnson}, \&
      {Rabus}}]{muscat3}
    {Narita}, N., {Fukui}, A., {Yamamuro}, T., {et~al.} 2020, in Society of
      Photo-Optical Instrumentation Engineers (SPIE) Conference Series, Vol. 11447,
      Society of Photo-Optical Instrumentation Engineers (SPIE) Conference Series,
      114475K, \dodoi{10.1117/12.2559947}
    
    \bibitem[{{Nelson} {et~al.}(2000){Nelson}, {Benz}, \&
      {Ruzmaikina}}]{Nelson2000}
    {Nelson}, A.~F., {Benz}, W., \& {Ruzmaikina}, T.~V. 2000, \apj, 529, 357,
      \dodoi{10.1086/308238}
    
    \bibitem[{{Newton} {et~al.}(2019){Newton}, {Mann}, {Tofflemire}, {Pearce},
      {Rizzuto}, {Vanderburg}, {Martinez}, {Wang}, {Ruffio}, {Kraus}, {Johnson},
      {Thao}, {Wood}, {Rampalli}, {Nielsen}, {Collins}, {Dragomir}, {Hellier},
      {Anderson}, {Barclay}, {Brown}, {Feiden}, {Hart}, {Isopi}, {Kielkopf},
      {Mallia}, {Nelson}, {Rodriguez}, {Stockdale}, {Waite}, {Wright}, {Lissauer},
      {Ricker}, {Vanderspek}, {Latham}, {Seager}, {Winn}, {Jenkins}, {Bouma},
      {Burke}, {Davies}, {Fausnaugh}, {Li}, {Morris}, {Mukai}, {Villase{\~n}or},
      {Villeneuva}, {De Rosa}, {Macintosh}, {Mengel}, {Okumura}, \&
      {Wittenmyer}}]{Netwon2019}
    {Newton}, E.~R., {Mann}, A.~W., {Tofflemire}, B.~M., {et~al.} 2019, \apjl, 880,
      L17, \dodoi{10.3847/2041-8213/ab2988}
    
    \bibitem[{{Ngo} {et~al.}(2016){Ngo}, {Knutson}, {Hinkley}, {Bryan}, {Crepp},
      {Batygin}, {Crossfield}, {Hansen}, {Howard}, {Johnson}, {Mawet}, {Morton},
      {Muirhead}, \& {Wang}}]{Ngo2016}
    {Ngo}, H., {Knutson}, H.~A., {Hinkley}, S., {et~al.} 2016, \apj, 827, 8,
      \dodoi{10.3847/0004-637X/827/1/8}
    
    \bibitem[{{Offner} {et~al.}(2016){Offner}, {Dunham}, {Lee}, {Arce}, \&
      {Fielding}}]{Offner2016}
    {Offner}, S. S.~R., {Dunham}, M.~M., {Lee}, K.~I., {Arce}, H.~G., \&
      {Fielding}, D.~B. 2016, \apjl, 827, L11, \dodoi{10.3847/2041-8205/827/1/L11}
    
    \bibitem[{{Offner} {et~al.}(2010){Offner}, {Kratter}, {Matzner}, {Krumholz}, \&
      {Klein}}]{Offner2010}
    {Offner}, S. S.~R., {Kratter}, K.~M., {Matzner}, C.~D., {Krumholz}, M.~R., \&
      {Klein}, R.~I. 2010, \apj, 725, 1485, \dodoi{10.1088/0004-637X/725/2/1485}
    
    \bibitem[{{Pearce} {et~al.}(2020){Pearce}, {Kraus}, {Dupuy}, {Mann}, {Newton},
      {Tofflemire}, \& {Vanderburg}}]{Pearce2020}
    {Pearce}, L.~A., {Kraus}, A.~L., {Dupuy}, T.~J., {et~al.} 2020, \apj, 894, 115,
      \dodoi{10.3847/1538-4357/ab8389}
    
    \bibitem[{{Petigura}(2015)}]{Petigura2015}
    {Petigura}, E. 2015, Phd Thesis, arXiv:1510.03902.
    \newblock \doarXiv{1510.03902}
    
    \bibitem[{{Petigura} {et~al.}(2017){Petigura}, {Howard}, {Marcy}, {Johnson},
      {Isaacson}, {Cargile}, {Hebb}, {Fulton}, {Weiss}, {Morton}, {Winn}, {Rogers},
      {Sinukoff}, {Hirsch}, \& {Crossfield}}]{Petigura2017}
    {Petigura}, E.~A., {Howard}, A.~W., {Marcy}, G.~W., {et~al.} 2017, \aj, 154,
      107, \dodoi{10.3847/1538-3881/aa80de}
    
    \bibitem[{{Petrovich}(2015)}]{Petrovich2015}
    {Petrovich}, C. 2015, \apj, 799, 27, \dodoi{10.1088/0004-637X/799/1/27}
    
    \bibitem[{{Price-Whelan} {et~al.}(2018){Price-Whelan}, {Sip{\H{o}}cz},
      {G{\"u}nther}, {Lim}, {Crawford}, {Conseil}, {Shupe}, {Craig}, {Dencheva},
      {Ginsburg}, {VanderPlas}, {Bradley}, {P{\'e}rez-Su{\'a}rez}, {de Val-Borro},
      {Paper Contributors}, {Aldcroft}, {Cruz}, {Robitaille}, {Tollerud},
      {Coordination Committee}, {Ardelean}, {Babej}, {Bach}, {Bachetti}, {Bakanov},
      {Bamford}, {Barentsen}, {Barmby}, {Baumbach}, {Berry}, {Biscani}, {Boquien},
      {Bostroem}, {Bouma}, {Brammer}, {Bray}, {Breytenbach}, {Buddelmeijer},
      {Burke}, {Calderone}, {Cano Rodr{\'\i}guez}, {Cara}, {Cardoso}, {Cheedella},
      {Copin}, {Corrales}, {Crichton}, {D{\textquoteright}Avella}, {Deil},
      {Depagne}, {Dietrich}, {Donath}, {Droettboom}, {Earl}, {Erben}, {Fabbro},
      {Ferreira}, {Finethy}, {Fox}, {Garrison}, {Gibbons}, {Goldstein}, {Gommers},
      {Greco}, {Greenfield}, {Groener}, {Grollier}, {Hagen}, {Hirst}, {Homeier},
      {Horton}, {Hosseinzadeh}, {Hu}, {Hunkeler}, {Ivezi{\'c}}, {Jain}, {Jenness},
      {Kanarek}, {Kendrew}, {Kern}, {Kerzendorf}, {Khvalko}, {King}, {Kirkby},
      {Kulkarni}, {Kumar}, {Lee}, {Lenz}, {Littlefair}, {Ma}, {Macleod},
      {Mastropietro}, {McCully}, {Montagnac}, {Morris}, {Mueller}, {Mumford},
      {Muna}, {Murphy}, {Nelson}, {Nguyen}, {Ninan}, {N{\"o}the}, {Ogaz}, {Oh},
      {Parejko}, {Parley}, {Pascual}, {Patil}, {Patil}, {Plunkett}, {Prochaska},
      {Rastogi}, {Reddy Janga}, {Sabater}, {Sakurikar}, {Seifert}, {Sherbert},
      {Sherwood-Taylor}, {Shih}, {Sick}, {Silbiger}, {Singanamalla}, {Singer},
      {Sladen}, {Sooley}, {Sornarajah}, {Streicher}, {Teuben}, {Thomas},
      {Tremblay}, {Turner}, {Terr{\'o}n}, {van Kerkwijk}, {de la Vega}, {Watkins},
      {Weaver}, {Whitmore}, {Woillez}, {Zabalza}, \& {Contributors}}]{astropy}
    {Price-Whelan}, A.~M., {Sip{\H{o}}cz}, B.~M., {G{\"u}nther}, H.~M., {et~al.}
      2018, \aj, 156, 123, \dodoi{10.3847/1538-3881/aabc4f}
    
    \bibitem[{{Raghavan} {et~al.}(2010){Raghavan}, {McAlister}, {Henry}, {Latham},
      {Marcy}, {Mason}, {Gies}, {White}, \& {ten Brummelaar}}]{Rahavan2010}
    {Raghavan}, D., {McAlister}, H.~A., {Henry}, T.~J., {et~al.} 2010, \apjs, 190,
      1, \dodoi{10.1088/0067-0049/190/1/1}
    
    \bibitem[{{Ricker} {et~al.}(2015){Ricker}, {Winn}, {Vanderspek}, {Latham},
      {Bakos}, {Bean}, {Berta-Thompson}, {Brown}, {Buchhave}, {Butler}, {Butler},
      {Chaplin}, {Charbonneau}, {Christensen-Dalsgaard}, {Clampin}, {Deming},
      {Doty}, {De Lee}, {Dressing}, {Dunham}, {Endl}, {Fressin}, {Ge}, {Henning},
      {Holman}, {Howard}, {Ida}, {Jenkins}, {Jernigan}, {Johnson}, {Kaltenegger},
      {Kawai}, {Kjeldsen}, {Laughlin}, {Levine}, {Lin}, {Lissauer}, {MacQueen},
      {Marcy}, {McCullough}, {Morton}, {Narita}, {Paegert}, {Palle}, {Pepe},
      {Pepper}, {Quirrenbach}, {Rinehart}, {Sasselov}, {Sato}, {Seager},
      {Sozzetti}, {Stassun}, {Sullivan}, {Szentgyorgyi}, {Torres}, {Udry}, \&
      {Villasenor}}]{ricker2015}
    {Ricker}, G.~R., {Winn}, J.~N., {Vanderspek}, R., {et~al.} 2015, Journal of
      Astronomical Telescopes, Instruments, and Systems, 1, 014003,
      \dodoi{10.1117/1.JATIS.1.1.014003}
    
    \bibitem[{Riddell {et~al.}(2018)Riddell, Hartikainen, Lee, riddell stan,
      Inacio, Chen, Arnold, Sutherland, Vehtari, SUZUKI, Kubo, Small, Erhardt,
      Hoover, Hoyer, Gerkin, Rings, Jackie, Ramsey, Darling, seantalts, Seabold,
      Shron, Brannigan, Foreman, Veron, Salwen, H, \& Rudiuk}]{pystan}
    Riddell, A., Hartikainen, A., Lee, D., {et~al.} 2018, stan-dev/pystan:
      v2.18.0.0, v2.18.0.0,  Zenodo, \dodoi{10.5281/zenodo.1456206}
    
    \bibitem[{{Santerne} {et~al.}(2012){Santerne}, {D{\'\i}az}, {Moutou}, {Bouchy},
      {H{\'e}brard}, {Almenara}, {Bonomo}, {Deleuil}, \& {Santos}}]{santerne2012}
    {Santerne}, A., {D{\'\i}az}, R.~F., {Moutou}, C., {et~al.} 2012, \aap, 545,
      A76, \dodoi{10.1051/0004-6361/201219608}
    
    \bibitem[{{Schwab} {et~al.}(2010){Schwab}, {Spronck}, {Tokovinin}, \&
      {Fischer}}]{CHIRON}
    {Schwab}, C., {Spronck}, J. F.~P., {Tokovinin}, A., \& {Fischer}, D.~A. 2010,
      in Society of Photo-Optical Instrumentation Engineers (SPIE) Conference
      Series, Vol. 7735, Ground-based and Airborne Instrumentation for Astronomy
      III, 77354G, \dodoi{10.1117/12.856709}
    
    \bibitem[{{Sigalotti} {et~al.}(2018){Sigalotti}, {Cruz}, {Gabbasov}, {Klapp},
      \& {Ram{\'\i}rez-Velasquez}}]{Sigalotti2018}
    {Sigalotti}, L. D.~G., {Cruz}, F., {Gabbasov}, R., {Klapp}, J., \&
      {Ram{\'\i}rez-Velasquez}, J. 2018, \apj, 857, 40,
      \dodoi{10.3847/1538-4357/aab619}
    
    \bibitem[{{Siverd} {et~al.}(2017){Siverd}, {Brown}, {Henderson}, {Hygelund},
      {Barnes}, {Bowman}, {De Vera}, {Eastman}, {Kirby}, {Norbury}, {Smith},
      {Taylor}, {Tufts}, \& {Van Eyken}}]{NRES}
    {Siverd}, R., {Brown}, T.~M., {Henderson}, T., {et~al.} 2017, in American
      Astronomical Society Meeting Abstracts, Vol. 230, American Astronomical
      Society Meeting Abstracts \#230, 102.07
    
    \bibitem[{Siverd {et~al.}(2018)Siverd, Brown, Barnes, Bowman, Vera, Foale,
      Harbeck, Henderson, Hygelund, Kirby, McCully, Nation, Smith, Taylor, \&
      Tufts}]{NRES2}
    Siverd, R.~J., Brown, T.~M., Barnes, S., {et~al.} 2018, in Ground-based and
      Airborne Instrumentation for Astronomy VII, ed. C.~J. Evans, L.~Simard, \&
      H.~Takami, Vol. 10702, International Society for Optics and Photonics (SPIE),
      1918 -- 1929, \dodoi{10.1117/12.2312800}
    
    \bibitem[{{Skrutskie} {et~al.}(2006){Skrutskie}, {Cutri}, {Stiening},
      {Weinberg}, {Schneider}, {Carpenter}, {Beichman}, {Capps}, {Chester},
      {Elias}, {Huchra}, {Liebert}, {Lonsdale}, {Monet}, {Price}, {Seitzer},
      {Jarrett}, {Kirkpatrick}, {Gizis}, {Howard}, {Evans}, {Fowler}, {Fullmer},
      {Hurt}, {Light}, {Kopan}, {Marsh}, {McCallon}, {Tam}, {Van Dyk}, \&
      {Wheelock}}]{twomass}
    {Skrutskie}, M.~F., {Cutri}, R.~M., {Stiening}, R., {et~al.} 2006, \aj, 131,
      1163, \dodoi{10.1086/498708}
    
    \bibitem[{{Spada} {et~al.}(2013){Spada}, {Demarque}, {Kim}, \&
      {Sills}}]{Spada2013}
    {Spada}, F., {Demarque}, P., {Kim}, Y.~C., \& {Sills}, A. 2013, \apj, 776, 87,
      \dodoi{10.1088/0004-637X/776/2/87}
    
    \bibitem[{{Stassun} {et~al.}(2018){Stassun}, {Corsaro}, {Pepper}, \&
      {Gaudi}}]{stassun2018}
    {Stassun}, K.~G., {Corsaro}, E., {Pepper}, J.~A., \& {Gaudi}, B.~S. 2018, \aj,
      155, 22, \dodoi{10.3847/1538-3881/aa998a}
    
    \bibitem[{{Stassun} {et~al.}(2019{\natexlab{a}}){Stassun}, {Oelkers},
      {Paegert}, {Torres}, {Pepper}, {De Lee}, {Collins}, {Latham}, {Muirhead},
      {Chittidi}, {Rojas-Ayala}, {Fleming}, {Rose}, {Tenenbaum}, {Ting}, {Kane},
      {Barclay}, {Bean}, {Brassuer}, {Charbonneau}, {Ge}, {Lissauer}, {Mann},
      {McLean}, {Mullally}, {Narita}, {Plavchan}, {Ricker}, {Sasselov}, {Seager},
      {Sharma}, {Shiao}, {Sozzetti}, {Stello}, {Vanderspek}, {Wallace}, \&
      {Winn}}]{TIC}
    {Stassun}, K.~G., {Oelkers}, R.~J., {Paegert}, M., {et~al.} 2019{\natexlab{a}},
      \aj, 158, 138, \dodoi{10.3847/1538-3881/ab3467}
    
    \bibitem[{{Stassun} {et~al.}(2019{\natexlab{b}}){Stassun}, {Oelkers},
      {Paegert}, {Torres}, {Pepper}, {De Lee}, {Collins}, {Latham}, {Muirhead},
      {Chittidi}, {Rojas-Ayala}, {Fleming}, {Rose}, {Tenenbaum}, {Ting}, {Kane},
      {Barclay}, {Bean}, {Brassuer}, {Charbonneau}, {Ge}, {Lissauer}, {Mann},
      {McLean}, {Mullally}, {Narita}, {Plavchan}, {Ricker}, {Sasselov}, {Seager},
      {Sharma}, {Shiao}, {Sozzetti}, {Stello}, {Vanderspek}, {Wallace}, \&
      {Winn}}]{Stassun2019}
    ---. 2019{\natexlab{b}}, \aj, 158, 138, \dodoi{10.3847/1538-3881/ab3467}
    
    \bibitem[{{Steinmetz}(2003)}]{Steinmetz2003}
    {Steinmetz}, M. 2003, in Astronomical Society of the Pacific Conference Series,
      Vol. 298, GAIA Spectroscopy: Science and Technology, ed. U.~{Munari}, 381.
    \newblock \doarXiv{astro-ph/0211417}
    
    \bibitem[{{Sterne}(1939)}]{Sterne1939}
    {Sterne}, T.~E. 1939, \mnras, 99, 451, \dodoi{10.1093/mnras/99.5.451}
    
    \bibitem[{{Taylor}(2006)}]{Taylor2006}
    {Taylor}, M.~B. 2006, in Astronomical Society of the Pacific Conference Series,
      Vol. 351, Astronomical Data Analysis Software and Systems XV, ed.
      C.~{Gabriel}, C.~{Arviset}, D.~{Ponz}, \& S.~{Enrique}, 666
    
    \bibitem[{{Telting} {et~al.}(2014){Telting}, {Avila}, {Buchhave}, {Frandsen},
      {Gandolfi}, {Lindberg}, {Stempels}, {Prins}, \& {NOT staff}}]{Telting2014}
    {Telting}, J.~H., {Avila}, G., {Buchhave}, L., {et~al.} 2014, Astronomische
      Nachrichten, 335, 41, \dodoi{10.1002/asna.201312007}
    
    \bibitem[{{Tobin} {et~al.}(2013){Tobin}, {Chandler}, {Wilner}, {Looney},
      {Loinard}, {Chiang}, {Hartmann}, {Calvet}, {D'Alessio}, {Bourke}, \&
      {Kwon}}]{Tobin2013}
    {Tobin}, J.~J., {Chandler}, C.~J., {Wilner}, D.~J., {et~al.} 2013, \apj, 779,
      93, \dodoi{10.1088/0004-637X/779/2/93}
    
    \bibitem[{{Tobin} {et~al.}(2016){Tobin}, {Looney}, {Li}, {Chandler}, {Dunham},
      {Segura-Cox}, {Sadavoy}, {Melis}, {Harris}, {Kratter}, \&
      {Perez}}]{Tobin2016}
    {Tobin}, J.~J., {Looney}, L.~W., {Li}, Z.-Y., {et~al.} 2016, \apj, 818, 73,
      \dodoi{10.3847/0004-637X/818/1/73}
    
    \bibitem[{{Tokovinin}(2017)}]{Tokovinin2017}
    {Tokovinin}, A. 2017, \mnras, 468, 3461, \dodoi{10.1093/mnras/stx707}
    
    \bibitem[{{Tokovinin} {et~al.}(2013){Tokovinin}, {Fischer}, {Bonati},
      {Giguere}, {Moore}, {Schwab}, {Spronck}, \& {Szymkowiak}}]{Tokovinin2013}
    {Tokovinin}, A., {Fischer}, D.~A., {Bonati}, M., {et~al.} 2013, \pasp, 125,
      1336, \dodoi{10.1086/674012}
    
    \bibitem[{{{\v{C}}otar} {et~al.}(2019){{\v{C}}otar}, {Zwitter}, {Kos},
      {Munari}, {Martell}, {Asplund}, {Bland-Hawthorn}, {Buder}, {de Silva},
      {Freeman}, {Sharma}, {Anguiano}, {Carollo}, {Horner}, {Lewis}, {Nataf},
      {Nordlander}, {Stello}, {Ting}, {Tinney}, {Traven}, {Wittenmyer}, \& {Galah
      Collaboration}}]{cotar2019}
    {{\v{C}}otar}, K., {Zwitter}, T., {Kos}, J., {et~al.} 2019, \mnras, 483, 3196,
      \dodoi{10.1093/mnras/sty3155}
    
    \bibitem[{{von Zeipel}(1910)}]{Vonzeipel1910}
    {von Zeipel}, H. 1910, Astronomische Nachrichten, 183, 345,
      \dodoi{10.1002/asna.19091832202}
    
    \bibitem[{Wallraff(1979)}]{Wallraff1979}
    Wallraff, H.~G. 1979, Behavioral Ecology and Sociobiology, 5, 201,
      \dodoi{10.1007/BF00293306}
    
    \bibitem[{Watson(1983)}]{watson1983statistics}
    Watson, G. 1983, Statistics on Spheres, Canadian Mathematical Society Series of
      Monographs and Advanced Texts (Wiley).
    \newblock \url{https://books.google.com/books?id=tBjvAAAAMAAJ}
    
    \bibitem[{{Weiss} {et~al.}(2017){Weiss}, {Deck}, {Sinukoff}, {Petigura},
      {Agol}, {Lee}, {Becker}, {Howard}, {Isaacson}, {Crossfield}, {Fulton},
      {Hirsch}, \& {Benneke}}]{Weiss2017}
    {Weiss}, L.~M., {Deck}, K.~M., {Sinukoff}, E., {et~al.} 2017, \aj, 153, 265,
      \dodoi{10.3847/1538-3881/aa6c29}
    
    \bibitem[{{Wu} \& {Murray}(2003)}]{Wu2003}
    {Wu}, Y., \& {Murray}, N. 2003, \apj, 589, 605, \dodoi{10.1086/374598}
    
    \bibitem[{{Zanazzi} \& {Lai}(2018)}]{Zanazzi2018}
    {Zanazzi}, J.~J., \& {Lai}, D. 2018, \mnras, 477, 5207,
      \dodoi{10.1093/mnras/sty951}
    
    \bibitem[{{Zhao} {et~al.}(2012){Zhao}, {Zhao}, {Chu}, {Jing}, \&
      {Deng}}]{Zhao2012}
    {Zhao}, G., {Zhao}, Y.-H., {Chu}, Y.-Q., {Jing}, Y.-P., \& {Deng}, L.-C. 2012,
      Research in Astronomy and Astrophysics, 12, 723,
      \dodoi{10.1088/1674-4527/12/7/002}
    
    \bibitem[{{Ziegler} {et~al.}(2020){Ziegler}, {Tokovinin}, {Brice{\~n}o},
      {Mang}, {Law}, \& {Mann}}]{Ziegler2020}
    {Ziegler}, C., {Tokovinin}, A., {Brice{\~n}o}, C., {et~al.} 2020, \aj, 159, 19,
      \dodoi{10.3847/1538-3881/ab55e9}
    
    \bibitem[{{Ziegler} {et~al.}(2021){Ziegler}, {Tokovinin}, {Latiolais},
      {Briceno}, {Law}, \& {Mann}}]{Ziegler2021}
    {Ziegler}, C., {Tokovinin}, A., {Latiolais}, M., {et~al.} 2021, arXiv e-prints,
      arXiv:2103.12076.
    \newblock \doarXiv{2103.12076}
    
    \end{thebibliography}
\end{document}